%% file: text_amiel.tex
\documentclass[twocolumn,twocolappendix]{aastex631}
\usepackage[utf8]{inputenc}
\usepackage{float}
\usepackage{amssymb,natbib,graphicx,color,ctable, amsmath,array}
\usepackage{hyperref}
\usepackage{mathtools}
\hypersetup{
    colorlinks=true,
    linkcolor=blue,
    citecolor=blue,
    filecolor=blue,
    urlcolor=blue
}
\usepackage[flushleft]{threeparttable} 
\usepackage{soul}

\usepackage[mathlines]{lineno}

\defcitealias{Bregman2022}{B+22}
\defcitealias{Das2023}{D+23}
\defcitealias{Planck2013}{{P+13}}
\defcitealias{Faerman2017}{FSM17}
\defcitealias{Faerman2020}{FSM20}
\defcitealias{Zhang2024}{Z+24}

\newcommand{\apth}      {a_{P,\rm th}}
\newcommand{\beq}       {\begin{equation}}
\newcommand{\eeq}       {\end{equation}}
\newcommand{\cgm}        {{\rm CGM}}
\newcommand{\cv}        {{\rm CGM,vir}}
\newcommand{\msun}     {M$_\odot$}
\newcommand{\msunm}    {{\rm M}_\odot}
\newcommand{\vir}       {{\rm vir}}
\def\cmc{\ifmmode {\rm cm^{-2}} \else $\rm cm^{-2}$\fi}
\def\cmv{\ifmmode {\rm cm^{-3} \:} \else $\rm cm^{-3}$\fi}

\newcommand{\oran}[1]{{\textcolor{black}{#1}}}

\shorttitle{SZ Effect in $L^*$ Galaxies}
\shortauthors{Oren et al.}

\graphicspath{{./}{figures/}}

\begin{document}

\title{Sunyaev-Zeldovich Signals from $L^*$ Galaxies: Observations, Analytics, and Simulations}

\author{Yossi Oren} 
\thanks{Email: orenyossi01@gmail.com}
\affiliation{School of Physics and Astronomy, Tel Aviv University, Ramat Aviv 69978, Israel}

\author{Amiel Sternberg}
\affiliation{School of Physics and Astronomy, Tel Aviv University, Ramat Aviv 69978, Israel}
\affiliation{Center for Computational Astrophysics, Flatiron Institute, 162 5th Ave., New York, NY, 10010, USA}
\affiliation{Max-Planck-Institut f\"ur extraterrestrische Physik (MPE), Giessenbachstr., 85748 Garching, Germany}

\author{Christopher F. McKee}
\affiliation{Department of Physics, University of California, Berkeley, CA 94720, USA}
\affiliation{Department of Astronomy, University of California, Berkeley, CA 94720, USA}

\author{Yakov Faerman}
\affiliation{Astronomy Department, University of Washington, Seattle, WA 98195, USA}

\author{Shy Genel}
\affiliation{Center for Computational Astrophysics, Flatiron Institute, 162 5th Ave., New York, NY, 10010, USA}
\affiliation{Columbia Astrophysics Laboratory, Columbia University, 550 West 120th Street, New York, NY 10027, USA}

\begin{abstract}

We analyze measurements of the thermal Sunyaev-Zeldovich (tSZ) effect arising in the circumgalactic medium (CGM) of $L^*$ galaxies, reported by \cite{Bregman2022} and \cite{Das2023}. In our analysis we use the \cite{Faerman2017,Faerman2020} CGM models, a new power-law model (PLM), and the TNG100 simulation. For a given $M_{\rm vir}$, our PLM has
four parameters; the fraction, $f_{\rm hCGM}$, of the halo baryon mass in hot CGM gas, the ratio, $\phi_T$, of the actual gas temperature at the virial radius to the virial temperature, and the power-law indices, $a_{P,{\rm th}}$ and $a_n$ for the thermal electron pressure and the hydrogen nucleon density. The \citetalias{Bregman2022} Compton-$y$ profile implies steep electron pressure slopes ($a_{P,{\rm th}}\simeq 2$). For isothermal conditions the temperature is at least $1.1\times 10^6$~K, with a hot CGM gas mass of up to $3.5\times 10^{11}$~M$_\odot$ for a virial mass of $2.75\times 10^{12}$~M$_\odot$. However, if isothermal the gas must be expanding out of the halos. An isentropic equation of state is favored for which hydrostatic equilibrium is possible. 
The \citetalias{Bregman2022} and \citetalias{Das2023} results are consistent with each other and with recent (0.5-2 keV) CGM X-ray observations \citep{Zhang2024} of Milky Way mass systems. 
For $M_{\rm vir}\simeq 3\times 10^{12}$~M$_\odot$, the scaled Compton pressure integrals, $E(z)^{-2/3}Y_{500}/M_{\rm vir,12}^{5/3}$, lie in the narrow range, $2.5\times 10^{-4}$ to $5.0\times 10^{-4}$~kpc$^2$, for all three sets of observations. 
TNG100 underpredicts the tSZ parameters by factors $\sim 0.5$ dex for the $L^*$ galaxies, suggesting that the feedback strengths and CGM gas losses are overestimated in the simulated halos at these mass scales.

\end{abstract}


\section{Introduction} \label{sec:intro}

The thermal Sunyaev-Zedovich (tSZ) effect 
\citep{Sunyaev1972}, in which the cosmic microwave background (CMB) radiation field is distorted by inverse Compton scattering, is a powerful diagnostic probe of the hot gas properties within galaxy clusters, and in the circumgalactic medium (CGM) surrounding individual galaxies \citep{Birkinshaw1999,carlstrom2002,Mroczkowski2019}. The tSZ effect is most readily observed, from the ground or from space, in massive X-ray luminous clusters where the scattering path lengths are large, and the electron temperatures and pressures are high \citep{Hasselfield2013,Bleem2015,Kitayama2016}.

\citet[hereafter \citetalias{Bregman2022}]{Bregman2022} recently reported weak but  statistically significant detections of tSZ signals in a sample of 12 nearby 
($\lesssim 10$~Mpc) $L^*$ galaxies, i.e.~at Milky Way 
scales, in a reanalysis of WMAP and {\it Planck}
{satellite data \citep[][hereafter \citetalias{Planck2013}]{Bennett2013,Planck2013}. 
\citetalias{Bregman2022} constructed multi-channel tSZ distortion maps, and derived radial profiles for the line-of-sight Compton-$y$ and volume integrated $Y$ parameters for the extended hot gas halos around the individual galaxies in their set.  They also constructed a galaxy stack for which the Compton tSZ parameters are more robustly inferred.  The measurements provide constraints on the electron pressures and gas masses, and \citetalias{Bregman2022} conclude that a significant fraction of the expected cosmic ``baryonic" mass within the halos is distributed through the CGM surrounding the inner galaxies. This conclusion is broadly consistent with {\oran{ high-ionization}} heavy-element absorption line observations of many similar systems \citep{Werk14,Tumlinson2017,Qu2018}.

\citet[hereafter \citetalias{Das2023}]{Das2023} have also presented results for tSZ detections in $L^*$ galaxies. \citetalias{Das2023} constructed Compton-$y$ maps for two patches of sky (456 and 1633 deg$^2$) combining \citetalias{Planck2013} data with higher resolution ground-based {\it Atacama Cosmology Telecope} (ACT) measurements \citep{Aiola2020}. They then cross-correlated the tSZ signals with galaxies in the infrared/optical WISE$\times$SuperCOSMOS photometric redshift catalog \citep{Bilicki2016} enabling estimates of the $y$ and $Y$ parameters for galaxy stacks spanning stellar masses from $2\times 10^{11}$ down to $6\times 10^9$~M$_\odot$.  

In this paper, we use the analytic models {\oran{for hot CGM gas}} developed in \citet{Faerman2017,Faerman2020}, hereafter 
\citetalias{Faerman2017} and \citetalias{Faerman2020}, and the TNG100 cosmological hydrodynamical simulation from the IllustrisTNG project \citep{Pillepich2018_2, Marinacci2018, Naiman2018, Nelson2018, Springel2018}, to study and interpret the \citetalias{Bregman2022} and \citetalias{Das2023} tSZ data.
For TNG100 we compute in postprocessing the Compton-$y$ and $Y$ parameters
for the wide range of halo virial masses found in the simulation. This enables us to also compare the predictions to the more massive systems observed by \citetalias{Planck2013}.

To facilitate our analysis we develop a power-law model (PLM) for the CGM hot gas content and density and pressure distributions as functions of virial mass, and we write down the scaling relations for $y$ and $Y$. 

We also discuss recent eROSITA X-ray detections of the hot CGM surrounding $L^*$ galaxies \citet[hereafter \citetalias{Zhang2024}]{Zhang2024} in a comparison to the observed \citetalias{Bregman2022} and \citetalias{Das2023} tSZ measurements that probe similar hot CGM environments.

The structure of our paper is as follows.
In \S~\ref{sec:fsm} we review the \citetalias{Faerman2017} and \citetalias{Faerman2020} models and introduce a parameter combination that we refer to as ``FSM20-maximal" (motivated by the \citetalias{Bregman2022} and \citetalias{Das2023} data). In \S~\ref{sec:PLMprofiles} we present our power-law model for the hot CGM gas. In \S~\ref{sec:CGMthermal} we discuss observational constraints on the cool, intermediate, and hot CGM gas fractions in $L^*$ galaxies.  In \S~\ref{sec:tszplm} we apply the PLM to the tSZ effect. {\oran{In \S~\ref{sec:PLMFSM} we fit PLMs to the \citetalias{Faerman2017} and \citetalias{Faerman2020} density, pressure, and Compton-$y$ and $Y$ profiles, and we describe our fitting procedure.}} In \S~\ref{sec:Bregman} we summarize the \citetalias{Bregman2022} data and results. {\oran{In \S~\ref{sec:plm-b22} and \S~\ref{sec:HSE} we fit the \citetalias{Bregman2022} {\oran{Compton-$y$}} data with our PLM and constrain the underlying pressure profiles}}, and in \S~\ref{sec:fsm-b22} we compare the data to the three FSM models. In \S~\ref{sec:tng} we describe how we use the TNG100 simulation outputs to generate Compton-$y$ maps, radial $y$ profiles, and $Y$ integrals, for a wide range of halo virial masses. In \S~\ref{subsec:maps} we introduce our TNG100 halo sample, in \S~\ref{sec:fracPLM} we parameterize the TNG100 CGM structures using the PLM, and in \S~\ref{sec:twoymaps} we construct TNG100 $y$-maps and profiles. In \S~\ref{sec:comparison} we compare our TNG100 computations to the \citetalias{Bregman2022}, \citetalias{Das2023}, and \citetalias{Planck2013} observations, and to our analytic FSM predictions. We also discuss the \citetalias{Zhang2024} X-ray observations in context of the tSZ measurements. We discuss and summarize our results in \S~\ref{sec:discussion}.

\section{The FSM Models} \label{sec:fsm}

\citetalias{Faerman2017} and \citetalias{Faerman2020}
presented two analytic phenomenological (forward) models for the {\oran{hot gas in the}} CGM around star-forming galaxies at the mass scale of the Milky Way. In these models we approximated the halo virial mass, $M_{\rm vir}$, as $M_{100}$, i.e.~the mass within which the mean over-density is 100, since \citet{Bryan1998} found this is very close to the density of a virialized halo at redshift $z=0$. For the Milky Way we adopted $M_{\rm vir}=M_{100}=1\times 10^{12}$~M$_\odot$. We note that \citet{Posti19} found $M_\vir=(1.3\pm0.3)\times 10^{12}$~\msun\ for the Milky Way.

The basic simplifying assumption of the FSM models is that the initially hot CGM gas is in hydrostatic equilibrium (HSE) in the dominating gravitational potential of the dark-matter halo and stellar disk. (The CGM gas is not self-gravitating.) {\oran{In the FSM models the CGM extends from an inner radius, $r_0=8.5$~kpc, comparable to twice the half-mass radius, $r_e$, of the Milky Way stellar disk \citep{BHG16}, out to near the virial radius.}} The modeled properties include the hot CGM gas mass, its spatial distribution, and the temperatures, metallicities, {\oran{dispersion measures}}, and ionization states of the gas. In addition to thermal pressure, there is (kinetic) pressure associated with turbulent motions, and non-thermal pressure due possibly to cosmic-rays and magnetic fields, all of which support the gas against gravity. The focus is on 
$T\gtrsim 10^5$~K gas, near to the characteristic virial temperature, and as constrained by UV/Xray spectral line absorptions and emissions of highly ionized species, mainly OVI, OVII, and OVIII, in the Galactic halo and around other galaxies.

The parameters and main outputs of the fiducial FSM models are summarized in Table~\ref{tab:tb_fsm}, along with a comparison to measured values of gas  densities, dispersion measures, and oxygen columns.

\input{tb_fsm.tex}

In any model (or simulation), a major challenge is to account for the simultaneous presence of the multiple ionization states observed within the CGM volume \citep{Faerman2023}. In \citetalias{Faerman2017} we developed a model in which hot gas with {\oran{a constant volume averaged mean temperature of $1.5\times 10^6$~K}} is hydrostatically supported, and gives rise to collisionally ionized OVII and OVIII. A separate 
$\sim 3.0\times 10^5$~K OVI component then condenses out of the hot gas via cooling instabilities in turbulent, isobaric density fluctuations.  The initial hot gas component is {\oran{ in HSE}} at a constant mean temperature throughout, and we refer to \citetalias{Faerman2017} as our ``isothermal model". Following the cooling out of the OVI component the thermal structure of the remaining hot gas is altered slightly, and it is no longer precisely isothermal. {\oran{Furthermore, the OVI component may not be isothermal nor is it precisely in HSE.}

{\oran{In the fiducial \citetalias{Faerman2017} model the hot CGM gas mass, $M_{\rm hCGM}=1.2 \times 10^{11}$~M$_\odot$ within the virial radius, corresponding to a hot gas fraction $f_{\rm hCGM}=0.77$ relative to the halo cosmic baryon budget of $1.56\times 10^{11}$~M$_\odot$ for a virial mass of $10^{12}$~M$_\odot$. For the Milky Way mass of $\sim 6.3\times 10^{10}$~M$_\odot$ \citep{Licquia2016_MW_M_star}, including stellar disk, bulge, and ISM, the fiducial \citetalias{Faerman2017} model exceeds the cosmic baryon budget by $\sim 20$\%.}}

In \citetalias{Faerman2020} we presented an alternate model in which we assumed a barotropic equation of state for the CGM gas. We refer to \citetalias{Faerman2020} as our ``isentropic model". We assumed that accreting intergalactic gas is heated by the virial shock to temperatures $\sim 3\times 10^5$~K as appropriate for a $10^{12}$~M$_\odot$ halo, and close to the OVI production peak for collisional ionization. The {\it Ansatz} is that feedback heating (by AGN and/or stellar feedback) drives the CGM gas to a convective equilibrium at constant entropy. We again assumed hydrostatic balance, with the (necessary) inclusion of non-thermal cosmic-ray and magnetic support. Due to the adiabatic compression at small radii, the gas is hotter closer to the inner galaxy. In this picture, the OVI extends out to the virial radius, as observed (and as in our isothermal model) but the OVII and OVIII, for which the Milky Way spatial distributions are as yet unknown, are instead produced in the inner parts of the CGM. In \citetalias{Faerman2020} (isentropic) the 
CGM is a single adiabatic structure with a temperature gradient. In \citetalias{Faerman2017} (isothermal) the 
OVI gas reflects a thermal instability and is fully mixed within the hotter OVII/OVIII medium.
In \citetalias{Faerman2020} we also included a radially declining metallicity distribution to account for possible enrichment by stellar winds and supernovae.  
{\oran{In \citetalias{Faerman2020} the outer boundary of the CGM is set at 1.1 times the virial radius. The inferred hot gas mass within the virial radius is $M_{\rm hCGM}=4.6 \times 10^{10}$~M$_\odot$, or $f_{\rm hGCM}=0.29$,}}
a factor $\approx 2$ lower than in \citetalias{Faerman2017}.
In \citetalias{Faerman2020} metagalactic photoionization of the oxygen ions plays a role, in addition to collisional ionization, because the gas densities are somewhat lower, as low as $\sim 10^{-5}$~cm$^{-3}$ near the virial radius.

For the Milky Way, and for similar star-forming $L^*$ galaxies, the \citetalias{Faerman2017} and \citetalias{Faerman2020} hot CGM gas masses are sufficient to account for the ``missing baryons” in such galaxies, and to power star-formation for longer than a Hubble time.}} The overall CGM gas metallicity must be high, $\gtrsim 0.5$ solar, indicative of significant enrichment.

In Table~\ref{tab:tb_fsm} we also list parameters and output quantities for a 
modified version of our \citetalias{Faerman2020} fiducial model in which the hot CGM gas mass within the virial radius is doubled to $9.2\times 10^{10}$~M$_\odot$, {\oran{so that $f_{\rm hCGM}=0.59$}}, and the overall baryon content within the halo (including the central galaxy) is {\oran{to within $\sim 1$\%}} cosmologically maximal. As we discuss further below, this is motivated by the high electron pressures indicated by the \citetalias{Bregman2022} measurements, and we refer to this version as FSM20-maximal. In it the metallicity is reduced by a corresponding factor to keep the predicted hot gas absorption/emission line constraints within the observed range. For our FSM20-maximal model there is no room for cool or intermediate temperature CGM gas without exceeding the halo cosmic baryon fraction. 

In \citetalias{Faerman2020} we computed the predicted 
Compton-$y$ and $Y$ parameters
for the fiducial CGM model (see Fig.~15 of that paper). In this paper we also present and discuss the predictions for the Compton-$y$ and $Y$ parameters in the \citetalias{Faerman2017} and FSM20-maximal models.

 An important finding of the FSM models is that the resulting hydrostatic CGM pressure profiles within the dark-matter halos are well approximated by simple radial power-laws, {\oran{from the assumed inner boundary at $8.5$~kpc}} near to the central galaxy, through the entire CGM, out to the virial radius.
 Both the total hydrostatic pressures and just the thermal pressures alone are well represented by power-laws in the FSM models.
 This behavior motivates the power-law model that we now describe.

\section{Power-Law Model} \label{sec:PLM}

\subsection{Pressure and density profiles}
\label{sec:PLMprofiles}

The thermal Sunyaev-Zeldovich (tSZ) distortions produced by the CGM of a galaxy depend primarily on the electron pressure profiles of the hot gas responsible for the Compton scattering \citep{Sunyaev1972,Reid2006,Nagai2007,Arnaud2010,Battaglia2012}.
In our power-law model (PLM)
we represent the hot component of the CGM, and the associated tSZ Compton parameters, assuming power-law distributions for the thermal electron pressures and 
{\oran{gas}} densities. Our PLM is motivated by our more detailed and observationally based FSM analytics, as well as the TNG100 cosmological simulations, as we discuss in \S~\ref{sec:tng}.

For a galaxy halo with virial mass $M_{\rm vir}$, the virial radius at any redshift $z$ is 
\beq
\begin{split}
    R_{\rm vir} \  & \equiv \  \bigl(\frac{4\pi}{3} \Delta \rho_c\bigr)^{-1/3}M_{\rm vir}^{1/3} \\
    & = \ 267 \ \Bigl(\frac{M_{\rm vir}}{10^{12} \ \msunm}\Bigr)^{1/3} E^{-2/3}(z) \ {\rm kpc} \ \ \ ,
    \label{eq:Rvir}
\end{split}
\eeq
where $\rho_c$ is the cosmological critical density, and $\Delta$ is the halo over-density factor. For the 
standard $\Lambda$CDM cosmology \citep{Planck2020b}
$\rho_{c,0}=8.53\times 10^{-30}\ {\rm g~cm^{-3}}$  at redshift $z=0$. The factor $E^2(z)\equiv \rho_c/\rho_{c,0} = \Omega_{\rm m}(1+z)^3+\Omega_\Lambda$ accounts for the increase in the critical density with redshift. We assume an overdensity factor $\Delta = 100$, which is appropriate for virialized halos at $z=0$ \citep{Bryan1998}.

For the hot gas electron pressure we write
\begin{equation}
    \frac{P_{e}(r)}{k_{\rm B}}  
     =  x_e n_{{\rm H,vir}}T(R_\vir)
    \left(\frac{r}{R_{\rm vir}}\right)^{-a_{P,{\rm th}}}
     \label{eq:epress}
\end{equation}
where $r$ is the radial distance from the central galaxy,
$x_e\equiv n_e/n_{\rm H}$ is the electron abundance, $n_{\rm H}$ is the hydrogen gas density, $n_{\rm H,vir}$ and $T(R_\vir)$ are the hot gas hydrogen density and temperature at $R_{\rm vir}$, and $k_{\rm B}$ is the Boltzmann constant. 
In this expression, $a_{P,{\rm th}}$ is the power-law index for the thermal pressure, and it is a basic free parameter.
We assume a constant metallicity throughout, with a helium abundance, $n_{\rm He}/n_{\rm H}=0.085$ \cite{Asplund2009}. For a fully ionized gas, $x_e=1.17$.

Following \citetalias{Faerman2020}, we define the characteristic virial temperature of the halo as
\begin{equation}
\begin{split}
    T_{\rm vir} & \ \equiv \ \frac{1}{3}\left(\frac{G {\bar m}}{k_{\rm B}}\right)\frac{M_{\rm vir}}{R_{\rm vir}},\\
     &=  \left(\frac{4\pi}{3}\Delta \rho_c\right)^{1/3}\left(\frac{G {\bar m}}{3k_{\rm B}}\right)M_{\rm vir}^{2/3},\\
     & =3.90\times 10^5\Bigl(\frac{M_{\rm vir}}{10^{12}\ \msunm}\Bigr)^{2/3} E^{2/3}(z)\ \ {\rm K} \\
     & = 3.90\times 10^5\Bigl(\frac{R_{\rm vir}}{267 \ {\rm kpc}}\Bigr)^2 E^2(z) \ \ \ {\rm K} \ ,
\end{split}
\label{eq:Tvir}
\end{equation}
where ${\bar m}=0.60m_p=1.00\times 10^{-24}$~g is the mean mass per gas particle (ions plus electrons), 
$m_p$ is the proton mass, and $G$ is the gravitational constant. The virial temperature is proportional to the surface area of the dark matter halo. 

{\oran{In the FSM models and TNG100 simulations, as well as for our PLM representations, the actual gas temperatures at the virial radii, $T({R_{\rm vir}})$, are not necessarily equal to the characteristic virial temperatures $T_{\rm vir}$ of the halos.}} We define the ratio of these temperatures as the parameter $\phi_T$,
\beq
T(R_\vir)\equiv \phi_T T_\vir \ \ \ .
\label{eq:phit}
\eeq

We express the total hot CGM gas mass within the virial radius as a fraction $f_{\rm hCGM}$ (nominally 50\%) of the total baryonic mass within the halo--i.e.,
\begin{equation}
\begin{split}
    M_{\rm hCGM,vir} &=  f_{\rm hCGM}\left(\frac{\Omega_b}{\Omega_m}\right) M_{\rm vir}\\
    &= 7.80\times 10^{10}\ \left(\frac{f_{\rm hCGM}}{0.5}\right)\Bigl(\frac{M_{\rm vir}}{10^{12} \ \msunm}\Bigr) \ \ {\rm M_\odot} \ \ \ ,
    \label{eq:Mcgm}
\end{split}
\end{equation}
where ${\Omega_b}/{\Omega_m}=0.156$ \citep{Planck2020b}
is the cosmic baryon fraction relative to the total matter content (dark plus baryonic). 

We also assume that the gas density alone obeys a power-law distribution, 
\begin{equation}
    n_{\rm H}(r) \ = \ n_{\rm H, vir}\Bigl(\frac{r}{R_{\rm vir}}\Bigr)^{-a_n} \ \ \ ,
    \label{eq:nH}
\end{equation}
{\oran{so that for $a_n<3$ the enclosed hot CGM gas mass is
\begin{equation}
M_{\rm h}(r) \ = \ \Bigl(\frac{r}{R_{\rm vir}}\Bigr)^{3-a_n}M_{\rm hCGM,vir} \ \ \ .
\label{eq:mass}
\end{equation}
}}Together with equations~(\ref{eq:epress}) and (\ref{eq:phit}), the gas temperature is then
\begin{equation}
    T(r) \ = \ \phi_T T_{\rm vir}\Bigl(\frac{r}{R_{\rm vir}}\Bigr)^{-a_T} \ \ \ ,
\end{equation}
with $a_T=a_{P,{\rm th}}-a_n$.
If ${\bar n_{\rm H}}$ is the mean {\it{hot}} CGM gas density within the halo, then
\begin{equation}
\begin{split}
    n_{\rm H, vir} \ & = \ \left(\frac{3-a_n}{3}\right){\bar n_{\rm H}} \\
     &= \left(\frac{3-a_n}{3}\right)f_{\rm hCGM}{\frac{\Omega_b}{\Omega_m}} \frac{\Delta \rho_c}{\mu_{\rm H}} \\
    & = 2.9\times 10^{-5}E^{2}(z) \left(\frac{3-a_n}{3}\right) \frac{f_{\rm hCGM}}{0.5}   \ \ {\rm cm}^{-3} \ \ \ .
    \label{eq:nHvir}
\end{split}
\end{equation}
Here $\mu_{\rm H} = 1.37m_p=2.29\times 10^{-24}$~g is the mean particle mass per hydrogen nucleus. 
It is also convenient to express $n_{{\rm H,\vir}}$ in terms of $M_{\rm hGGM,vir}$:
\beq
n_{\rm{H,\vir}}=\frac{(3-a_n)M_{\rm hGGM,vir}}{4\pi\mu_{\rm H} R_\vir^3}\ \ \ .
\eeq 
For a finite $M_{\rm hCGM}$ we must have $a_n<3$. {\oran{ For a given hot CGM mass fraction the mean hot gas density, 
${\bar n_{\rm H}}$, is independent of the virial mass, and is a constant proportional to the universal virial overdensity $\Delta \rho_c$.}} For example, for $f_{\rm hCGM}=0.5$, ${\bar n_{\rm H}}=2.9\times 10^{-5}E^2(z)$~cm$^{-3}$ from equation (\ref{eq:nHvir}) (independent of halo mass). This is the characteristic hot CGM gas density surrounding galaxies at any redshift.

It follows from Eqs.~(\ref{eq:epress}), (\ref{eq:Tvir}), and (\ref{eq:nHvir}) that the electron pressure is
\begin{equation}
\begin{split}
    \frac{P_{\rm e}(r)}{k_{\rm B}}  =  13.2 E^{8/3}(z)\phi_T&\left(\frac{3-a_n}{3}\right) \frac{f_{\rm hCGM}}{0.5} 
      \Bigl(\frac{M_{\rm vir}}{10^{12}\ \msunm}\Bigr)^{2/3}\ \\
    &~~~~\times \Bigl(\frac{r}{R_{\rm vir}}\Bigr)^{-a_{P,{\rm th}}} \ \ {\rm cm}^{-3} \ {\rm K} \ \ \ .
\label{eq:epresscgs}
\end{split}
\end{equation}
In the PLM, the electron pressure at the virial radius depends on just the gas density power-law index $a_n$, the hot gas CGM baryon fraction $f_{\rm hCGM}$, the temperature parameter $\phi_T$, and 
the halo virial mass $M_{\rm vir}$ (which is the primary dimensional parameter of the PLM).

Two limiting cases may be considered, isothermal and constant entropy (isentropic).  For isothermal gas, 
$T$ is the same everywhere, and $a_n=a_{P,{\rm th}}$. For constant entropy, $Tn_{\rm H}^{-2/3}$ is invariant, and $a_n=(3/5)a_{P,{\rm th}}$. {\oran{We consider these limiting cases in \S~\ref{sec:PLMFSM} when fitting PLMs to the FSM models.}}


{\oran{\subsection{Hydrostatic Equilibrium}}}

The FSM models assume 
hydrostatic equilibrium (HSE), but with the inclusion of non-thermal components. In these models the resulting thermal pressure profiles are well approximated by power-laws {\oran{(as we discuss in \S~\ref{sec:PLMFSM})}}. However, an arbitrary power-law thermal pressure profile alone is not necessarily in HSE within a given gravitational potential. The degree to which just thermal pressure can maintain HSE is measured by the ratio of the gravitational force to the pressure force,
\beq
\eta\equiv\frac{\rho g}{dP_{\rm th}/dr},
\eeq
where $\rho=\mu_{\rm H} n_{\rm H}(r)$ is the gas density, {\oran{ $P_{\rm th}$ is the thermal pressure}}, and the gravitational acceleration $g=-GM(r)/r^2$ where $M(r)$ is the total gravitating mass within $r$. In other words, $\eta$ is the ratio of the pressure gradient required for HSE to the actual thermal pressure gradient. If $\eta > 1$ additional pressure components must be present to maintain HSE. If $\eta<1$ the large thermal pressure gradients alone will cause the system to expand.

For a power-law pressure profile, $P_{\rm th}\propto r^{-a_{P,{\rm th}}}$, the HSE parameter at the virial radius is
\beq
\eta_\vir=\frac{3}{a_{P,{\rm th}}\phi_T}
\label{eq:etav}
\eeq
with the aid of Eqs.~(\ref{eq:Tvir}) and (\ref{eq:phit}). {\oran{As just noted, if $\eta_{\rm vir} > 1$, additional pressure components must be present to maintain HSE. If $\eta_{\rm vir}<1$, thermal pressure gradients will cause the system to expand.}}

\vspace{0.5cm}

\subsection{Thermal Components}
\label{sec:CGMthermal}

Our PLM is for the hot component of the CGM. In general the CGM consists of several distinct thermal components from cool to hot as we now discuss. 

In this paper, ``hot" refers to gas for which $T>0.4 \times T_{\rm vir}$, where $T_{\rm vir}$ is the characteristic virial temperature of the halo (eq.~[\ref{eq:Tvir}]). {\oran{As we discuss in \S~\ref{sec:fracPLM} more than $\sim 90$\% of the tSZ effect in the TNG100 halos is produced in such hot gas 
for any $M_{\rm vir}$.}} For $M_{\rm vir}=1\times 10^{12}$~M$_\odot$, this includes the $T\gtrsim 10^5$~K gas considered in the FSM models described above, including the OVI component in \citetalias{Faerman2017}.

In contrast, ``cool" refers to gas 
visible in low-ionization absorption lines, including those of CII, CIII, SiII, SiIII, and MgII.  
The temperature of the cool CGM is set by thermal balance between photoionization heating by metagalactic radiation and self-consistent cooling by ionic forbidden line emission \citep[e.g.][]{Gnat2017}, and for a given spectrum and intensity, it is a function of the gas density and metallicity. For densities above $10^{-4}~{\rm cm^{-3}}$ and metallicities above $0.1$ solar, motivated by the analysis in \cite{Faerman2023}, and for the \cite{KS19} metagalactic field, this temperature is $T<3\times 10^4$~K;
we therefore define the cool CGM as gas with $T<3\times 10^4$~K. The third component is ``intermediate temperature" gas between cool and hot.

If a galaxy has a baryonic mass $10^{10.8}M_{\rm gal,10.8}$~\msun\ (including interstellar medium gas), then the maximum value of the hot gas fraction, $f_{\rm hCGM}$, is
\beq
f_{\rm hCGM,\,max}=1-  0.40\left(\frac{M_{\rm gal,10.8}}{M_{\vir,12}}\right) - f_{\rm cCGM} - f_{\rm itCGM}  \ \ \ ,
\label{eq:fCGMmax}
\eeq
where $f_{\rm cCGM}$ and $f_{\rm itCGM}$ are the fractions in any cool and intermediate temperature components of the CGM gas, and $10^{12}M_{\rm vir,12}$~M$_\odot$ is the halo virial mass. In this expression we are assuming that for a given $M_{\rm vir}$ the total baryonic mass within $R_{\rm vir}$ does not exceed the cosmic mean.

For $L^*$ galaxies the fraction of CGM gas in the cool and intermediate temperature phases is not well constrained observationally. Two models have been proposed: A patchy model, in which the cool CGM is very inhomogeneous so that a significant fraction of the cool gas is detectable only along a small fraction of the sight lines \citep{Werk14,Prochaska17}, and a smooth model, in which the cool CGM is relatively homogeneous so that the mass of the cool CGM in a galaxy can be estimated from a single line of sight \citep{Faerman2023}. \citet{Prochaska17} analyzed spectra of 13 quasars from the COS-Halos survey and estimated the average mass of cool gas in the galaxies along the lines of sight. Some of these galaxies had $M_*> 9 \times 10^{10}$~\msun, corresponding to $M_\vir>5 \times 10^{12}$~\msun\ with the \citet{Moster2010} stellar-halo mass relation. Excluding these massive galaxies, their analysis implies an average cool-gas mass of $4.2 \times 10^{10}$~\msun~(J. Werk and J. Prochaska, personal communication). This is significantly less than the $9.2\times 10^{10}$~\msun\ found by \citet{Prochaska17} because the galaxy that contributed almost half the total cool-gas mass is very massive, $M_\vir\simeq 2 \times 10^{13} $~\msun. Since the average virial mass of the non-massive galaxies is $1.2 \times 10^{12}$~\msun, it follows that the patchy model implies a cool-gas baryon fraction of $f_{\rm cCGM}\simeq 22\%$.
By contrast, \citet{Faerman2023} used the smooth model and found that the HI data for 2/3 of their sources from the COS-Halos survey can be explained with $M_{\rm cCGM}<10^{10}$ \msun. Adopting this mass estimate and using virial masses calculated and reported in \cite{Werk13}, we find that the smooth model leads to a cool-gas baryon fraction of $f_{\rm cCGM}\simeq 5\%$. There is thus at least a factor $\approx 4$ uncertainty in $f_{\rm cCGM}$, which has important consequences for permissible values of the mass of the hot CGM. We again stress that in \citetalias{Faerman2017} and \citetalias{Faerman2020} only hot gas is included\footnote{In the FSM papers we use the terminology ``warm/hot" that in this paper we subsume as simply ``hot".}.

\section{The \MakeLowercase{t}SZ effect}
\label{sec:tszplm}

\begin{figure*}[]
        \makebox[\textwidth][c]{\includegraphics[width= 1.1 \textwidth] {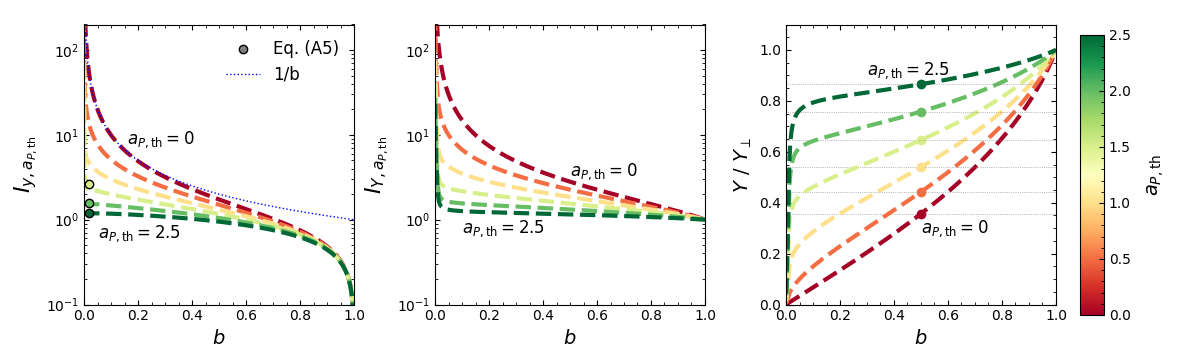}}
        \caption{The dimensionless integrals, $I_{y,a_{P,{\rm th}}}(b)$ (left panel), $I_{Y,a_{P,{\rm th}}}(b)$ (middle panel), and the ratio $Y/Y_\perp\equiv 1/I_{Y,a_{P,{\rm th}}}(b)$ (right panel). These integrals are defined by Eqs.~(\ref{eq:dimy}) and (\ref{eq:dim_Yp}) respectively. Curves are shown for the pressure power-law index $a_{P,{\rm th}}$ ranging from 0 to 2.5.  In the right panel the gray lines indicate the values of $1/I_{Y,a_{P,{\rm th}}}$ at impact parameter $b=0.5$. {\oran{In the left panel the three points are the values of $I_{y,a_{P,{\rm th}}}$ for $b\rightarrow 0$, as given by Eq.~(\ref{eq:iyz}) in Appendix A.}}}
        \label{fig:dimensionless_y}
    \end{figure*}

In the PLM, the Compton-$y$ produced within the CGM along a line-of-sight at projected radius $r_\perp$ relative to the origin, and with impact parameter $b\equiv r_\perp/R_{\rm vir}$, is

\begin{equation}
\begin{split}
    y(b)   &  =  2  \left(\frac{\sigma_T k_{\rm B}}{m_ec^2}\right) \int_0^{R_{\rm vir}(1-b^2)^{1/2}}n_eT \ dZ, \\
   &  =  2 \left(\frac{\sigma_T k_{\rm B}}{m_ec^2}\right) \phi_T T_{\rm vir} x_e n_{H,{\rm vir}}R_{\rm vir}b^{1-a_{P,{\rm th}}} I_{y,a_{P,{\rm th}}}(b),\\ 
   & = 2.59\times 10^{-9} E^2(z)\phi_T \Bigl(\frac{3-a_n}{3}\Bigr)
   \frac{f_{\rm hCGM}}{0.5}\Bigl(\frac{M_{\rm vir}}{10^{12} \ \msunm}\Bigr)\\
   & \ \ \times \ b^{1-a_{P,{\rm th}}} \ I_{y,a_{P,{\rm th}}}(b) \ \ \ ,
   \label{eq:Compton_y}
\end{split}
\end{equation}
where the dimensionless integral\footnote{This integral is commonly referred to as the hypergeometric function, and given our parameters it is written as $_2 F_1 (1/2, a_{P,{\rm th}}/2, 3/2, 1 - b^{-2})$}
\begin{equation}
    I_{y,a_{P,{\rm th}}}(b) \equiv \int_0^{\sqrt{1/b^2-1}}\frac{1}{ (1+u^2)^{a_{P,{\rm th}}/2}} \ du \ \ \ ,
    \label{eq:dimy}
\end{equation}
$u\equiv Z/r_\perp$ and $Z$ is the coordinate along the line of sight. For any $b$, $y$ scales linearly with the halo virial mass. For $3>a_{P,{\rm th}}>0$, 
the $y$ parameter is always larger for isentropic gas than for isothermal gas\footnote{For $0 < a_{P,{\rm th}} < 3$, the isentropic $y$ is a factor $(15-3a_{P,{\rm th}})/(15-5a_{P,{\rm th}})$ larger than for isothermal. E.g., for $a_{P,{\rm th}}=2.5$, by a factor of 3.}.

In Equation~(\ref{eq:Compton_y}) we have included the redshift factor $E^2(z)$. For a given set of PLM parameters, the Compton-$y$ is larger at higher redshift because the CGM spheres are hotter and denser for a given $M_{\rm vir}$: The path lengths are reduced by the factor $E(z)^{-2/3}$, but the CGM gas pressures are larger by $E(z)^{8/3}$.

In Fig.~\ref{fig:dimensionless_y} (left panel) we plot curves for $I_{y,a_{P,{\rm th}}}(b)$ for indices $a_{P,{\rm th}}$ ranging from 0 to 2.5.  For $a_{P,{\rm th}}\gtrsim 2$, $I_{y,a_{P,{\rm th}}}$ remains close to unity from $b=0$ to large radii (see Fig.~\ref{fig:dimensionless_y}). For example, for $a_{P,{\rm th}}=2$, $I_{y,a_{P,{\rm th}}}$ ranges from $\sim 1.5$ to $\sim 0.7$ for $b=0$ to 0.8. The Compton parameter is then well approximated as a power-law 
\beq
y(b) \propto b^{1-a_{P,{\rm th}}}
\label{eq:ypower}
\eeq
from $b=0$ to large radii, with a geometrical cutoff near $b=1$.

The volume-integrated $Y$ parameters can be defined 
for spheres or in projection. Within spheres, 
\begin{equation}
    Y(r)  \equiv \left(\frac{\sigma_T k_{\rm B}}{m_ec^2}\right)4\pi \int_0^r n_eT r^{\prime 2}\ dr^\prime  =     \Bigl(\frac{r}{R_{\rm vir}}\Bigr)^{3-a_{P,{\rm th}}}Y_{\rm vir}
    \label{eq:Ysint}
\end{equation}
where
\beq
Y_\vir=\left(\frac{\sigma_T x_e k_{\rm B}}{\mu_{\rm H} m_ec^2}\right)\langle T\rangle M_{\rm hCGM,vir}
\eeq
and $\langle T\rangle$ is the mass-weighted temperature of the hot gas. 
For the PLM
\beq
\langle T\rangle = \left(\frac{3-a_n}{3-a_{P,{\rm th}}}\right)\phi_T T_\vir,
\eeq
so that
\begin{equation}
\begin{split}
    Y_{\rm vir} &  =  \left(\frac{\sigma_T k_{\rm B}}{m_ec^2}\right)\phi_T T_{\rm vir}x_e\left(\frac{3-a_n}{3-a_{P,{\rm th}}}\right) \frac{M_{\rm hCGM,vir}}{\mu_{\rm H}},\\
    & = 3.64\times 10^{-4} E^{2/3}(z) \phi_T \left(\frac{3-a_n}{3-a_{P,{\rm th}}}\right)\frac{f_{\rm hCGM}}{0.5}\\
    &~~~~~~~~~~\times\Bigl(\frac{M_{\rm vir}}{10^{12} \ \msunm}\Bigr)^{5/3} \  {\rm kpc}^2 \ \ \ .
    \end{split}
    \label{eq:Yvir}
\end{equation}
$Y_{\rm vir}$ is proportional to the total integrated thermal pressure within the CGM sphere and scales as $M_{\rm vir}^{5/3}$. This is the well-known mass scaling for ``self-similar" clusters and galaxies \citep[e.g.,][]{Kaiser1986,DaSilva2004}. {\oran{The PLM extends from a vanishing inner radius, $r_0$, at the origin to the virial radius. For a finite $Y_{\rm vir}$ that does not diverge at the origin we must have a pressure slope $a_{P,{\rm th}}<3$.}}
We have included the redshift factor $E^{2/3}(z)$ that enters because for any $M_{\rm vir}$ the CGM volume scales as $E^{-2}$ and the pressure as $E^{8/3}$.

In projection, but only for elements within the virial radius, 
the volume-integrated $Y$ parameter is
\begin{equation}
\begin{split}
Y_\perp(b) & \ =2\pi \int_0^b y(b')b' db',\\
    & \ =Y_\vir \,b^{3-a_{P,{\rm th}}}I_{Y,a_{P,{\rm th}}}(b),  
    \end{split}
    \label{eq:Ypint}
\end{equation}
where
\begin{equation} \label{eq:dim_Yp}
    I_{Y,a_{P,{\rm th}}}(b) \ \equiv \frac{ (3-a_{P,{\rm th}})}{ b^{3-a_{P,{\rm th}}}} \int_0^b b^{\prime 2-a_{P,{\rm th}}}I_{y,a_{P,{\rm th}}}(b^\prime) db^\prime \ \ \ ,
\end{equation}
and $I_{y,a_{P,{\rm th}}}$ is defined in equation~(\ref{eq:dimy}). We plot curves for $I_{Y,a_{P,{\rm th}}}(b)$ in Figure \ref{fig:dimensionless_y} (middle panel). For any $r<R_{\rm vir}$ (i.e., $b<1$), the spherical value is less than the projected one, i.e.~$Y < Y_\perp$. We plot the ratio
\begin{equation}
    Y/Y_\perp \ \equiv \ 1/I_{Y,a_{P,{\rm th}}}(b)
\end{equation}
as a function of $b$ for the various values of $a_{P,{\rm th}}$ in Fig.~\ref{fig:dimensionless_y} (right panel). At $b=1$ both integrals (eqs.[\ref{eq:Ysint}] and [\ref{eq:Ypint}]) are over the entire CGM sphere, so that
$I_{Y,a_{P,{\rm th}}}(1)=1$ for all $a_{P,{\rm th}}$, and $Y(R_{\rm vir}) = Y_\perp(R_{\rm vir})=Y_{\rm vir}$. At $b=0.5$, (i.e. at $\sim R_{500}$; see below), $Y/Y_\perp$ ranges from 0.36 to 0.87 for $a_{P,{\rm th}}$ between 0 and 2.5.

Observationally, the tSZ distortions and the associated Compton parameters may include contributions from additional halos (often referred to as the ``2-halo" term) and intergalactic backgrounds from outside the CGM of a specific galaxy. Estimating the intrinsic spherical $Y$ within the CGM of a galaxy requires removal of the 2-halo terms and backgrounds from the observed projected $Y_\perp$. {\oran{(The spherical $Y$ is not directly observable.)}} Estimates for the external contributions depend on the assumed masses and sizes of the galaxy halos, and the pressure distributions within them, which determine the intrinsic ``1-halo" contributions. Our expressions for $y$ and $Y_\perp$ (Eqs.~[\ref{eq:dimy}] and [\ref{eq:Ypint}]) are for the intrinsic CGM of a given galaxy halo (aka the 1-halo term).

The $Y$-parameters have units of area, and the total SZ distortion flux is proportional to $Y_{\rm vir}/d_A^2$, where $d_A$ is the (angular diameter) distance to the CGM sphere. In the CMB literature the observed distortion fluxes are often stated in terms of
\beq
{\tilde Y} \ \equiv \ E^{-2/3}(z) \frac{Y}{(500 \ {\rm Mpc})^2}
\label{eq:tildeY}
\eeq
in arcmin$^2$, assuming a common distance of 500~Mpc, and also rescaling the $Y$-parameters to $z=0$ by multiplying through by the redshift factor $E^{-2/3}(z)$. Doing this gives
\begin{equation}
\begin{split}
    {\tilde Y}_{\rm vir} = & 1.72\times 10^{-8} \phi_T \\ 
    & \times \Bigl(\frac{3-a_n}{3-a_{P,{\rm th}}}\Bigr)\frac{f_{\rm hCGM}}{0.5}\Bigl(\frac{M_{\rm vir}}{10^{12} \, \msunm}\Bigr)^{\frac 53}  {\rm arcmin^2} 
    \label{eq:Yarcmin}
\end{split}
\end{equation}
for the PLM.

\begin{figure}[]
        \includegraphics[width=0.5\textwidth]{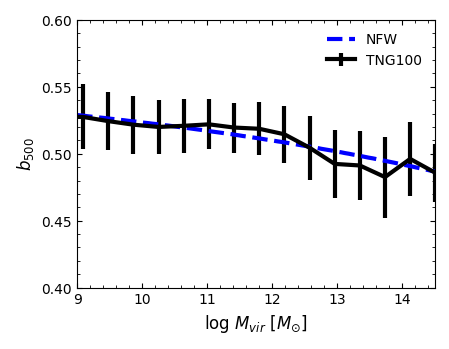}
        \caption{TNG100 results (black curve) for $b_{500}\equiv R_{500}/R_{\rm vir}$ versus $M_{\rm vir}$ for redshift $z=0$ halos. Error bars are 1$\sigma$. The dashed blue curve is for NFW halos with (median) cosmological concentration parameters (see text).}
        \label{fig:r_delta}
\end{figure}

Many of the tSZ observations reported in the literature are restricted to the inner denser parts of the hot halos, and the measurements are often stated in terms of the spherical pressure integrals $Y_{500}$ within which the mean mass over-density is $\Delta=500$. 
For the PLM it follows from Eqs.~(\ref{eq:Ysint}), (\ref{eq:Yvir}) and (\ref{eq:Ypint}) that
\begin{equation}
Y_{500} \equiv Y(R_{500}) \ = \ b_{500}^{3-a_{P,{\rm th}}}\ Y_{\rm vir} \ \ \ ,
\label{eq:Y500}
\end{equation}
and
\begin{equation}
Y_{\perp,500} \ = \ Y_\perp(R_{500}) = b_{500}^{3-a_{P,{\rm th}}} \ I_{Y,a_{P,{\rm th}}}(b_{500}) \ Y_{\rm vir}  \ \ \ ,
\end{equation}
where $b_{500}\equiv R_{500}/R_{\rm vir}$, and $R_{500}$ is the radius at which $\Delta=500$. For NFW halos \citep{Navarro1997} with $M_{\rm vir}$ from $10^9$ to 10$^{15}$~$M_\odot$, and for which the cosmological concentration parameters, $c$, range from 21.5 to 5.6 \citep{Dutton2014}, $b_{500}$ ranges from 0.53 to 0.48. In  Fig.~\ref{fig:r_delta} we plot $b_{500}$ versus $M_{\rm vir}$ as computed in the TNG100 simulation for redshift $z=0$ halos. The results match the analytic NFW estimates for $b_{500}$. (We discuss our TNG100 computations in \S~\ref{sec:tng}.)

{\oran{\section{PLM fits to the FSM models}
\label{sec:PLMFSM}}}

\begin{figure*}[]
        \makebox[\textwidth][c]{\includegraphics[width= 1.1 \textwidth] {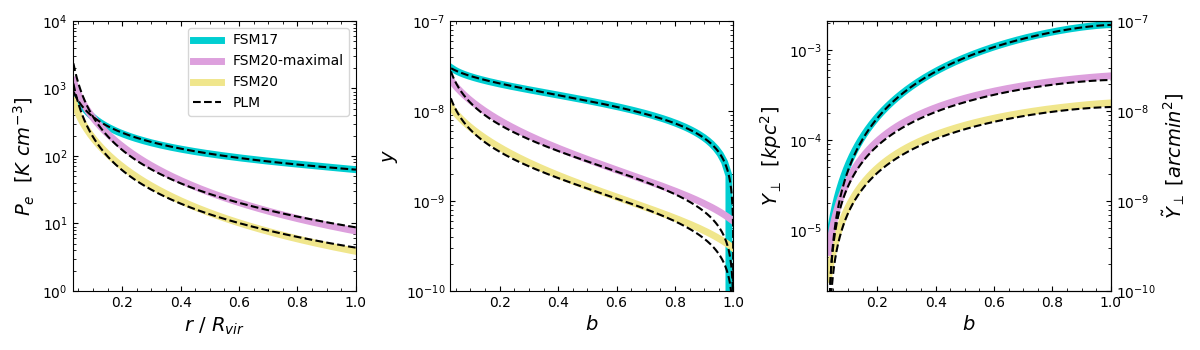}}
        \caption{Electron pressure (left panel) and Compton parameter profiles (middle and right panels) for the FSM models and PLM representations. In each panel the solid curves (yellow, purple, and turquoise) are for the FSM20 isentropic (fiducial and maximal) and {\oran{for the FSM17 model after the formation of the OVI component (see text).}} The dashed curves are the {\oran{fitted}} PLM profiles, see Eqs.~(\ref{eq:epresscgs}), (\ref{eq:Compton_y}), and (\ref{eq:Ypint}). {\oran{For FSM20, the PLM parameters are $a_{P,{\rm th}}=1.64$, $a_n=0.98$, and $\phi_T=0.71$, with $f_{\rm hCGM}=0.29$ and $\phi_T^*=0.83$ for fiducial, and $f_{\rm hCGM}=0.59$ and $\phi_T^*=1.67$ for maximal. 
        For FSM17, $f_{\rm hCGM}=0.77$, and $a_{P,{\rm th}}= 0.78$, $a_n=0.72$, and $\phi_T=3.05$ ($\phi_T^*=10.62$)}}. For all three FSM models $M_{\rm vir}=1.0\times 10^{12}$~M$_\odot$. In drawing the PLM curves we account for the different overdensities, $\Delta$, of 120 and 110 assumed in \citetalias{Faerman2017} and \citetalias{Faerman2020}. In Equation~\ref{eq:epresscgs}, $\Delta=100$.}
        
        \label{fig:FSM_PLM}
\end{figure*}

{\oran{Our PLM expressions are motivated by the FSM models in which the resulting radially dependent thermal electron pressures are well approximated by simple power laws.
The PLM parameters for \citetalias{Faerman2017} (post OVI formation) and for \citetalias{Faerman2020} (fiducial and maximal) are summarized in Table~\ref{tab:PLM_parameters}. The FSM versus PLM curves for the electron pressures, and the Compton-$y$ and $Y_\perp$ parameters, are shown in Fig.~\ref{fig:FSM_PLM}.
}}

{\oran{\subsection{Fitting Procedure}}}

{\oran{We construct our PLM fits to the FSM results as follows. First, we find the best fitting (best $\chi^2$) power-law index, $a_n$, for the shapes of the FSM gas density profiles, from the assumed inner CGM radius, $r_0$, to $R_{\rm vir}$, and subject to the integral constraint that the gas masses exactly equal the FSM hot gas masses $M_{\rm hCGM}$ (as given by the various FSM values for $f_{\rm hCGM}$).
This determines the PLM gas densities at the virial radii, $n_{\rm H,vir}$, as given by Equation~(\ref{eq:nHvir}), so that the gas densities are fully specified. Second, we find the best fitting index, $a_{P,{\rm th}}$, for the shapes of the thermal pressure curves. We then impose the normalizing integral constraint that the PLM $Y_{\rm vir}$ as given by equation~(\ref{eq:Yvir})
exactly equals the FSM values for $Y_{\rm vir}$. 
For the PLM $r_0=0$ by definition, whereas for the FSM models $r_0=8.5$~kpc. We neglect this difference, and for the $a_{P,{\rm th}}$ values that we find below the error in this approximation is less than 1\% from Equation 
(\ref{eq:Ysint}). As seen from equation~(\ref{eq:Yvir}), given the fits for $a_n$ and $a_{P,{\rm th}}$, and with $f_{\rm hCGM}$, this constraint determines the PLM value for $\phi_T$. 
The PLM $\phi_T$ will in general differ somewhat from the actual FSM $\phi_T$. }}

{\oran{When density and pressure data are both available, as for the FSM models or for the TNG100 simulations we discuss in \S~\ref{sec:tng} below, we recommend our two-integral-constraint procedure for extracting the PLM parameters. The derived PLM parameters can then be used to generate the Compton-$y$ and $Y$ curves (Eqs.~[\ref{eq:ypower}], [\ref{eq:Ysint}], and [\ref{eq:Ypint}]).}}

{\oran{We caution that the PLM breaks down and our fitting procedure becomes inaccurate if the density or pressure profiles become too steep, with $a_n$ or $a_{P,{\rm th}}$ approaching 3, since then the PLM gas masses and integrated pressures diverge at small radii, where conditions are uncertain and are strongly affected by the central galaxy.}}

{\oran{\subsection{PLM Pressures}}

{\oran{For \citetalias{Faerman2017}, we fit PLMs to both the initial isobarically fluctuating $T=1.5\times 10^6$~K gas, and to the final two-component gas distribution after the OVI component has cooled out. We stress that the OVI component is included in our definition of hot gas ($T>0.4\times T_{\rm vir}$). Thus, the hot CGM gas fractions are unaltered in the transition from initial to final, with $f_{\rm hCGM}=0.77$ for the FSM17 fiducial model (and the PLM representations).}}

{\oran{The thermal pressure profile for the initial isothermal gas distribution in \citetalias{Faerman2017} is shown in Fig.~3 of that paper. 
It extends from an assumed inner CGM radius, $r_0=8.5$~kpc, to an outer radius of 250~kpc, close to the nominal virial radius, $R_{\rm vir}=267$~kpc, so that $r_0/R_{\rm vir}\approx 0.03$.  This is similar to the inner CGM radii, $r_0/R_{\rm vir}\approx 0.04$, we find for the TNG100 halos (see \S~{\ref{sec:tng}})  and as also found by \cite{Karmakar2023}.
 The best-fitting PLM indicies are $a_{P,{\rm th}}=a_n=0.71$, and given
 $f_{\rm hCGM}=0.77$ and $Y_{\rm vir}=2.17 \times 10^{-3}$~kpc$^2$, the PLM $\phi_T=3.54$. The PLM density at the virial radius (Eq.~[\ref{eq:nHvir}]) is $4.16 \times 1 0^{-5}$~cm$^{-3}$, and the thermal electron pressure there is $72.8$~cm$^{-3}$~K (Eq.~[\ref{eq:epress}]). These are close to the FSM17 values of $4.78 \times 10^{-5}$~cm$^{-3}$ and $78.1$~cm$^{-3}$~K for the initial gas distribution. The maximal error between the initial FSM17 curve and the PLM is $\Delta P_e/P_e=20 \%$, and occurs at the inner boundary.}}

{\oran{As discussed in \S~\ref{sec:fsm}, in the \citetalias{Faerman2017} model, gas parcels in the initial isobarically fluctuating gas, with pressure $P_h(r)$, and with temperatures below a radially dependent cooling limit, $T_{\rm cool}(r)$, cool isochorically, and in place, to produce the OVI component. It is assumed that the OVI component is well mixed with the hotter gas so that they act as a single fluid. The median temperature of the OVI component is $T_{\rm OVI}=3.0\times 10^5$~K. The cooling is isochoric and the mean gas density distribution, $\langle n \rangle(r)$, of the final two-component system remains equal to the initial density distribution. However, the overall gas pressure is reduced. In \citetalias{Faerman2017} we did not derive the mean pressure curve, $\langle P \rangle (r)$, for the final two-component system, and we do that here now.}}

{\oran{If $p(T)dT$ is the volume filling factor\footnote{{\oran{In FSM17 the volume weighted filling factors for the isobarically fluctuating gas (with a median temperature of $1.5\times 10^6$~K) are given by a log-normal distribution in temperature (see Eq.~10 of \citetalias{Faerman2017} ). At any radius, $T_{\rm cool}(r)$ is the temperature below which the cooling time is shorter than the dynamical time.}}} of initial fluctuations in temperature range $dT$, the volume filling factor of the cooled out OVI component is
\begin{equation}
f_{\rm OVI}(r) \ = \ \int_0^{T_{\rm cool}(r)} p(T)dT \ \ \ ,
\end{equation}
and the mean pressure of the OVI component is
\begin{equation}
\langle P_{\rm OVI} \rangle (r) \ = \ \frac{P_h(r)}{f_{\rm OVI}(r)}\int_0^{T_{\rm cool}(r)} \biggr(\frac{T_{\rm OVI}}{T}\biggl)\ p(T)\ dT \ \ \ .
\end{equation}
The mean pressure of the combined two-component system is then
\begin{equation}
    \label{eq:warmhot}
    \langle P \rangle(r) \ = \ \langle P_{\rm OVI} \rangle f_{\rm OVI} + P_h(1-f_{\rm OVI}) \ \ \ .
\end{equation}
In Fig.~\ref{fig:FSM_PLM} (left panel) we plot the mean thermal electron pressure (solid turquoise curve) for the two-component \citetalias{Faerman2017} system, computed using Equation~(\ref{eq:warmhot}). For this pressure curve, $Y_{\rm vir}=1.86 \times 10^{-3}$~kpc$^2$. The best-fitting PLM indicies are $a_{P,{\rm th}}=0.78$ and $a_n=0.72$, and with $f_{\rm hCGM}=0.77$ and $Y_{\rm vir}$ the PLM $\phi_T=3.05$. The overlying dashed curve is the PLM pressure, and the agreement is excellent, with a maximal error $\Delta P_e/P_e=12$\% occurring at the inner FSM boundary, $r_0=8.5$~kpc. At the virial radius, the PLM density and electron pressure are $4.16 \times 10^{-5}$~cm$^{-3}$, and  $62.7$~cm$^{-3}$~K, compared to $4.71 \times 10^{-5}$~cm$^{-3}$, and  $62.6$~cm$^{-3}$~K in for the intrinsic \citetalias{Faerman2017} model following the formation of the OVI component. In our consideration of the tSZ effect for \citetalias{Faerman2017} we focus on the final electron pressure profile as given by Eq.~(\ref{eq:warmhot}) and displayed in Fig.~\ref{fig:FSM_PLM}, together with the associated PLM parameters.}}

{\oran{In Fig.~\ref{fig:FSM_PLM} (left panel) we also plot the parallel pressure curves for our two isentropic \citetalias{Faerman2020} models, fiducial (yellow curve) with $f_{\rm hCGM}=0.29$, and maximal (purple) with $f_{\rm hCGM}=0.59$. The PLM parameters are identical, with
 $\phi_T=0.71$, $a_{P,{\rm th}}=1.64$, and $a_n=0.98$, with $a_n\approx (3/5)a_{P,{\rm th}}$ as expected for isentropic gas.
The maximal error between the \citetalias{Faerman2020} pressure curves and the PLM fits occurs at the inner boundary and is $\Delta P_e/P_e=74$\%. By $r/R_{\rm vir}=0.1$ the error falls to less than 13\% out to the virial radii. This example shows that $r_0\approx 2R_e\approx 0.04R_{\rm vir}$ can be too small for the inner radius of the PLM; setting $r_0=0.1R_{\rm vir}$ is much safer. At the virial radius, and for $f_{\rm hCGM}=0.29$, the PLM density and electron pressure are $1.29 \times 10^{-5}$~cm$^{-3}$, and  $4.38$~cm$^{-3}$~K, compared to $1.28 \times 10^{-5}$~cm$^{-3}$, and  $3.85$~cm$^{-3}$~K for the intrinsic fiducial model. By construction, the outer thermal pressures in the \citetalias{Faerman2020} model is much lower than in \citetalias{Faerman2017}.}}


{\oran{\subsection{HSE and the PLM parameters}}}

{\oran{By assumption, the {\it initial} CGM gas state in \citetalias{Faerman2017} (i.e., before the cooling that leads to the creation of the OVI gas), and the single-component gas distributions in the \citetalias{Faerman2020} models are in HSE. For example, for the FSM20 PLM parameters, $\eta_\vir\simeq 2.6$, as given by Eq.~(\ref{eq:etav}).
Indeed, virial equilibrium in \citetalias{Faerman2020} is achieved by including nonthermal pressure due to cosmic rays, magnetic fields and turbulence so that the total pressure is about 3 times the thermal pressure.}} 

{\oran{However, for the initial CGM gas state in \citetalias{Faerman2017}, there is a significant variation of $a_{P,{\rm th}}$ with radius; at the virial radius, $a_{P,{\rm th}}=0.47$, and with $\phi_T=3.57$  we have $\eta_\vir\simeq 1.8$. The pressure gradient required for HSE is thus almost twice the thermal pressure gradient, which is consistent with the fact that a nonthermal pressure comparable to the thermal pressure was included in \citetalias{Faerman2017}.}}

{\oran{\subsection{$y$ and $Y_\perp$}}}

{\oran{In the middle and right panels of Fig.~\ref{fig:FSM_PLM}, we plot $y(b)$ and $Y_\perp(b)$ for our three FSM models (solid curves) together with our PLM representations (dashed curves) as given by Eqs.~(\ref{eq:Compton_y}) and (\ref{eq:Ypint}). The agreement reflects the accuracy of the power-law forms for the pressure profiles in the FSM models. For \citetalias{Faerman2017}  
the OVI component contributes $\sim 5\%$ throughout. In the fiducial \citetalias{Faerman2020} model the CGM extends to an outer boundary of $1.1\times R_{\rm vir}$ whereas the PLM cuts off at $R_{\rm vir}$ as can be seen for for the $y(b)$ curves in Fig.~\ref{fig:FSM_PLM}. }}

{\oran{In the TNG100 computations we present below (in \S~\ref{sec:tng} and \S~\ref{sec:comparison}) we shall be focusing on $y$ values at an impact parameter $b=0.2$, and also $Y_{500}$, in comparison to observations. In Table~2 we assemble the PLM parameters for the three FSM models, and also list the $y(0.2)$ and $Y_{500}$ values as given by the PLM (Eqs.~[\ref{eq:Compton_y}] and [\ref{eq:Y500}]). These are in good agreement with the intrinsic Compton parameters listed in Table~3. For \citetalias{Faerman2017} the percentage errors for $y(0.2)$ and $Y_{\rm 500}$ are 1.1\% and 6.3\%. For \citetalias{Faerman2020} the errors are 9.5\% and 6.1\%.}}


\begin{center}
\input{tb_FSM_B_TNG_PLM_v2.tex}
\end{center}

\vspace{0.5cm}

\section{Bregman SZ Stack} 
\label{sec:Bregman}

\begin{figure}[]
        \includegraphics[width=0.45\textwidth]{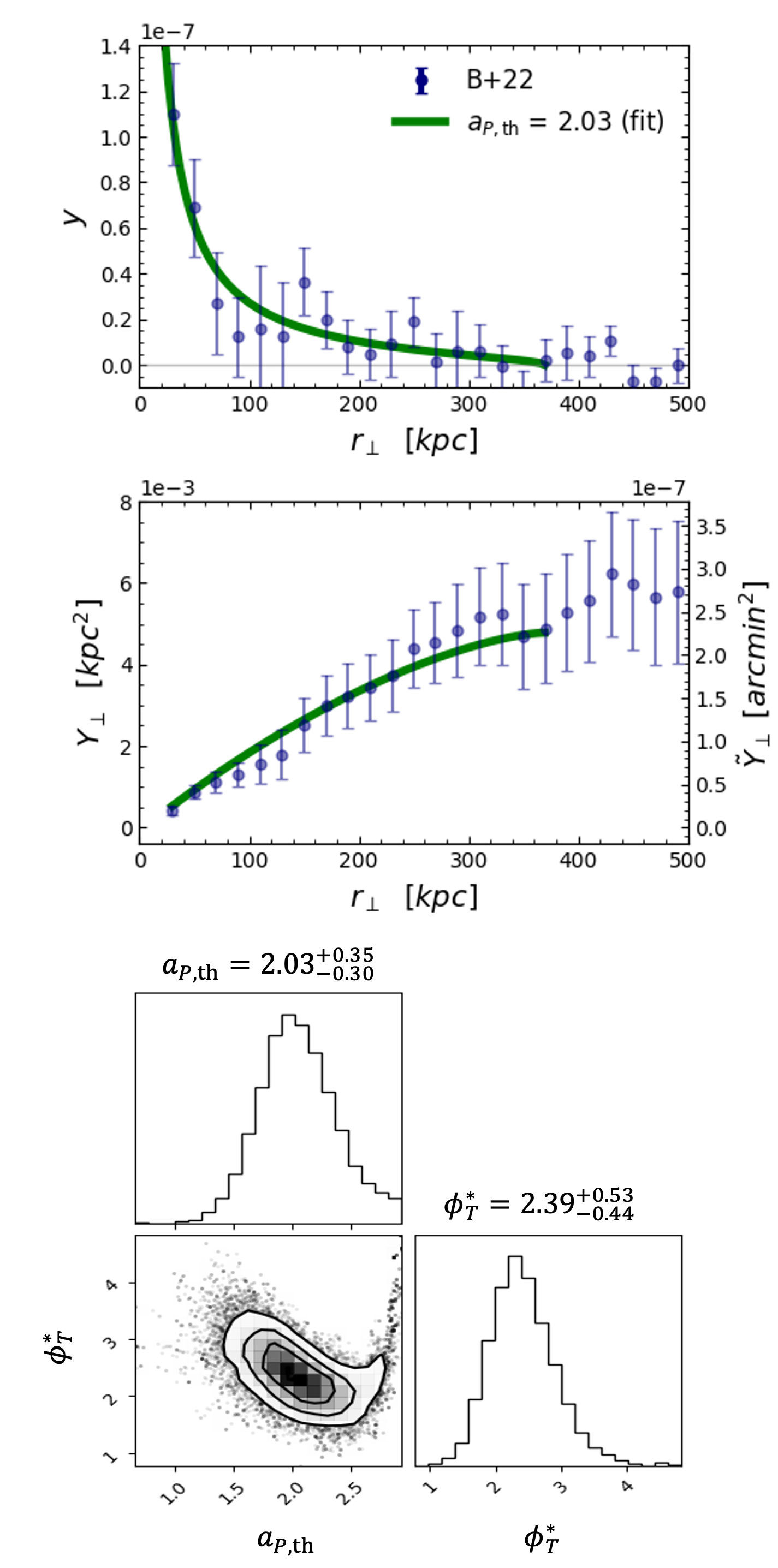}
        \caption{The Compton parameter $y(r_\perp)$ (top panel) and projected $Y_\perp(r_\perp)$ (middle panel) reported by \citetalias{Bregman2022} for their 11-galaxy stack, including their statistical error bars. The blue dashed curves are the \citetalias{Bregman2022} $\beta$-model fits to the data (Eq.~[\ref{eq:B22fit}]).
        The green curves show our 2D PLM fit, assuming $M_{\rm vir}=2.75\times 10^{12}$~M$_\odot$, with best fitting parameters $a_{P,{\rm th}}=2.03$, and $\phi_T^*=2.39$ (see text). The bottom panel is the corner plot for the $a_{P,{\rm th}}$ and $\phi_T^*$ posteriors. {\oran{To view the data on logarithmic scales, see Fig. \ref{fig:p_comparison} in Appendix~\ref{app:p_comparison}}}.}
        \label{fig:PLM_bregman}
\end{figure}

\citetalias{Bregman2022} extracted thermal
SZ distortion signals for a set of nearby 
$L^*$ galaxies from multi-channel ($30-857$ GHz) {\it Planck} and WMAP cosmic microwave background (CMB) data \citep{Bennett2013,Planck2020}. When constructing their SZ maps, \citetalias{Bregman2022} removed point sources, such as individual background galaxies, clusters, or radio-loud AGN (see their Figs. 1 and 2). They also excluded regions within $\sim 20$~kpc of the central galaxies to eliminate dust-contamination of the SZ signals. They argued that any 2-halo corrections are negligible because their galaxies are nearby with large angular extents. Errors were estimated via bootstrap resampling among the galaxies in the set. 

\citetalias{Bregman2022} created data stacks within which the SZ fluxes and associated Compton parameters are detected with high significance ($\sim 4\sigma$).  The primary \citetalias{Bregman2022} stack consists of co-added SZ maps (degree-sized, and at the 10$^\prime$ {\it Planck} spatial resolution) of the well-known near-field  galaxies NGC 253, 1291, 3031, 3627, 4258, 4736, 4826, 5055, 5194, 5236, and 5457. Because the SZ distortions scale as $M_{\rm vir}^{5/3}/d^2_A$ (see above) these relatively low mass galaxies were selected to be nearby, $\lesssim 10$~Mpc, with potentially resolvable hot halos surrounding their inner optical/stellar and dusty components. At 10~Mpc, a spatial resolution of 10$^\prime$ corresponds to 29~kpc, so the halos are well resolved. Using K-band luminosities, \citetalias{Bregman2022} estimate a median stellar mass of $\log M_*$=10.83 for their galaxy stack, with a range log$M_*$=0.38 between the minimum and maximum values.

In the upper panel of Fig.~\ref{fig:PLM_bregman} we show the $y(r_\perp)$ data reported by \citetalias{Bregman2022} for their 11-galaxy stack, including the statistical error bars. {\oran{The data extend from 29 to 490 kpc.}} The blue dashed curve is the \citetalias{Bregman2022} truncated, projected ``$\beta$ model" fit to the data \citep[see also][]{Plagge2010}\footnote{The original $\beta$ model \citep{Cavaliere1976} was for the density in terms of the spherical radius, $r$; we refer to a $\beta$ model as a function of $r_\perp$ as a ``projected $\beta$ model." The condition for $y(b)$ to be about the same for the $\beta$ model and a PLM is that $r_c/R_\vir \ll b \lesssim r_{\rm cut}/R_\vir$--there is no constraint on the value of $I_{y,a_{P,\rm th}}$. The condition for $Y(r)$ and $Y_\perp(b)$ to be about the same for the $\beta$ model and a PLM is $(r_c/r)^{3-a_{P,\rm th}}\ll 1$, and here there is no constraint on the value of
 $I_{Y,a_{P,\rm th}}$.},
\begin{equation}
\begin{split}
y(r_\perp) \ = \ & S_1[1+({r_\perp}/r_c)^2]^{-3\beta/2+1/2}\\
& \times \exp[-({r_\perp}/r_{\rm cut})^d] \ + \ S_0 \ \ \ .
\end{split}
\label{eq:B22fit}
\end{equation}
Their fit has $\beta=0.6$, a core radius $r_{\rm c}=10$~kpc, and a mean for the entire stack of $S_1=2.4\pm 0.64\times 10^{-7}$. The \citetalias{Bregman2022} fit includes a cutoff term with $r_{\rm cut}=300$~kpc, and $d=2$. $S_0=0.18\pm 2.08\times 10^{-9}$ is their estimate for the background. For $r_c \ll r_\perp \lesssim r_{\rm cut}$ the projected $\beta$ model implies $y(b) \propto b^{1-3\beta}$ (where $b=r_\perp/R_{\rm vir}$ is the normalized impact parameter). 
This has the same scaling with $b$ as our PLM to within a factor $I_{y,a_{P,\rm th}}$,
with $a_{P,{\rm th}}=3\beta=1.8$ in this case. For steep ($a_{P,{\rm th}} \gtrsim 2$) pressure power laws and $0.8\ga 
b>0.1$, the factor $I_{y,a_{P,{\rm th}}}\sim 1$. Similar conditions are needed for the projected $Y_\perp(b)$ to be about the same for the projected $\beta$ model and the PLM (with $I_{y,a_{P,\rm th}}$ replaced by $I_{Y,a_{P,\rm th}}$). However, in addition, one requires $(r_c/r_\perp)^{3-a_{P,\rm th}}\ll 1$ so that the SZ effect due to the core is negligible.

The middle panel in Fig.~\ref{fig:PLM_bregman} shows the projected $Y_\perp(r_\perp)$ data found by  \citetalias{Bregman2022} for their 11-galaxy stack. We have added the dashed blue curve, which we computed as the projected integral of their fit for $y(r_\perp)$. 

The median $\log M_*$=10.83 stellar mass of the stack, corresponds to a virial halo mass of $2.75\times 10^{12}$~M$_\odot$ using the \cite{Moster2010} abundance matching stellar-mass to halo-mass (SMHM) relation between $M_*$ and $M_{\rm vir}$\footnote{\citet{Moster2010} relied on the results of \citet{Bryan1998} in defining their halos, just as we have done, requiring an overdensity at $z=0$ of at least $\Delta\simeq 100$.}. For this mass, the virial temperature is $T_{\rm vir}=7.7\times 10^5$~K (as given by Eq.~[\ref{eq:Tvir}]), and the virial radius is $R_{\rm vir}=376$~kpc (as given by Eq.~[\ref{eq:Rvir}]). Remarkably, this radius is where the observed $Y_\perp$ flattens and begins to saturate (see Fig.~\ref{fig:PLM_bregman}), suggesting that most of the SZ signal is indeed produced within the virial radii of the galaxies.

\subsection{PLM fit to B+22}
\label{sec:plm-b22}

We have carried out an MCMC analysis \citep{Hastings1970,Foreman-Mackey2013} to fit the \citetalias{Bregman2022} data using a PLM (at $z=0$). 
{\oran{Here we are fitting a PLM to the directly observed $y$-profile, rather than to given density and pressure profiles, as in the analytic FSM models or the TNG100 simulations. The fitting procedure is therefore different. First, we assume a stellar mass-halo mass relation \citep{Moster2010} to infer the virial mass from the observed stellar mass of the galaxy. For the \citetalias{Bregman2022} stack, we set $M_{\rm vir}=2.75\times 10^{12}$~M$_\odot$. Then, given an observed $y$-profile we}}
determine two free parameters. As seen in  Equation~(\ref{eq:Compton_y}), the shape of the Compton-$y$ profile depends on the pressure power-law index, $a_{P,{\rm th}}$, and this is our first parameter. 
The magnitude of the profile depends on the {\oran{normalization}} factor
\begin{equation}
    \phi_T^* \ \equiv \phi_T \times (2f_{\rm hCGM})\times (3-a_n) \ \ \ ,
    \label{eq:phiT*}
\end{equation}
and this is our second parameter. For an assumed CGM gas fraction, the resulting $\phi_T$ depends on the equation-of-state, with $a_n=a_{P,{\rm th}}$ for isothermal, or $a_n=(3/5)a_{P,{\rm th}}$ for isentropic.

In our MCMC procedure we assume uniform priors for $0 \le a_{P,{\rm th}} \le 3$ and $\phi_T^* > 0$, with Gaussian likelihoods. We explored the 2D parameter space with 256 walkers spanning 256 steps. The resulting distribution for the $a_{P,{\rm th}}$ and $\phi_T^*$ posteriors is shown in the corner plot in the bottom panel of Fig.~\ref{fig:PLM_bregman}. The best fit is for $a_{P,{\rm th}}=2.03$ and $\phi_T^*=2.39$. The $1\sigma$ ranges for these parameters are $[1.73,2.38]$ and $[1.95,2.92]$ respectively.

The green curves in the upper and middle panels of Fig.~\ref{fig:PLM_bregman} show show our best 2D PLM fits for $y(r_\perp)$ and $Y_\perp$. For the PLM, and for $a_{P,{\rm th}} \gtrsim 2$,  $y(r_\perp)$ is well approximated as a power-law varying as $b^{1-a_{P,{\rm th}}}$ (Eq.~[\ref{eq:ypower}]), i.e.~as $b^{-1}$ for our best fitting $a_{P,{\rm th}}=2.03$.  (This is close to the \citetalias{Bregman2022} fit for $y(r_\perp)$ that varies as $b^{-0.8}$.) We conclude that the measurements are indeed consistent with a simple power law distribution for the electron pressure in the mean CGM of the galaxy stack.

For our 2D fit, and for any equation-of-state, $Y_{\rm vir}=4.8\times 10^{-3}$~kpc$^2$, or ${\tilde Y}_{\rm vir}=2.26\times 10^{-7}$~arcmin$^2$.
For $M_{\rm vir}=2.75\times 10^{12}$~M$_\odot$, {\oran{$R_{\rm vir}=376$~kpc, and {$R_{500}=0.51 R_{\rm vir}$}} (see Fig.~\ref{fig:r_delta}). For this radius $Y/Y_\perp=0.76$ for $a_{P,{\rm th}}=2.03$.  For our 2D fit we then have $Y_{\perp,500}=3.24\times 10^{-3}$~kpc$^2$ and $Y_{500}=2.46\times 10^{-3}$~kpc$^2$ (or ${\tilde Y}_{500}=1.21\times 10^{-7}$~arcmin$^2$). {\oran{For our 2D fit, 8\% of the total $Y_{\rm vir}$ is produced in the unobserved sphere within 29~kpc (or within $0.08R_{\rm vir}$).}}

Alternatively, we can set $Y_{\rm vir}=5.5\times 10^{-3}$~kpc$^2$, as estimated by \citetalias{Bregman2022} for the value of $Y_\perp$ at 376~kpc that corresponds to $R_{\rm vir}$ for the mean $M_{\rm vir}=2.75\times 10^{12}$~M$_\odot$ for the stack. This requires $\phi_T^*=2.75$. Fitting for just the pressure power-law index then gives $a_{P,{\rm th}}=1.91$, with a 1$\sigma$ error range [2.05,2.10], very close to our 2D fitting result. 

In their analysis, \citetalias{Bregman2022} assumed isothermal conditions to estimate the hot gas mass. Their $Y_{\rm vir} = 5.5\times 10^{-3}$~kpc$^2$ then gives $\phi_T f_{\rm hCGM}=1.41$ {\oran{(see Eq.~\ref{eq:Yvir})}}, independent of the pressure slope for isothermal gas. \citetalias{Bregman2022} adopted a temperature $T=3\times 10^6$~K, corresponding to $\phi_T=3.9$. This implies $f_{\rm hCGM}=0.36$, or a hot gas mass of $1.54\times 10^{11}$~M$_\odot$ within the virial radius. For the \citetalias{Bregman2022} $\beta$-model, with $a_{P,{\rm th}}\approx 3\beta=1.8$, the implied hot gas mass within 250~kpc is $9.44\times 10^{10}$~M$_\odot$, consistent with their result for $T=3\times 10^6$~K (see their Eq.~6). {\oran{For our best 2D PLM fit, with $\phi_T^*=2.39$, and assuming isothermal conditions, so that $a_n=a_{P,{\rm th}}=2.03$, Eq.~(\ref{eq:phiT*}) implies $\phi_Tf_{\rm hCGM}=1.23$.  For $\phi_T=3.9$, we have $f_{\rm hCGM}=0.32$, and the total hot gas mass is $1.37\times 10^{11}$~M$_\odot$. Within 250~kpc it is $9.12 \times 10^{10}$~M$_\odot$. A significant hot gas content for the CGM of the galaxy stack is implied, even for an assumed high temperature of $3\times 10^6~$~K.}}

{\oran{\subsection{HSE?}
\label{sec:HSE}}}

Is HSE possible for our best fitting PLM? First, for HSE, and with the inclusion of non-thermal pressure, $\eta_{\rm vir} = 3/(\phi_T a_{P,{\rm th}}) \ge 1$ (Eq.~[\ref{eq:etav}]). For our best-fitting $a_{P,{\rm th}}=2.03$ we obtain the {\it upper} limit $\phi_T\le 1.48$. For a larger $\phi_T$ the thermal pressure alone is too large for HSE and the gas must expand. Thus, for the \citetalias{Bregman2022} interpretation (isothermal, $a_{P,{\rm th}}=1.8$, and $\phi_T=3.9$) the hot tSZ gas cannot be in HSE. 

Second, as given by Equation~(\ref{eq:phiT*}), $\phi_T^* \le 2 \phi_Tf_{\rm hCGM,max}(3-a_n)$, where $f_{\rm hCGM,max}$ is the maximal hot gas fraction. Our best-fitting $\phi_T^*=2.39$ then gives the {\it lower} limit $\phi_T \ge 1.20/[f_{\rm hCGM,max}(3-a_n)]$. For the \citetalias{Bregman2022} stack, we estimate $f_{\rm hCGM,max}=0.83$, if cool and intermediate gas is negligible with $f_{\rm cCGM}=0$. This follows from equation~(\ref{eq:fCGMmax}) assuming a stack stellar mass $\log M_*=10.83$, plus an additional $\sim 10\%$ for gas within the ISM of the stellar disks, and with $M_{\rm vir,12}=2.75$. We then obtain the lower limits $\phi_T \ge 1.48$ for isothermal gas $(a_n=2.03)$ or $\phi_T \ge 0.81$ for isentropic gas $(a_n=1.22)$.
For $\phi_T$ below these limits the gas is too cool to produce the observed Compton-$y$ profile even for the maximal CGM gas mass. 

Together with our upper limit $\phi_T \le 1.48$ we conclude that HSE is possible for isentropic gas, but only marginally if the hot gas is isothermal. With just a small inclusion of cool and intermediate temperature gas, $f_{\rm hCGM,max}<0.83$ and isothermal HSE is not possible for the hot gas.
Thus, stable isothermal 
conditions appear unlikely for the \citetalias{Bregman2022} stack for any temperature. For isentropic gas, HSE is possible for hot gas fractions $f_{\rm hCGM}$ between 0.45 and 0.83, corresponding to $\phi_T$ between 1.48 and 0.81, $\eta_\vir$ between 1 and 1.8, and $T(R_{\rm vir})$ between $1.1\times 10^6$ and $6.2\times 10^5$~K. Nonthermal pressure components would be needed to maintain HSE for $\eta_\vir>1$. It is beyond the scope of this paper to discuss such pressure components here. However, this analysis suggests that the tSZ observations favor isentropic conditions, as we now argue in comparing our FSM models directly to the \citetalias{Bregman2022} data. Our \citetalias{Bregman2022} PLM parameters for the isothermal and isentropic options for which HSE is possible are summarized in Table~\ref{tab:PLM_parameters}.

\vspace{0.5cm}

\subsection{FSM and B+22} 
\label{sec:fsm-b22}

\begin{figure*}[]
        \makebox[\textwidth][c]{\includegraphics[width= 0.9 \textwidth] {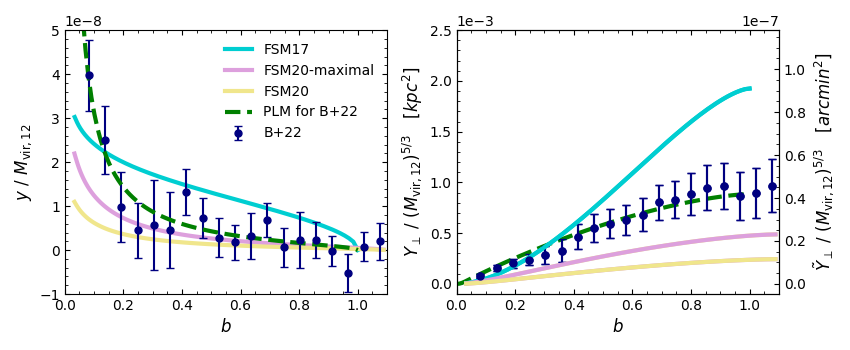}}
        \caption{\citetalias{Bregman2022} stack data and error bars (blue), normalized by a stack mass of $M_{\rm vir,12}=2.75$, in comparison to the three models, \citetalias{Faerman2020} (yellow), FSM20-maximal (purple) and \citetalias{Faerman2017} (turquoise). {\oran{The \citetalias{Faerman2017} model includes the OVI component (see text).}} The FSM models are unaffected by the renormalization since for these $M_{\rm vir,12}=1.$ The green dashed curves show our best fitting PLM for the \citetalias{Bregman2022} data.}
        \label{fig:FSM_bregman}
\end{figure*}

In Fig.~\ref{fig:FSM_bregman} we again plot the \citetalias{Bregman2022} stack data but now together with $y(r_\perp)$ and $Y_\perp(r_\perp)$ for our three models, \citetalias{Faerman2017}-fiducial, \citetalias{Faerman2020}-fiducial, and FSM20-maximal.  For this comparison we plot the data and model curves as functions of impact parameter $b=r_\perp/R_{\rm vir}$, rather than $r_\perp$ alone as in Fig.~\ref{fig:PLM_bregman}. For the stack, $M_{\rm vir}=2.75\times 10^{12}$~M$_\odot$ and $R_{\rm vir}=376$~kpc. For the FSM models, $M_{\rm vir}=10^{12}$~M$_{\odot}$, and $R_{\rm vir}\approx 267$~kpc {\oran{(see Table~1)}}. To scale out the mass dependence we divide $y$ by $M_{\rm vir, 12}\equiv M_{\rm vir}/{10^{12}}$M$_{\odot}$, and $Y_\perp$ by $M_{\rm vir, 12}^{5/3}$. 
In Fig.~\ref{fig:FSM_bregman} we also show our best-fitting PLM (green dashed curves) to the \citetalias{Bregman2022} data.

It is apparent that the rescaled \citetalias{Bregman2022} data lie mostly between the higher pressure \citetalias{Faerman2017} and lower pressure \citetalias{Faerman2020} curves for $y(b)$ and $Y_\perp(b)$, except for the inner two data points that are above even the FSM17 curve. For further comparisons, in Table~\ref{tab:fsm_B22} we list the values for $y(0.2)/M_{\rm vir,12}$, $Y_{\perp,500}/M_{\rm vir,12}^{5/3}$, and $Y_{\rm vir}/M_{\rm vir,12}^{5/3}$, for the \citetalias{Bregman2022} stack, and for the three FSM models. The \citetalias{Bregman2022} values are bracketed by \citetalias{Faerman2017} and FSM20-maximal.


\input{tb_fsm_B22.tex}

We have computed the $\chi^2$ deviations for all three models in comparison to the data. The best matching model is isentropic FSM20-maximal, even if the two inner data points are included. If these are excluded the fit with FSM20-maximal is even better (see left panel in Fig.~\ref{fig:FSM_bregman}). With the exclusion of the two inner points the $\chi^2$ values for the three FSM17, FSM20-maximal, and FSM20 models are {\oran{41.3}}, 8.5, and 11.9 respectively, and for the best-fitting PLM the $\chi^2$ is 8.0. The inner two stack points, within $0.2R_{\rm vir}$, are a factor $\sim 2.5$ higher than the FSM values, suggesting that the CGM pressures close to the galaxies are higher than predicted in the FSM models.  Indeed, the pressure profile in our best fitting PLM, with pressure index $a_{P,{\rm th}}=2.03$, and which does fit the inner data points (see Fig.~\ref{fig:PLM_bregman}) is steeper than for FSM20-maximal (and FSM20) for which $a_{P,{\rm th}}\approx 1.66$. But as indicated above in \S~\ref{sec:plm-b22}, the PLM suggests that isentropic models such as FSM20 or FSM20-maximal are preferred. Further refinement of the FSM models could involve the inclusion of a specific cool gas fraction, but we do not attempt that here.

\begin{figure}[]
        \includegraphics[width=0.45\textwidth]{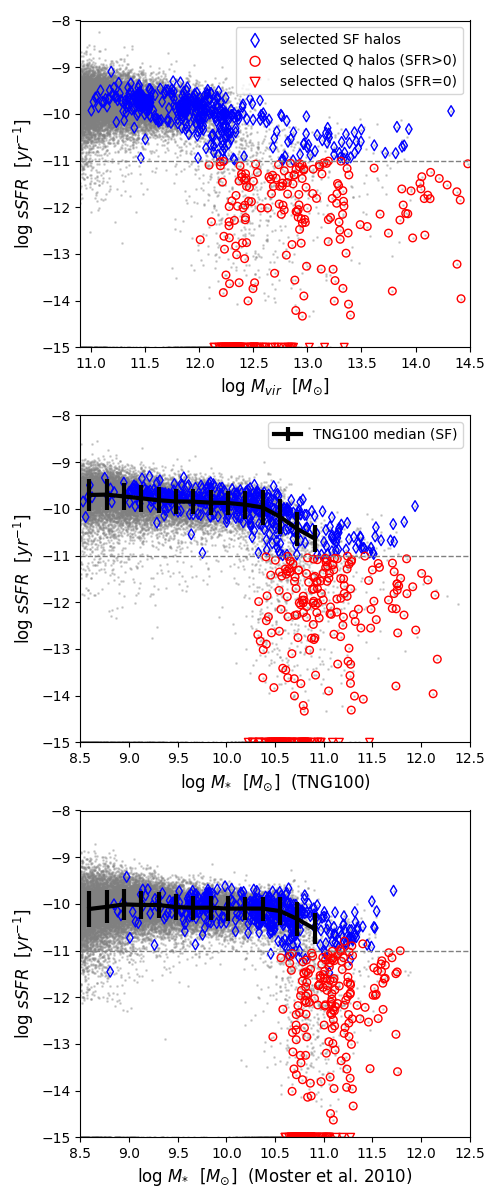}
        \caption{$sSFR$ vs.~$M_{\rm vir}$ in the TNG100 simulation (top panel), converted to sSFR vs.~the TNG100 $M_{*}$ estimates (middle panel), or sSFR vs.~$M_{*}$ using the \cite{Moster2010} SMHM relation (bottom panel). The small gray dots are for the entire set of TNG100 halos at redshift $z=0$. The blue and red markers are for our subset of TNG halos for which we have constructed maps and have computed Compton-$y$ profiles.  Blue diamonds are star-forming (SF) halos, with specific star-formation rates $sSFR > 10^{-11}$~yr$^{-1}$. The red circles are quenched (Q) halos, with $sSFR < 10^{-11}$~yr$^{-1}$. In the middle and bottom panels, the black curves show the median $sSFR$ for all TNG100 SF halos at redshift $z=0$.}
        \label{fig:whitaker}
    \end{figure}

\section{\MakeLowercase{t}SZ in TNG100} \label{sec:tng}

We now turn to cosmological galaxy evolution simulations, and for this purpose we make use of IllustrisTNG. We postprocess the simulation outputs to compute line-of-sight Compton-$y$ maps and $Y$ pressure integrals for a large set of galaxy halos spanning a wide range of virial masses.
In \S~\ref{sec:comparison} we compare the simulation results to the tSZ observations, and to our FSM models.

We focus on the intermediate resolution TNG100 simulation. Briefly, for TNG100 the hydrodynamics box size is 75~$h^{-1}$~Mpc (co-moving) on a side, and includes $2\times 1820^3$ resolution elements, with a spatial resolution of $\gtrsim 1$ kpc, a baryonic mass resolution of $1.4\times 10^6$ M\textsubscript{$\odot$}, and a dark matter mass resolution of $7.5\times 10^6$ M\textsubscript{$\odot$}. By redshift $z=0$ over 4 million dark matter halos (FOF groups) have formed within the large scale structure, spanning $2\times 10^8 < M_{\rm vir}/\msunm <4\times10^{14}$. The simulation follows five types of particles: gas, dark matter, star/wind particles, black holes, and passive tracers. The gas particles in particular are relevant for the tSZ parameters that we compute. 

  The TNG100 simulation is a powerful tool for analyzing the intergalactic medium and the CGM around individual galaxies. Its implementation of various feedback mechanisms, via AGN and stellar winds, models for chemical enrichment, and tracking of gas heating and cooling enable it to emulate the complex environment that comprises the CGM. The large box size and large number of halos enable exploration of the dynamic and complicated evolutionary processes involving inflows from the IGM, outflows from the ISM, heating and cooling, and evolution of the multiphase gas itself over time.

\subsection{Halo sample} 
\label{subsec:maps}

We wish to compute the tSZ Compton-$y$ and $Y$ integrals formed within the CGM of individual halos in the TNG100 simulation, as well as contributions from the surrounding unbound IGM. We consider halos with $M_{\rm vir}=1\times 10^{11}$ to $3\times 10^{14}$~M$_\odot.$

Generating $y$-maps is computationally expensive, and we focus {\oran{our analysis}} on a subset of 648 uniformly selected $z=0$ halos across the $1.0\times 10^{11}$ to $3.0\times 10^{14}$ M$_\odot$ virial mass range.
We display these in Figure~\ref{fig:whitaker}, in plots of specific star formation rate ({\it sSFR}) versus $M_{\rm vir}$ (upper panel), and versus stellar mass $M_*$ (lower two panels). The gray dots in Figure~\ref{fig:whitaker} are for the most massive 10,000 $z=0$ halos in the TNG100 catalog. Blue diamonds indicate our selected halos containing star-forming (SF) galaxies ($sSFR > 10^{-11}$~yr$^{-1}$). Red circles are our selected quenched (Q) galaxies ($sSFR < 10^{-11}$~yr$^{-1}$). In our halo/galaxy set, 398 are SF, and 250 are Q.
In the middle panel we convert $M_{\rm vir}$ to $M_*$ using the TNG100 results for the galaxy stellar masses. In the lower panel we instead use the \cite{Moster2010} SMHM relation to assign a stellar mass $M_*$ for a given TNG100 $M_{\rm vir}$. {\oran{The black curves show the median $sSFR$ versus $M_*$ relations for the SF halos.}} We illustrate both $M_*$ versus $M_{\rm vir}$ relations in Fig.~\ref{fig:moster}, which shows the generally good agreement between both conversions, taking into account that the mass definitions differ in terms of the considered effective aperture between the observations underlying the \cite{Moster2010} relation and the measurements from the simulation\footnote{Our comparison of TNG100 versus \cite{Moster2010} differs from \cite{Pillepich2018_2} (their Fig.~11). They used $M_{200}$ in the abscissa, whereas in an apples-to-apples comparison with \cite{Moster2010} we plot $M_*$ versus $M_{100}$.}.
 At $M_{\rm vir}\sim 10^{12}$~M$_\odot$ the Moster relation implies stellar masses larger by a factor $\sim 2$ compared to TNG100.  Overall, the differences between the TNG100 and \cite{Moster2010} SMHM relations are comparable to the differences with and amongst the \cite{Moster2013,Moster2018} formulations as well as amongst different choices of profiles and apertures for the measurement of the stellar masses. The yellow point in Fig.~\ref{fig:moster} is for the Milky Way assuming $M_{\rm vir}=1.3 \pm 0.3 \times 10^{12}$ M$_\odot$ (\citealt{Posti19}) and $M_* = 5.7^{+ 1.5}_{- 1.1} \times 10^{10}$~M$_\odot$ (\citealt{Licquia2016_MW_M_star}).

{\oran{\subsection{CGM gas and thermal fractions}
\label{sec:fracPLM}}}

To compute the Compton-$y$ and $Y$ parameters for any halo the first step is to characterize CGM gas particles.  We define these as all gas particles that (a) are not star forming, (b) are outside twice the stellar half mass radius, {\oran{$r_e$}}, of any galaxy, including that of satellite galaxies (subhalos), and (c) are within the virial radius. We consider all particles within $R_{\rm vir}$, not just those that are part of the parent halo FOF group.

For each halo 
we construct line-of-sight $y$ maps, and compute spherical and projected $Y$ integrals including only CGM gas as defined above. To estimate background contributions we also compute maps that include gas out to (arbitrarily) large distances of {\oran{$10$~Mpc}} from the central galaxies. In computing the $Y$ integrals and $y$ maps we include all thermal components for the CGM gas; ``cool" ($T<3\times 10^4$~K), ``intermediate", and ``hot" ($T>0.4\times T_{\rm vir}$) (see our discussion in \S~\ref{sec:fsm}). 
As we demonstrate below the hot component (as we have defined it) dominates the production of the $Y$ and $y$ parameters and the associated tSZ signals.  

In Fig.~\ref{fig:f_cgm} we plot the median CGM mass fractions versus virial mass, for each of the three thermal components. Black curves are for all of the halos in our sample. The blue and red curves are for the SF and Q halos separately. The upper left panel is $f_{\rm cCGM}$ for cool gas. Upper right is $f_{\rm itCGM}$ for intermediate temperature gas. Lower left shows the hot gas fractions $f_{\rm hCGM}$. In this panel we also display the hot gas fractions for each of our FSM models (only hot gas is included in these). Lower right shows the total CGM gas fractions $f_{\rm CGM}$ for our TNG100 sample. It is evident that hot gas dominates the CGM for all virial masses, although at low $M_{\rm vir}$ a non-negligible fraction is cool. At $M_{\rm vir}=10^{12}$~M$_\odot$, the cool, intermediate, and hot gas fractions are 0.12, 0.07, and 0.37 respectively, with a total CGM gas fraction $f_{\rm CGM}=0.56$.  Similar gas partitions have been found in the recent study by \cite{Wright2024}. The cool gas fractions are within the range (0.05 to 0.22) implied by the smooth versus patchy interpretations of the observed low-ion CGM probes discussed in \S~\ref{sec:CGMthermal}. However, in this paper we have not attempted to determine if the 
spatial distribution of the cold gas in the TNG100 simulations is closer to the patchy or smooth absorption model.

There is a noticeable dip in the gas fraction, primarily in the dominant hot gas component, at $\sim 3\times 10^{12}$~M$_\odot$, {\oran{epsecially for the quenched galaxies. This feature is associated with the onset of kinetic AGN feedback. In the TNG100 subgrid model, younger, lower mass AGN produce feedback as pure thermal energy that heats the surrounding gas and lowers star formation. Older, more massive AGN inject feedback into the surrounding gas in the form of kinetic energy (see \citealp{Weinberger2017}). The typical transition virial mass from thermal to kinetic feedback is near the mass where the dip in $f_{\rm CGM}$ occurs. As shown in \cite{Davies2020}, lower SFRs are correlated with more massive central black holes, which in turn inject more energy as kinetic feedback into the halo, driving more gas out of the CGM. This accounts for the difference between the CGM gas fractions in halos surrounding star forming galaxies compared to the fractions in halos surrounding quenched galaxies.}}


\begin{figure}[]
        \includegraphics[width=0.5\textwidth]{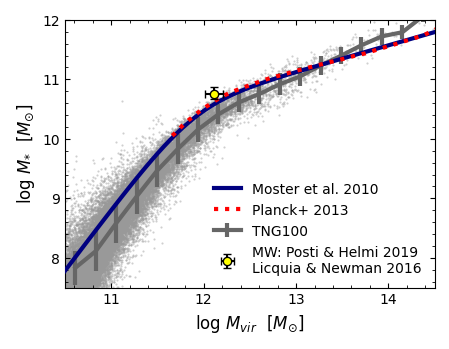}
        \caption{Stellar mass, $M_*$, vs.~virial mass, $M_{\rm vir}$ (the SMHM relation). The gray dots are the TNG100 results for all halos at redshift $z=0$. The gray curve shows the TNG100 medians and standard deviations. The blue curve is the observationally based \cite{Moster2010} SMHM relation. The red dotted curve shows the relation used by \citetalias{Planck2013}. The yellow dot is for the Milky Way.}
        \label{fig:moster}
\end{figure}

\begin{figure*}[]
        \makebox[\textwidth][c]{\includegraphics[width= 0.8 \textwidth] {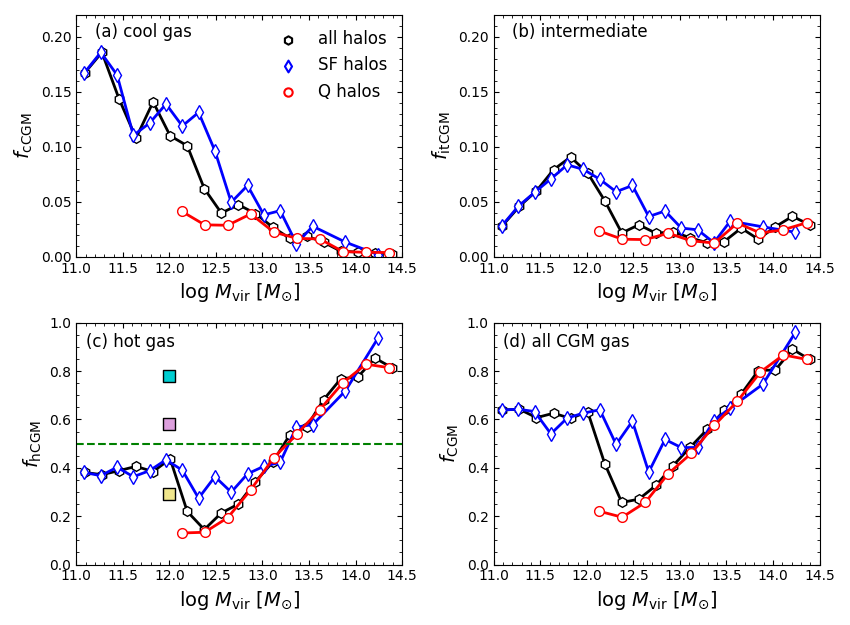}}
        \caption{
        CGM gas fractions for all of our 648 selected TNG halos (see Fig.~\ref{fig:whitaker}). (a) cool gas ($T<3\times 10^4$~K); (b) Intermediate temperature; (c) Hot gas ($T>0.4\times T_{\rm vir}$; and (d) Total CGM gas mass. We display the medians for all halos (black curves) and the star forming (SF) quenched (Q) halos separately (blue and red curves).  The nominal PLM value $f_{\rm hCGM}=0.5$ is the green dashed line. We also display (squares) the FSM values for $f_{\rm hCGM}$, \citetalias{Faerman2017} (turquoise), FSM20-maximal (purple) and \citetalias{Faerman2017} (yellow).
        }
        \label{fig:f_cgm}
\end{figure*}

\begin{figure*}[]
        \makebox[\textwidth][c]{\includegraphics[width= 0.9 \textwidth] {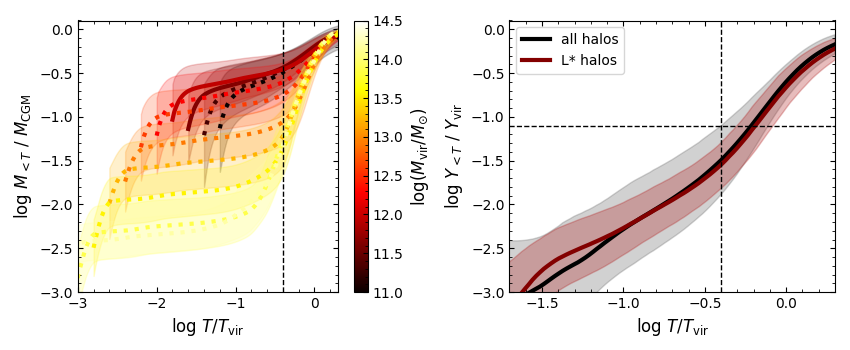}}
        \caption{Cumulative distributions for the CGM gas masses, $M_{<T}/M_{\rm CGM}$ (left panel), and Compton-$Y$ pressure integrals, $Y_{<T}/Y_{\rm vir}$ (right panel), as functions of $T/T_{\rm vir}$ for TNG100 halos with $M_{\rm vir}$ from 10$^{11}$ to 10$^{14.4}$~M$_\odot$. The vertical dashed lines mark the temperatures $0.4\times T_{\rm vir}$ above we which we define the gas as ``hot".}
        \label{fig:Y_M_of_T}
\end{figure*}

Which thermal components of the CGM gas contribute to the tSZ parameters in the TNG100 simulations? In Fig.~\ref{fig:Y_M_of_T} we plot the cumulative distributions $Y_{<T}/Y_{\rm vir}$ (right panel) and $M_{<T}/M_{\rm CGM}$ (left panel) both as functions of $T/T_{\rm vir}$, and for $M_{\rm vir}$ from $10^{11}$ to $10^{14.4}$~M$_\odot$. Here $Y_{<T}$ is the spherical pressure integral out to the virial radius (Eq.~[\ref{eq:Ysint}]) but including only CGM gas particles with temperatures less than $T$. Similarly, $M_{<T}$ is the CGM mass within $R_{\rm vir}$ with gas temperatures less than $T$. For sufficiently large $T$ all of the particles are included and $Y_{<T}\rightarrow Y_{\rm vir}$, and $M_{<T}\rightarrow M_{\rm CGM}$.
We compute $Y_{<T}/Y_{\rm vir}$ for our entire range of halo masses and for the more restricted set of $L^*$ halos with $M_{\rm vir}$ between $10^{12}$ and $3 \times 10^{12}$ $M_\odot$. The purple and black curves are the resulting medians, and the bands show the $1\sigma$ dispersions. It is evident that hot gas, with $T > 0.4 \times T_{\rm vir}$, accounts for at least 92\% of the total $Y_{\rm vir}$ for 84\% of the halo population, or at least 80\% of $Y_{\rm vir}$ for 98\% of the population.  For $L^*$ systems, the hot gas accounts for at least 95\% of the total $Y_{\rm vir}$ for 84\% of these halos, or at least 90\% of the total $Y$ for 98\% of them. {\oran{Almost all of the remaining contribution to $Y$ is from intermediate temperature gas.}}

The lefthand panel of Fig.~\ref{fig:Y_M_of_T} for $M_{<T}/M_{\rm CGM}$ shows that for low virial masses significant portions of the CGM are in the cool or intermediate components. For example, for $M_{\rm vir}=1\times 10^{12}$~M$_\odot$ cool refers to  $T/T_{\rm vir}<7.7\times 10^{-2}$, and this accounts for 21\% of the CGM mass. Hot gas, with $T>0.4 \times T_{\rm vir}$ is 65\%. These percentages are consistent with the halo baryon fractions shown in Fig.~\ref{fig:f_cgm}.

{\oran{\subsection{PLM fits}
\label{sec:PLMfits}}}

Next we construct power-law model fits for the hot ($T>0.4 T_{\rm vir}$) gas distributions in each of our 648 halos. 
{\oran{We use the same two-integral-constraint procedure we used in \S~\ref{sec:PLMFSM} to fit PLMs to the analytic FSM models.}}




\begin{figure*}[]
        \makebox[\textwidth][c]{\includegraphics[width= \textwidth] {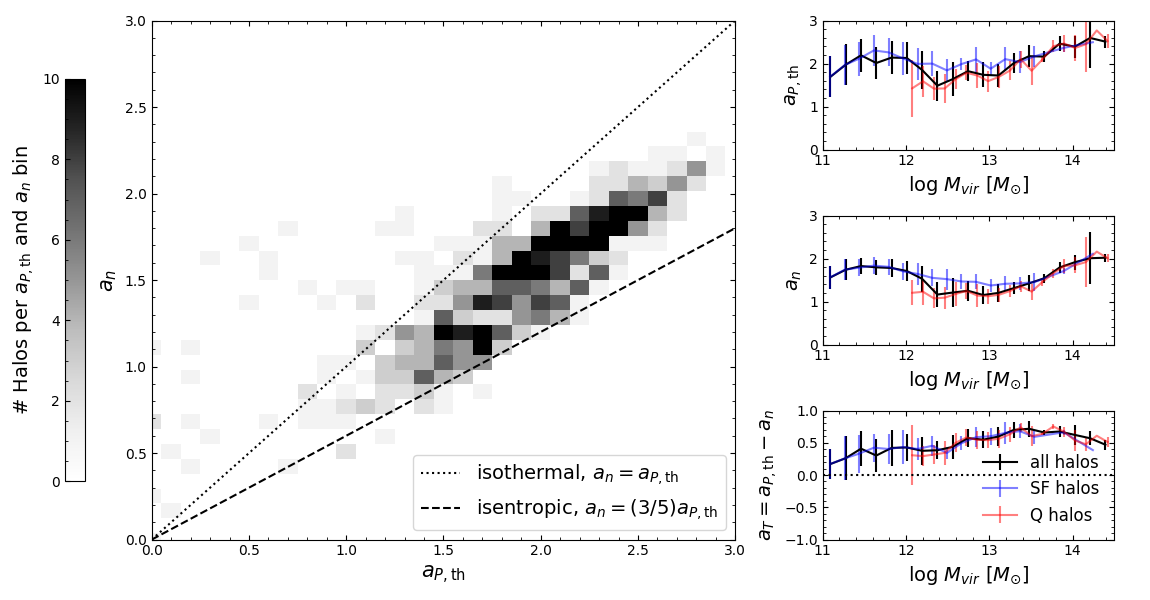}}
        \caption{PLM fit results for our selected TNG100 halos.  Left panel: Distribution of PLM pressure and density indices, $a_{P,{\rm th}}$ and $a_n$, for our TNG100 halo sample. The dotted line is for isothermal ($a_{P,{\rm th}}=a_n$) gas. Dashed is isentropic ($a_n=3a_{P,{\rm th}}/5)$. Right panel: Pressure, density, and temperature indices, $a_{P,{\rm th}}$, $a_n$, and $a_T=a_{P,{\rm th}}-a_n$, (top, middle, and bottom) for SF, Q, and all halos, as functions of $M_{\rm vir}$.}
        \label{fig:TNG_alpha_beta}
\end{figure*}

\begin{figure*}[]
        \makebox[\textwidth][c]{\includegraphics[width= 1.1 \textwidth] {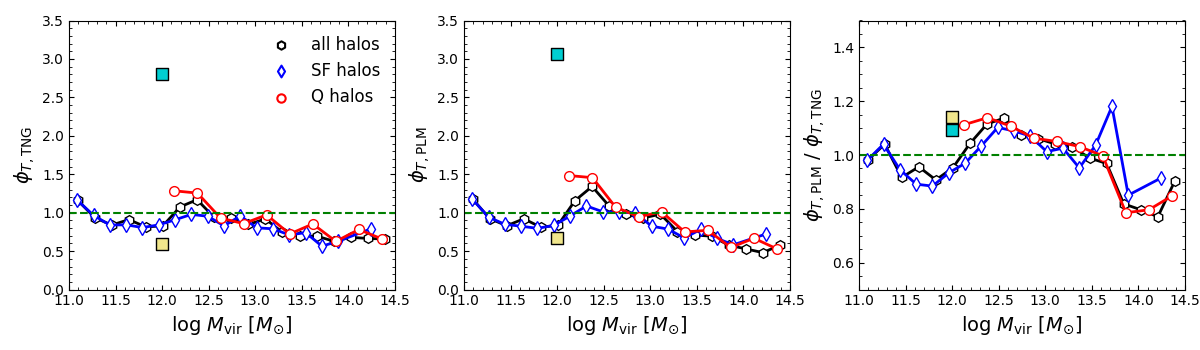}}
        \caption{\oran{Left panel: The intrinsic temperature parameter, $\phi_{T,{\rm TNG}}$, as functions of virial mass, for all (black), SF (blue), and Q (red) halos in our TNG100 sample. Middle panel: The fitted PLM parameters $\phi_{T,{\rm PLM}}$, for these halos. Right panel: The ratios $\phi_{T,{\rm PLM}}/\phi_{T,{\rm TNG}}$ versus halo mass. The turquoise and yellow squares, are for the \citetalias{Faerman2017}, and \citetalias{Faerman2020} models (fiducial and maximal).}}

        \label{fig:phi_T}
\end{figure*}

\begin{figure*}[]
        \makebox[\textwidth][c]{\includegraphics[width= 0.8 \textwidth] {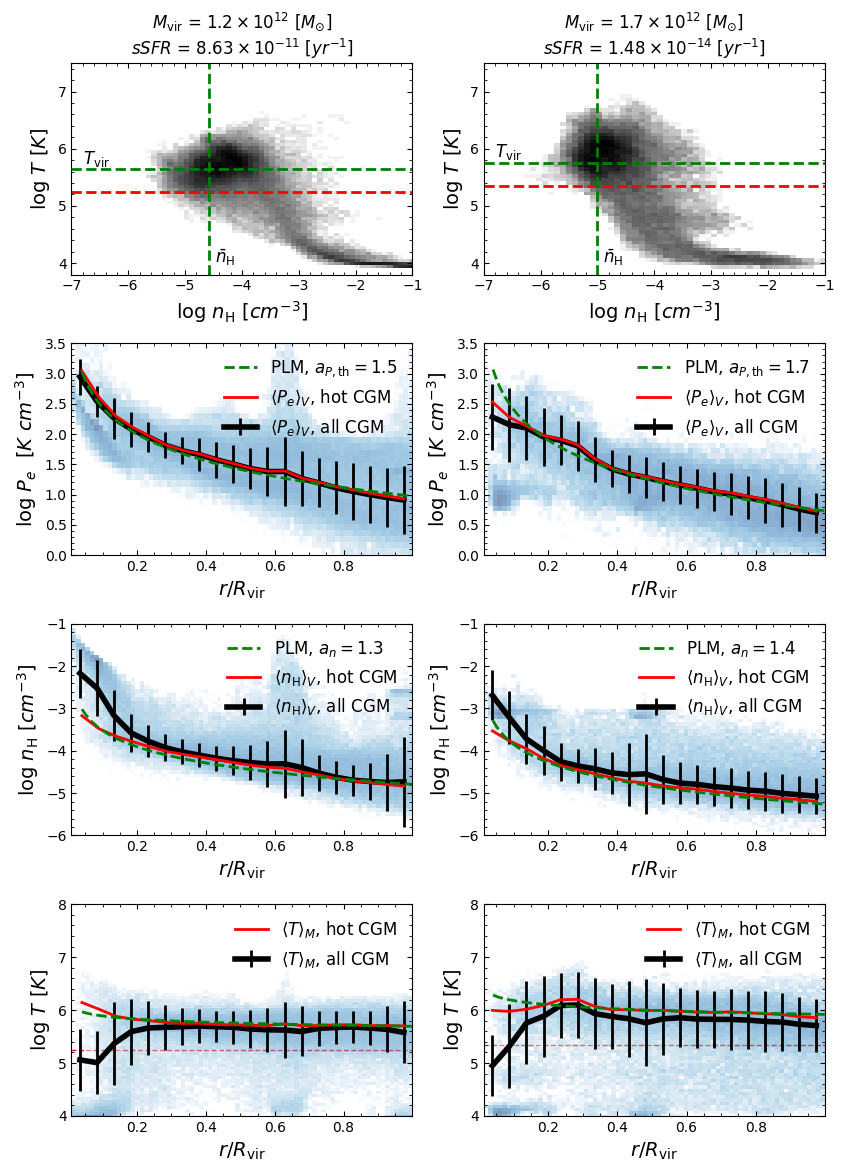}}
        \caption{CGM gas properties for two illustrative TNG100 $L^*$ halos. Left column is for a star forming (SF) halo with $M_{\rm vir}=1.2\times 10^{12}$~M$_\odot$, and $sSFR=8.63\times 10^{-11}$~yr$^{-1}$. Right column is for a quenched (Q) halo with $M_{\rm vir}=1.7\times 10^{12}$~M$_\odot$ and $sSFR=1.48\times 10^{-14}$~yr$^{-1}$. 
        Upper row: thermal phase plots for the CGM gas in the density ($n_{\rm H}$) vs.~temperature ($T$) plane. The horizontal green dashed lines mark the virial temperatures $T_{\rm vir}=4.2\times10^{5}$~K (left) and $5.6\times10^{5}$~K (right). The red dashed lines are at $0.4\times T_{\rm vir}$ above which we define any gas as ``hot" (see text). The hot gas mass fractions are $f_{\rm hCGM}=0.46$ (left) and 0.17 (right). The vertical green dashed lines mark the mean hot CGM gas densities, ${\bar n}_{\rm H}=2.7\times 10^{-5}$~cm$^{-3}$ (left) and $1.0 \times 10^{-5}$~cm$^{-3}$ (right). 
        Second row: volume weighted electron pressure within $R_{\vir}$ for all TNG100 particles, cool, intermediate, and hot, (solid black curve with 1$\sigma$ bars), and the hot phase electron pressure alone (solid red curve). The best fitting PLMs for the hot gas are also plotted (green dashed curves) for which $\phi_T=1.1$, $a_{P,{\rm th}}=1.5$, and $a_n=1.3$ (left), and  $\phi_T=1.48$, $a_{P,{\rm th}}=1.7$, and $a_n=1.4$ (right).
        Third row: volume weighted hydrogen density for all TNG100 particles (solid black curves), hot particles (solid red curves), and the PLM fits for the hot gas (green dashed). 
        Bottom row: mass weighted gas temperature for all TNG100 particles (solid black curves), hot TNG100 particles (solid red curves), and the PLM fits for which $a_T=a_{P,{\rm th}}-a_n=0.2$ (left) and $0.3$ (right). Gas is hot above the red dashed lines. The background blue regions are 2D histograms showing the radially dependent simulation CGM particle values for the pressure, density, and temperature, within the two halos.}
        \label{fig:phase}
\end{figure*}

\begin{figure*}[]
        \makebox[\textwidth][c]{\includegraphics[width= 0.8 \textwidth] {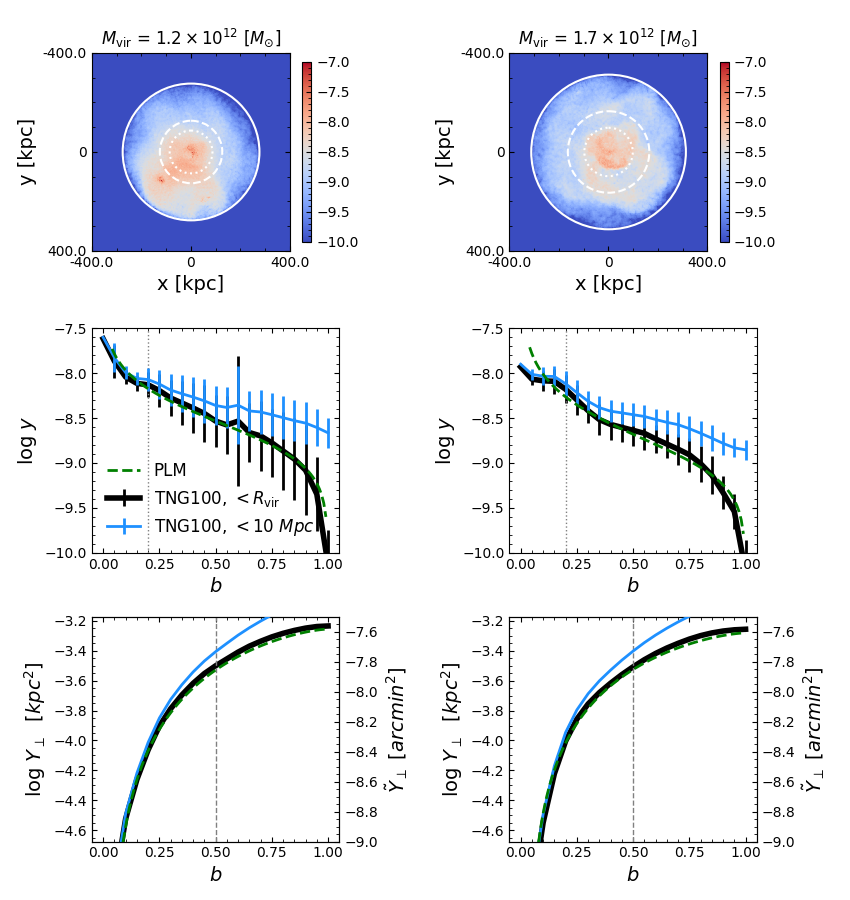}}
        \caption{Compton-$y$ maps and profiles for the same two TNG100 SF and Q halos in Fig.~\ref{fig:phase}. Top row: Compton-$y$ maps due to gas entirely within the virial radii indicated by the outer white circles (impact parameter $b=1$). The dashed circles are at $b_{500}\approx 0.5$ and the dotted circles are at $b=0.2$. Middle row: The black curves (with 1$\sigma$ bars) are the radially averaged TNG100 Compton-$y$-parameter profiles extracted from the above TNG100 maps. The blue curves are Compton-$y$ profiles that also include particles beyond the virial radii and out to $10$ Mpc from the halo centers, the dashed green curves are the PLM forms for $y(b)$ (Eq.~[(\ref{eq:dimy}]) for the best fitting PLM parameters shown in Fig.~\ref{fig:phase}. 
        Bottom row: The projected TNG100 $Y_\perp(b)$ integrals including gas only within the virial radii (black), out to 10~Mpc (blue), the dashed green curves are the PLM forms for $Y_\perp(b)$ (Eq.~[\ref{eq:Ypint}]) for the best fitting PLM parameters in Fig.~\ref{fig:phase}. }
        \label{fig:y_map}
\end{figure*}

For each halo we extract radial profiles for the median electron pressures, ${\bar P}_e(r)$, and densities, ${\bar n}_e(r)$, 
from the inner CGM {\oran{at twice the stellar half-mass radii, $r_e$, of the central galaxies out to the virial radii. As found by \cite{Karmakar2023} for IllustrisTNG $r_e\approx 0.02R_{\rm vir}$ for redshift $z=0$ halos. The inner radii for our TNG100 halos are therefore at $r_0\approx 0.04R_{\rm vir}$ similar to the inner radii $r_0=0.03R_{\rm vir}$ for the FSM models (see \S~\ref{sec:fsm}.)}} 
{\oran{We determine the best-fitting PLM indices $a_{P,{\rm th}}$ and $a_n$ 
for the shapes of the density and electron pressure profiles.
We normalize the PLM densities and pressures such that the PLM hot gas fractions (or masses) and $Y_{\rm vir}$ integrals
equal the TNG100 halo values. The resulting indices}} are shown in Fig.~\ref{fig:TNG_alpha_beta}. Remarkably, most of the halos lie within the isothermal ($a_n=a_{P,{\rm th}}$) and isentropic ($a_n=3a_{P,{\rm th}}/5$) boundaries, as seen in the lefthand panel. In the three righthand panels, we show the fitted indices, $a_{P,{\rm th}}$, $a_n$, and $a_T=a_{P,{\rm th}}-a_n$, as functions of $M_{\rm vir}$ for all of the selected halos, and for the SF and Q halos separately. 

{\oran{It is of interest to compare the intrinsic TNG100 hot gas temperatures to the fitted PLM temperatures, at the virial radii. We express both in terms of our parameter
$\phi_T\equiv T(R_{\rm vir})/T_{\rm vir}$ (Eq.~[\ref{eq:phit}]). For the intrinsic TNG value, $\phi_{T,{\rm TNG}}$, we compute the mass-weighted temperature for the hot gas between 0.95 and 1.05 $R_{\rm vir}$ for each halo. The results are plotted in Fig.~\ref{fig:phi_T} (left panel) for all halos, and the SF and Q halos separately. For the PLM, we compute $\phi_{T,{\rm PLM}}$ using Equation~(\ref{eq:Yvir}), together with the fitted indices $a_{P,{\rm th}}$ and $a_n$, and the TNG100 halo values for $Y_{\rm vir}$ and $f_{\rm hCGM}$. The results are plotted in Fig.~\ref{fig:phi_T} (middle panel). The intrinsic and fitted temperatures agree to within a factor 1.3 as shown in Figure~\ref{fig:phi_T} (right panel).
}}
 
 Our total sample of 648 halos, contains 354 halos with virial masses from $3\times 10^{11}$ to $3\times 10^{12}$~M$_\odot$. Of these 267 are star-forming $L^*$ halos, and the remaining are quenched (Q). For the SF L$^*$ halos, $1.9 < a_{P,{\rm th}} < 2.3$, and $1.5 < a_n < 1.8$. We also find that $0.8 < \phi_T < 1.0$, i.e., $\phi_T$ close to unity, supporting our definition of the virial temperature (Eq.~[\ref{eq:Tvir}]). These PLM parameters are summarized in Table~\ref{tab:PLM_parameters}.
For the TNG100 $L^*$ galaxies the pressure profiles are slightly steeper than in the isentropic FSM20 model ($a_{P,{\rm th}} = 1.64$), and much steeper than in the isothermal FSM17 model ($a_{P,{\rm th}} = 0.78$). 
At $2.75\times 10^{12}$~M$_\odot$ for the \citetalias{Bregman2022} stack, the median TNG100 pressure index for SF halos is $a_{P,{\rm th}}=1.9$, remarkably close to our best fitting PLM pressure index of 2.03 at this mass.

{\oran{
\subsection{Two examples}
\label{sec:twoexamples} }}

{\oran{ As two specific examples we consider two $L^*$ halos in detail in Fig.~\ref{fig:phase}. First (left) is a SF halo with $M_{\rm vir}=1.2\times 10^{12}$~M$_\odot$, and $sSFR=7.54 \times 10^{-11}$ yr\textsuperscript{-1}. Second (right) is a Q halo with $M_{\rm vir}=1.7\times 10^{12}$~M$_\odot$ and $sSFR= 1.48 \times 10^{-14}$ yr$^{-1}$. The results we show are similar to those presented by \cite{Ramesh2023} in their TNG50 study of $L^*$ galaxies, and \cite{Lim2021} in simulations of more massive halos.}}

{\oran{The upper panels in Fig.~\ref{fig:phase} are $n_{\rm H}$ versus $T$ phase diagrams for the total CGM gas within the virial radii, including all three thermal components, cool, intermediate, and hot. The horizontal green dashed lines indicate the virial temperatures $T_{\rm vir}=4.4 \times 10^{5}$~K and $5.6 \times 10^{5}$~K for the two halos. 
Hot gas lies above the red dashed lines at $0.4\times T_{\rm vir}$ for each halo. The hot CGM gas masses are $M_{\rm hCGM} = 8.3 \times 10^{10} ~M_\odot$ and $4.5 \times 10^{10}~M_\odot$, for the SF and Q halos respectively. The corresponding hot CGM gas mass fractions are $f_{\rm hCGM}=0.46$ and 0.17.
The vertical dashed green lines indicate the mean hot gas densities ${\bar n}_H=2.7\times 10^{-5}$~cm$^{-3}$ and $1.0\times 10^{-5}$~cm$^{-3}$.  Cool gas ($T<3\times 10^4$~K) is visible in the phase plots in the tails that extend to relatively high densities.}}
 
{\oran{The next three rows in Fig.~\ref{fig:phase} show 
the 1D radial distributions for the electron pressures, gas densities, and temperatures, averaged over spherical shells within each halo.  The pressures and densities are volume averages, ${\langle P_e \rangle}_V$ and ${\langle n_{\rm H} \rangle}_V$. The temperatures, ${\langle T \rangle}_M$, are mass weighted averages\footnote{{\oran{The volume averaged electron pressure ${\langle P_e \rangle}_V\equiv x_e{\langle n_{\rm H} T \rangle}_V = x_e{\langle n_{\rm H} \rangle}_V {\langle T \rangle}_M$, where ${\langle n_{\rm H} \rangle}_V$ is the volume averaged hydrogen gas density and ${\langle T \rangle}_M$ is the mass weighted temperature (see Eq.~27 in \citetalias{Faerman2017}). For hot fully ionized gas the electron fraction $x_e=1.17$, and is constant.}}}. The black curves include all components, cool, intermediate, and hot. The red curves are for the hot gas only. For all gas (black curves) the increases in density and the corresponding drops in temperature in the inner regions indicate the presence of the cooler components at these radii. But the generally good match between the averages for the total and hot gas pressures indicates that the volume filling factors of the cool and intermediate temperature gas are small.}}

{\oran{
The green dashed curves in the lower three panels in Fig.~\ref{fig:phase} are our PLM fits for the hot gas pressure, density and temperature distributions. For the SF halo we find a PLM $\phi_T=1.13$ (the intrinsic $\phi_T=1.12$) and indicies $a_{P,{\rm th}}=1.5$, $a_n=1.3$ and $a_T=a_{P,{\rm th}}-a_n=0.2$.  For the Q halo, the PLM $\phi_T=1.48$ (intrinsic is 1.24) and $a_{P,{\rm th}}=1.7$ $a_n=1.4$, and $a_T=0.3$.}}

{\oran{For the Q halo, the PLM curve rises above the actual hot gas distribution at small radii.  Based on spot checks over a wide range of masses in our halo sample we find that when central deviations appear, especially for the steepest pressure slopes, these generally occur at radii $r/R_{\rm vir}\lesssim 0.1$, as for the Q halo in Fig.~\ref{fig:phase}.
}}

\vspace{0.5cm}

\subsection{Compton-y maps}
\label{sec:twoymaps}

We have constructed individual maps for the line-of-sight Compton-$y$ parameters for all of our 648 selected SF and Q halos, spanning $1\times 10^{11}$ to $3\times 10^{14}$~M$_\odot$. 
When constructing the $y$-maps, we consider sightlines parallel to the arbitrary $Z$-axis of the simulation (i.e., we do not reorient the halos). We consider map sizes large enough to encompass the virial radii for any mass while maintaining pixel sizes of $1.6\times 1.6$~kpc$^2$. We place the map origin at the position of the central galaxy, and we compute
\begin{equation}
    y_{\rm pixel} = \frac{\sigma_Tk_{\rm B}}{m_ec^2}\int n_eT\ dZ \ \ \ 
    \label{eq:yTNG}
\end{equation}
for each pixel in the map.
For our basic CGM $y$-maps, we compute the integrals through just the CGM within $R_{\rm vir}$ at any impact parameter. For each map we compute projected radial profiles, $y({r_\perp})$, for the median CGM-only $y$ values within circular annuli, from the central galaxies to the virial radii. We also construct maps that include possible background contributions by integrating through sightlines spanning $\pm 10$~Mpc relative to the central galaxy.
The AREPO-based IllustrisTNG hydrodynamics follows a moving Voronoi mesh defined by the evolving particle distribution. 
For each CGM gas particle, with electron density $n_e$ and temperature $T$, we approximate the Voronoi cell as a sphere with equal volume, and for each line-of-sight we sum over the chords, $dZ$, that traverse each spherical cell. 

In Fig.~\ref{fig:y_map}, we again focus on the same SF and Q halos presented in Fig.~\ref{fig:phase}. The top row shows the resulting CGM $y$-maps for these two halos. Background contributions from outside the virial radii of 276 and 312~kpc are excluded. In the middle rows we plot the medians for $\log(y)$ as functions of the impact parameter $b$, as the solid black curves. The blue curves include background contributions from outside the virial radii along the $\pm 10$~Mpc sight-lines. The lower panels show the corresponding curves for the projected integrals $Y_\perp(b)$. At $b=0.2$ the backgrounds contributes less than 10\% to the Compton-$y$ for both halos. At $b=0.5$, which corresponds to $R_{500}$ (see Fig.\ref{fig:r_delta}), the backgrounds contribute $\sim 40\%$ and $\sim 50\%$ for the SF and Q halos respectively.  {\oran{For the integrals $Y_\perp(b)$ the backgrounds contribute 8\% and 14\% at $b=0.2$, and 24\% and 27\% at $b=0.5$, for the the SF and Q halos.  }}

The dashed green curves in the middle and bottom rows are the PLM representations for $y(b)$ and $Y_\perp(b)$, as given by Eqs.~[\ref{eq:dimy}] and [\ref{eq:Ypint}], and using the PLM parameters $f_{\rm hCGM}$, $\phi_T$, $a_{P,{\rm th}}$, and $a_n$ found for these halos, as shown in Fig.~\ref{fig:phase}. In as much as the pressure and density profiles for the hot gas are well represented by simple power-laws, the Compton-$y$ and $Y$ parameters are well described by our PLM expressions too. {\oran{For example, at $b=0.2$ the PLM errors on $y$ are 11\%, and 19\%, and on the spherical $Y_{\rm 500}$ are 11\%, and 6\%, for the SF and Q halos respectively. (The $Y_{500}$ values for these two cases are $1.81 \times 10^{-4}$ and $2.32 \times 10^{-4}$~kpc$^2$.)}}

\section{TNG100 and the \MakeLowercase{t}SZ Observations}
\label{sec:comparison}

We now compare our TNG100 simulation results to the \citetalias{Bregman2022}, \citetalias{Das2023} and \citetalias{Planck2013} tSZ observations with reference also to the analytic FSM models and the PLM representations. Our comparisons are shown in Figs.~\ref{fig:TNG_full_plus_B22} through \ref{fig:as_m_vir}. We focus on the (spherical) $Y_{500}$ pressure integrals, and the median line-of-sight $y$ values at impact parameters $b=0.2$.
This is sufficiently far to avoid contamination from the central galaxies (in both the observations and simulations) but not too far out where the tSZ signals weaken substantially.

In Fig.~\ref{fig:TNG_full_plus_B22} we plot $E^{-2/3}(z)Y_{500}/M_{\rm vir}^{5/3}$ versus $M_{\rm vir}$ for our TNG100 sample. We divide by $M_{\rm vir,12}^{5/3}$ to remove the simple overall mass scaling for $Y$ (see Eq.~[\ref{eq:Yvir}]), and for focus on our PLM parameters $a_{P,{\rm th}}$, $a_n$, $\phi_T$ and $f_{\rm CGM}$ (see Eq.~[\ref{eq:Yvir}]). We multiply by the redshift factor $E^{-2/3}(z)$ to enable scaling of observations at any redshift to our $z=0$ computations\footnote{Recall that $Y$ and $y$ depend on redshift because $R_{\rm vir}$ and the CGM spheres are smaller and hotter at higher $z$ at a given $M_{\rm vir}$ (see Eqs.~[\ref{eq:dimy}] and [\ref{eq:Yvir}]). By convention ${\tilde Y}$ already includes the scaling to $z=0$ (see Eq.~[\ref{eq:tildeY}]).}.
We plot (black curve) the medians for our entire TNG100 sample, and for the SF and Q subsamples separately (blue and red curves). The bars are $1\sigma$ deviations. For the Q halos, we show results starting at $10^{12}$~M$_\odot$. Q halos are very rare below this mass. The noticeable dip in the TNG100 curves near log($M_{\rm vir})=12.5$ is associated with AGN feedback and the reduction in the CGM gas fractions near this mass \citep[see Fig.~\ref{fig:f_cgm} and also][]{Davies2019}. 

The green square in Fig.~\ref{fig:TNG_full_plus_B22} is the \citetalias{Bregman2022} result for their 11-galaxy stack,
for which they inferred $Y_{500}=2.3 ~\pm~ 0.8 \times 10^{-3}$~kpc$^2$  or ${\tilde Y}_{500}=~1.1 ~\pm~ 0.4 \times 10^{-7}$ arcmin$^2$.  In their procedure \citetalias{Bregman2022} removed point sources from within the individual galaxy maps and corrected for possible dust contamination from the central galaxies. However, because their sample is of nearby ($\sim 10$~Mpc) isolated galaxies with large angular extents they argued that any 2-halo terms are likely small. 

In their analysis \citetalias{Bregman2022} assume a projection ratio $Y/Y_\perp = 0.76$, at $R_{500}$, as we also find using our PLM with pressure index $a_{P,{\rm th}}=2$ (see Fig.~\ref{fig:dimensionless_y}).
The \citetalias{Bregman2022} galaxies are mostly star-forming, and $Y_{500}$ for the stack is 0.5 dex larger than the TNG100 predictions at this $L^*$ mass scale. This is where the dip occurs in the TNG100 curves due to reduced CGM gas fractions at this mass scale. The discrepancy with the \citetalias{Bregman2022} observations may be due in part to overestimates of the {\oran{kinetic}} feedback strengths in IllustrisTNG, and excessive mass loss from the simulated CGM into the intergalactic medium at these virial masses. It may also indicate higher gas temperatures and thermal pressures in the observed galaxy stack compared to the simulation predictions.

In Fig.~\ref{fig:observations} we again plot $E^{-2/3}Y_{500}/M_{\rm vir}^{5/3}$ versus $M_{\rm vir}$, but without our TNG100 results. We retain the \citetalias{Bregman2022} point, and also include the \citetalias{Das2023} data. The dashed green line is for our best fitting PLM, ($a_{P,{\rm th}}=2.03$, $\phi_T^*=2.39$) to the \citetalias{Bregman2022} Compton-$y$ profile (see Fig.~\ref{fig:PLM_bregman}). The PLM line naturally passes through the \citetalias{Bregman2022} data point. The yellow, purple, and turquoise squares in Fig.~\ref{fig:observations} are the FSM20, FSM20-maximal and FSM17 predictions (as also listed in Table~\ref{tab:fsm_B22}). 

As we discussed in \S~\ref{sec:fsm-b22}  the isentropic FSM20 models better match the directly observed \citetalias{Bregman2022} Compton-$y$ profile (see Fig.~\ref{fig:FSM_bregman}), whereas in Fig.~\ref{fig:observations} it is FSM17 that appears better aligned with the \citetalias{Bregman2022} point. However, this is just an artificial alignment. It arises because the projection ratio $Y/Y_\perp$ for \citetalias{Faerman2017} (0.51) is smaller than the comparable ratios (0.74-0.76) for FSM20 and \citetalias{Bregman2022}. We remark that this is an example of how the commonly used $Y_{500}$ can be highly misleading, where a model-dependent adjustment brings a discrepant model into apparent harmony with observations. Indeed, for the {\it projected} $Y_{\perp,500}$ the observed \citetalias{Bregman2022} value is midway between FSM20-maximal and \citetalias{Faerman2017} (see Table \ref{tab:fsm_B22}).

\citetalias{Das2023} used the \citet{Zu2015} SMHM relation to estimate $M_{\rm vir}$, and $R_{500}$, for the galaxies in their set. They separated the intrinsic $y(b)$ and $Y_\perp(b)$ profiles from the 2-halo and background contributions assuming the \citet{Nagai2007} pressure profile (which differs from a PLM, {\oran{see Appendix \ref{app:p_comparison}}}). For consistency with our results we apply the Moster SMHM relation to the \citetalias{Das2023} stellar mass data to estimate the virial masses. We have verified that the $R_{500}$ and hence the resulting $Y_{500}$ values are hardly affected by this switch (for their assumed pressure profile). The correction to $Y_{500}$ is at most 20\% at the high-mass end, and negligible at the low-mass end of the \citetalias{Das2023} data shown in Fig.~\ref{fig:observations}.
The \citetalias{Das2023} results are remarkably close to the \citetalias{Bregman2022} point, with a peak in relative signal strength at slightly higher masses, followed by a decline towards the highest masses. The observed decline may be indicative of feedback-induced mass loss of the kind predicted by TNG100, but occurring at somewhat higher virial masses than in the simulations.

\begin{figure}[]
        \includegraphics[width=0.5\textwidth]{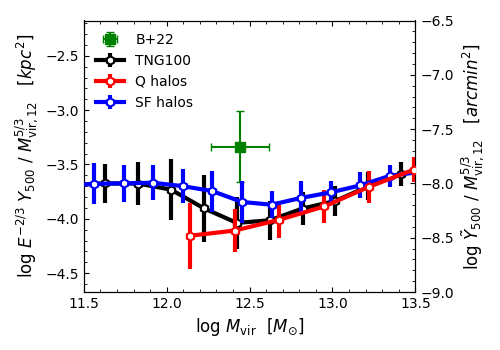}
        \caption{Spherical Compton pressure integrals, $Y_{500}$ (kpc$^2$) or ${\tilde Y}_{500}$ (arcmin$^2$), normalized by $M_{\rm vir,12}^{5/3}$ ($M_{\rm vir}\equiv 10^{12}M_{\rm vir,12})$ for all of our selected TNG100 halos (black), and for the star forming (SF, blue) and quenched(Q, red) halos separately. The green square is the  \citetalias{Bregman2022} stack measurement.}
        \label{fig:TNG_full_plus_B22}
\end{figure}

\begin{figure}[]
        \includegraphics[width=0.45\textwidth]{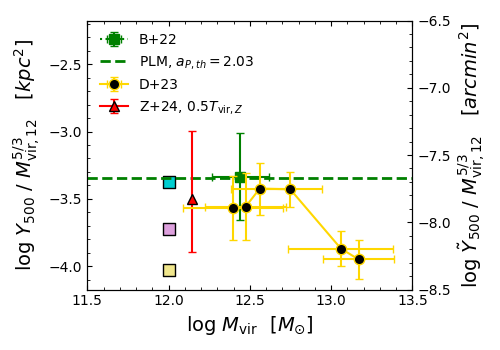}
        \caption{Spherical Compton pressure integrals, $Y_{500}$ (kpc$^2$) or ${\tilde Y}_{500}$ (arcmin$^2$), normalized by $M_{\rm vir,12}^{5/3}$, for the FSM20, FSM20-maximal, and FSM17 models (yellow, pink, and turquoise squares), together with the \citetalias{Bregman2022} stack (green square), and \citetalias{Das2023} measurements. For the \citetalias{Das2023} data we have used the \citet{Moster2010} relation to convert stellar mass to virial mass (see text). The dashed green line is for our best fitting PLM ($a_{P,{\rm th}}=2.03$, $\phi_T^*=2.39$) for the \citetalias{Bregman2022} Compton-$y$ profile in Fig.~\ref{fig:PLM_bregman}. 
        The red triangle is the $Y_{500}$ we infer from the recent \cite{Zhang2024} X-ray observations of $L^*$ galaxies (see \S~\ref{sec:Xrays}).}
        \label{fig:observations}
\end{figure}

   \begin{figure*}[]
        \makebox[\textwidth][c]{\includegraphics[width= 0.9 \textwidth] {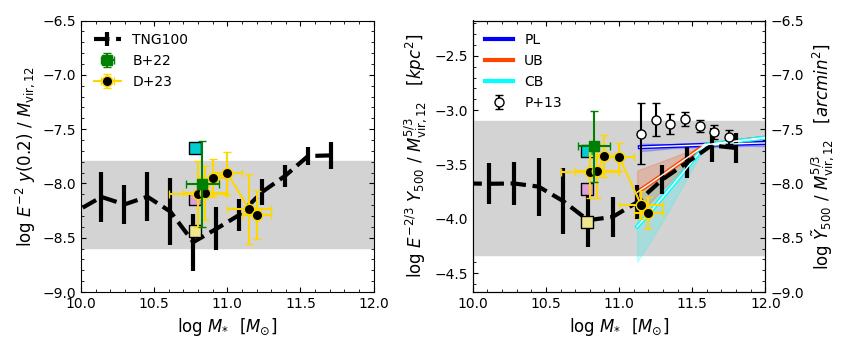}}
        \caption{
        The normalized Compton parameters, $y(0.2 R_{\rm vir})/M_{\rm vir,12}$ (left panel) and $Y_{500}/M_{\rm vir,12}^{5/3}$  (right panel) as functions of central stellar mass, $M_*$. The gray strips show the predicted ranges for our PLM (see text). The black curves (with 1$\sigma$ bars) are the medians for our set of TNG100 halos (SF+Q), and using the \citet{Moster2010} SMHM relation to convert the TNG100 virial masses to stellar masses.
        The vertically aligned squares (yellow, purple, and turquoise) are the FSM20, FSM20-maximal and FSM17 predictions, and at the Galactic stellar mass of $6\times 10^{10}$~M$_\odot$ (the Galactic stellar mass departs from the \citet{Moster2010} relation, see text). The green square is the \citetalias{Bregman2022} stack measurement, and the black filled gold circles are the \citetalias{Das2023} data. In the right panel we include the original \citetalias{Planck2013} data for massive galaxies and clusters {\oran{(open black)}}. We also show the three \citet{Hill2018} model estimates (PL blue, UB red, and CB cyan) for the intrinsic (1-halo) contributions to the \citetalias{Planck2013} observations.}        
        \label{fig:Y_500_vs_M*}
    \end{figure*}

\begin{figure*}[]
        \makebox[\textwidth][c]{\includegraphics[width= 0.9 \textwidth] {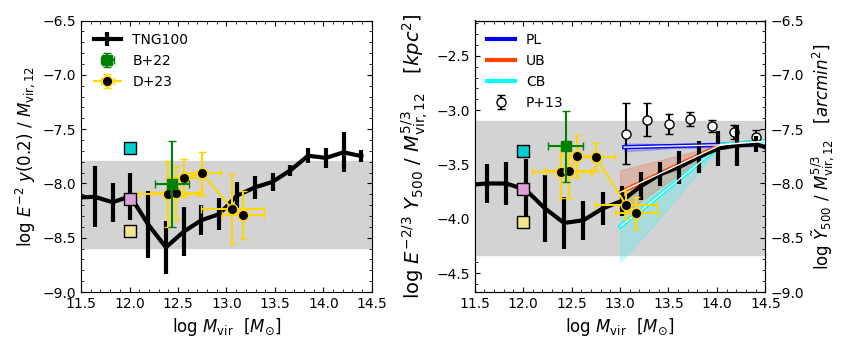}}
         \caption{The normalized Compton parameters, $y(0.2 R_{\rm vir})/M_{\rm vir,12}$
        (left panel), and  $Y_{500,s}/M_{\rm vir,12}^{5/3}$ (right panel) as functions of halo virial mass. The symbols and curves are for the same quantities as in Fig.~\ref{fig:Y_500_vs_M*}. For the \citetalias{Das2023} data we have used the \citet{Moster2010} SMHM relation to convert the observed stellar masses to virial masses (see text).}   
        \label{fig:as_m_vir}
    \end{figure*}

In Figs.~\ref{fig:Y_500_vs_M*} and \ref{fig:as_m_vir} we gather the \citetalias{Bregman2022}, \citetalias{Das2023} and \citetalias{Planck2013} observational data, together with our PLM, FSM, and TNG100 theoretical predictions, in plots of $E^{-2}y(0.2)/M_{\rm vir,12}$ (left panels) and $E^{-2/3}Y_{500}/M_{\rm vir,12}^{5/3}$ (right panels), as functions of $M_*$ and $M_{\rm vir}$.  We divide $y$ by $M_{\rm vir,12}$ to remove the overall mass scaling (Eq.~[\ref{eq:dimy}]), and multiply by $E^{-2}$ to remove the redshift dependence.

The gray strips show our PLM predictions, assuming $f_{\rm hCGM}=0.5$ and $\phi_T=1$, and for a range of $a_{P,{\rm th}}$ and $a_n$. The lower boundary is for isothermal $a_{P,{\rm th}}=a_n=0$ uniform pressure systems, for which $E^{-2}y(0.2)/M_{\rm vir,12}=2.55 \times 10^{-9}$, and $E^{-2/3}Y_{500}/M_{\rm vir,12}^{5/3}=5.18 \times 10^{-5}$~kpc$^2$. The upper boundary is for isentropic systems with steep pressure profiles with $a_{P,{\rm th}}=2.5$ and $a_n=1.5$, for which the scaled Compton parameters are equal to $1.65 \times 10^{-8}$ and $6.52 \times 10^{-4}$~kpc$^2$ respectively. The position of the gray strip shifts vertically and linearly with either $f_{\rm hCGM}$ or $\phi_T$.

The black curves in Figs.~\ref{fig:Y_500_vs_M*} and \ref{fig:as_m_vir} are the medians (with 1$\sigma$ bars) for our full TNG100 sample. We do not plot the curves for the SF or Q halos separately as in Fig.~\ref{fig:TNG_full_plus_B22}. In Fig.~\ref{fig:Y_500_vs_M*} we used the Moster SMHM relation (Fig.~\ref{fig:moster}) to assign stellar masses to the virial masses computed in the simulation. The dip in the TNG100 curve near log($M^*$)=10.8 is the same dip seen near log($M_{\rm vir})=12.5$ in Fig.~\ref{fig:TNG_full_plus_B22}.
We stress that the variations of the TNG100 results with $M_*$ or $M_{\rm vir}$ are due mainly to the variations in the gas fractions shown in Fig.~\ref{fig:f_cgm}, and only secondarily to the spread in $a_{P,{\rm th}}$ and $a_n$ illustrated in Fig.~\ref{fig:TNG_alpha_beta}. The overlap of the TNG100 curves with the gray strips demonstrates the overall validity of our PLM in capturing the behavior of the TNG100 simulation results.

For the three FSM points in Fig.~\ref{fig:Y_500_vs_M*} we set $M_*$ to the Milky Way value of $6\times 10^{10}$~M$_\odot$, which is a factor 2.03 larger than the Moster value for a virial mass of $1.3\times 10^{12}$~M$_\odot$ (see Fig.~\ref{fig:moster}). For the \citetalias{Bregman2022}, \citetalias{Das2023} and \citetalias{Planck2013} data the stellar masses are as directly observed. For the FSM models, $E^{-2}y(0.2)/M_{\rm vir,12}$ equals $3.9\times 10^{-9}$, $7.7\times 10^{-9}$, and $2.0\times 10^{-8}$. The \citetalias{Bregman2022} point is at $9.8\times 10^{-9}$. The line-of-sight Compton-$y$ parameter is not affected by projection corrections, and it is evident that our FSM20-maximal point is closest to the \citetalias{Bregman2022} observation, consistent with our results in \S~\ref{sec:fsm-b22}.

As for the \citetalias{Bregman2022} point, the \citetalias{Das2023} data lie well above the TNG100 predictions, by up to 0.5~dex, for $M_*<10^{11}$~M$_\odot$. As we have stated, this may be evidence for unrealistic CGM gas loss in the TNG100 simulation in this mass range, and/or higher pressures in the observed galaxy halos than predicted by the simulation. At larger stellar masses the \citetalias{Das2023} measurements do finally drop down and lie close to the TNG100 predictions.

In the righthand panel of Fig.~\ref{fig:Y_500_vs_M*} 
we include (as large open circles) the original \citetalias{Planck2013} data for the high mass galaxies and clusters in their tSZ survey. 
\citetalias{Planck2013} tabulated the Compton-${\tilde Y}_{500}$ (arcmin$^2$), as functions of stellar mass, $M_*$, from $1.0\times 10^{12}$ down to their $3\sigma$ limit of $1.8\times 10^{11}$~M$_\odot$. They used the \cite{Guo2011} semi-analytic model to compute a stellar-mass to halo-mass conversion that is essentially identical\footnote{In their Fig.~3, \cite{Planck2013} plot $M_{200}$ versus $M_*$. In our Fig.~\ref{fig:moster} we have converted their halo masses to $M_{100}=1.15M_{200}$ for consistency with the \cite{Moster2010} (and our) approximation for the virial mass, $M_{\rm vir}\simeq M_{100}$.} 
to the \cite{Moster2010} relation (see Fig.~\ref{fig:moster}). 
They then estimated $R_{500}$ for each $M_*$ assuming NFW mass profiles, and used the ``universal" \citet{Arnaud2010} pressure profile to convert the observed projected ${\tilde Y}_{\perp,500}$ to a spherical ${\tilde Y}_{500}$. \citetalias{Planck2013} did not correct for 2-halo or background contributions. 

However, as noted in several studies \citep{Vikram2017,Hill2018} the \citetalias{Planck2013} results are likely influenced or even dominated by the 2-halo contributions from nearby galaxies, especially at the low mass end of the survey. For example, for $M_{\rm vir}$ between $10^{13}$ and $3\times 10^{13}$~M$_\odot$, the 2-halo contribution may be $\sim 40\%$ at a mean projected radius $R_{500}\approx 350$~kpc \citep[see Fig.~3 in][]{Vikram2017}. In their reanalysis of the \citetalias{Planck2013} data, \citet{Hill2018} considered three models for the electron pressures (dubbed ``PL", ``UB" and ``CB") to infer the relative contributions of the intrinsic (1-halo) versus external (2-halo) terms to the observed $y$ parameters\footnote{The three \citet{Hill2018} models are for variations on the \citet{Battaglia2012} pressure profile. For PL (power-law) a pressure normalization term is included that varies as a power-law in mass. For UB (uncompensated-break) the normalization term is included only below a specified mass break-point. In CB (compensated break) the shape of the profile is further adjusted to account for gas redistribution due to feedback effects.}. In Fig.~\ref{fig:Y_500_vs_M*} the blue, red, and cyan lines are the \citet{Hill2018} estimates for the intrinsic 1-halo contributions for their three models. Below $M_*=1.3\times 10^{11}$~$M_\odot$ the intrinsic halo contributions to the \citetalias{Planck2013} results are small, and all three \citet{Hill2018} models are viable given the \citetalias{Planck2013} data alone. However, we remark that UB and CB are consistent with the TNG100 predictions for $M_*\gtrsim 1.5\times 10^{11}$~M$_\odot$. Furthermore, the UB and CB corrections are consistent with the \citet{Das2023} results near $M_*=1.5\times 10^{11}$~M$_\odot$, but PL is not.

In Fig.~\ref{fig:as_m_vir}, we plot the same data for $y(0.2)$ and $Y_{500}$ as in Fig.~\ref{fig:Y_500_vs_M*}, but now versus $M_{\rm vir}$ rather than $M_*$. 
For the \citetalias{Das2023} observations we again use the \citet{Moster2010} SMHM relation to convert the observed stellar masses to virial masses. The black curves are the TNG100 results, and as we have already emphasized the overall shapes, including the dips near log$(M_{\rm vir})=12.5$ and the rise towards higher masses, are due mainly to variations in the hot CGM gas fractions (Fig.~\ref{fig:f_cgm}). 

Interestingly, the FSM20-maximal and TNG100 predictions at $1.0\times 10^{12}$~M$_\odot$ are identical, as can be seen by considering the PLM parameters. For TNG100 at this mass, $\phi_T=0.80$, $f_{\rm hCGM}=0.45$, and a cool gas fraction $f_{\rm cCGM}=0.13$). {\oran{The FSM20 models are not quite as hot, both with $\phi_T=0.71$, but the FSM20-maximal hot gas fraction $f_{\rm hCGM}=0.59$ (with no room for cool gas) is larger than for TNG100.}} Together with the appropriate $a_{P,{\rm th}}$ and $a_n$ values the PLM expressions give $y(0.2)=5.5 \times 10^{-9}$ and $Y_{\rm 500}= 1.8 \times 10^{-4}$ ~kpc$^2$ for TNG100, and $3.5 \times 10^{-9}$ and $2.0 \times 10^{-4}$~kpc$^2$, for FSM-maximal, consistent with the results in Fig.~\ref{fig:as_m_vir}.

{\oran{\subsection{X-rays}
\label{sec:Xrays} }}

Finally, we remark on some important recent X-ray observations. As is well known, the hot gas that gives rise to the tSZ effect will also produce CGM X-ray emission \citep[e.g.,][]{Arnaud2010,Singh2018}, although the X-rays are more complicated to interpret because the emission efficiencies are sensitive to gas metallicity and density fluctuations, in addition to the overall gas pressures. \cite{Zhang2024} (\citetalias{Zhang2024}) have reported  eROSITA detections of 0.5-2~keV X-ray CGM emission at Milky Way mass scales via galaxy stacking methods \citep[see also][]{Oppenheimer2020}. They provide projected $\beta$-model fits for the X-ray surface brightness profiles (similar to the \citetalias{Bregman2022} projected $\beta$ model for the Compton-$y$ parameter, see \S~\ref{sec:Bregman}). They assume 
\beq
\nonumber
S_X \ = \ S_{X,0} [1 + (r_\perp/r_c)^2]^{-3\beta + 1/2}\ \ \ \ {\rm erg} \ {\rm s}^{-1} \ {\rm kpc}^{-2},  \ \,
\label{Zhang}
\eeq
and for a virial mass of $1.4\times 10^{12}$~M$_\odot$ they find log($S_{X,0}$)=35.5$^{+0.5}_{-0.5}$, $r_c=30^{+40}_{-25}$~kpc, and $\beta=0.42^{+0.12}_{-0.11}$ (see their Table 4).  \citetalias{Zhang2024} infer hot gas CGM baryon fractions for a range of metallicities and temperatures, assuming isothermal conditions. For a metallicity of 0.3$\times$Solar, and a temperature $T=1.3\times 10^6$~K, they infer a hot CGM gas fraction $f_{\rm hCGM}=0.51$ (see their Fig.~9). For our PLM, their assumed temperature corresponds to $\phi_T=2.56$, and with $a_{P,{\rm th}}=a_n=3\beta=1.26$. Our equations~(\ref{eq:Yvir}) and (\ref{eq:Y500}) then give $Y_{\rm vir}/M_{\rm vir}^{5/3}=9.53\times 10^{-4}$~kpc$^2$, and $Y_{500}/M_{\rm vir}^{5/3}=2.9\times 10^{-4}$~kpc$^2$ for the \citetalias{Zhang2024} X-ray galaxies. We include these quantities in Tables~\ref{tab:PLM_parameters} and \ref{tab:fsm_B22}. For the isothermal conditions \citetalias{Zhang2024} assumed, the relatively flat density profile implies that the X-ray luminosity is dominated by the outer regions of the CGM.

We show our inferred tSZ $Y_{500}/M_{\rm vir,12}^{5/3}$ value for the \citetalias{Zhang2024} galaxies in our Fig.~\ref{fig:observations} as the red triangle{\oran{, with standard errors estimated by sampling the parameter space for $r_c$ and $\beta$ as given by \citetalias{Zhang2024}}}. It is satisfying that the predicted tSZ signal inferred from the X-rays is very close to the directly observed tSZ parameters found by \citetalias{Bregman2022} and \citetalias{Das2023}, lending further credence to all three sets of these challenging observations.

\vspace{0.5cm}
 
\section{Discussion and Summary} \label{sec:discussion}

Our objective is to develop physically motivated models for the gaseous circumgalactic medium (CGM) of star-forming $L^*$ galaxies that are consistent with observations of the thermal Sunyaev-Zeldovich (tSZ) effect. 

In this paper we analyze recent observations, reported by \cite{Bregman2022} and \cite{Das2023} (\citetalias{Bregman2022} and \citetalias{Das2023}), of tSZ cosmic-microwave-background (CMB) distortion signals arising in the CGM surrounding $L^*$ galaxies, or at virial dark-matter halo mass scales down to $10^{12}$~M$_\odot$. Both observational studies make use of galaxy stacks to detect the tSZ signatures, and to extract the line-of-sight Compton-$y$ parameters and $Y$ pressure integrals that are the basic measures of the tSZ effect.  For the CGM of an individual galaxy with virial mass $M_{\rm vir}\sim 3\times 10^{12}$~M$_\odot$, \citetalias{Bregman2022} and \citetalias{Das2023} report characteristic (1-halo) Compton-$y$ parameters (or CMB distortions $\Delta T/T)$ of order a few times 10$^{-8}$. It is remarkable that such small tSZ signals can now be detected. These measurements provide new constraints on the properties of the hot Compton-scattering gas and the CGM baryon content and distributions for individual galaxies at Milky Way mass scales.  We compare the tSZ measurements to recent (0.5-2 keV) X-ray observations of galaxies at these mass scales \cite{Zhang2024} (\citetalias{Zhang2024}) and show that they are consistent with the tSZ measurements.

We analyze the tSZ observations using three interrelated theoretical tools. First are the phenomenological \cite{Faerman2017} and \cite{Faerman2020} hydrostatic equilibrium models for the hot CGM around Milky Way type galaxies. These models reproduce UV observations of OVI absorption lines in the COS-Halos sample of galaxies and X-ray observations of OVII and OVIII absorption in the Milky Way. Second is an analytic power-law model (PLM) for the thermal pressures and density distributions for the hot CGM gas, with analytic expressions for the radially dependent Compton-$y$ and $Y$ profiles. Third is the TNG100 cosmological galaxy evolution hydrodynamics simulation, which we use to construct Compton-$y$ maps and to compute $Y$ pressure integrals for a subset of the simulated halos with a
wide range of virial masses.

In \S~\ref{sec:fsm} we review the FSM models, dubbed ``\citetalias{Faerman2017} isothermal" and ``\citetalias{Faerman2020} isentropic,"  
where we also introduce a gas-rich version of the isentropic model that we name FSM20-maximal. These three models provide specific predictions for the thermal electron pressures, and the Compton-$y$ and $Y$ integrals, for halos with virial mass $M_{\rm vir}= M_{100}=1.0\times 10^{12}$~M$_\odot$, as illustrated in Fig.~\ref{fig:FSM_PLM}.  FSM20-maximal is an ``upper-limit" isentropic model: It contains a cosmologically maximal amount of hot halo gas to enhance the tSZ signals, but with little room for any cool CGM gas. 

An important property of the FSM models is that the resulting hydrostatic thermal electron pressure profiles vary approximately as simple radial power-laws within the CGM spheres. This motivates our power-law model described in \S~\ref{sec:PLMprofiles}. The PLM is for the hot CGM component, and can be used to describe the FSM models, the hot gas distributions in the TNG100 halos, and the observations. For a given $M_{\rm vir}$, the PLM structures depend on the hot gas mass fraction, $f_{\rm hCGM}$, relative to the expected cosmic baryon mass within the halo; the ratio $\phi_T\equiv T(R_{\rm vir})/T_{\rm vir}$ of the actual gas temperature at the virial radius, $T(R_{\rm vir})$, to the halo virial temperature $T_{\rm vir}$; and the power-law indices, $a_{P,{\rm th}}$ and $a_n$, for the radially dependent electron pressures and hot hydrogen gas densities through the halo out to the virial radii. We focus on isothermal ($a_n=a_{P,{\rm th}}$) or isentropic ($a_n=3a_{P,{\rm th}}/5)$ options. A given thermal gas pressure profile may or may not be in hydrostatic equilibrium (HSE) within the gravitational potential of the parent halo, and we state a condition for HSE in terms of $a_{P,{\rm th}}$ and $\phi_T$.  

Alternate analytic models for the CGM may also be well represented by PLMs \citep[see Fig.~2 of][]{Singh2024}. This includes cooling flows \citep{Stern2019}, precipitation models \citep{Sharma2012,Voit2019} and baryon pasting models \citep{Ostriker2005,Osato2023}.

In \S~\ref{sec:CGMthermal} we define and discuss observational constraints on the cool ($T < 3\times 10^4$~K), intermediate, and hot ($T>0.4\times T_{\rm vir}$) gas contents of the CGM surrounding $L^*$ galaxies. 
Depending on how the low ions that trace the cool gas are interpreted, 
as patchy, with most of the variations in observed column densities occurring within individual galaxies, or smooth, with most of the variation occurring among galaxies,
the inferred mean cold gas fractions are $\sim 20\%$ or $\sim 5\%$. 
Our sample of TNG100 $L^*$ galaxies contains cool and intermediate temperature gas fractions of $\sim 20\%$, but we have not determined whether the spatial distribution and volume filling factors of this gas favor the patchy or smooth interpretation.


In \S~\ref{sec:tszplm} we write down our PLM expressions for the Compton-$y$ parameter, and the spherical and projected pressure integrals $Y$ and $Y_\perp$ in terms of the parameters $f_{\rm hCGM}$, $\phi_T$, $a_{P,{\rm th}}$, and $a_n$.  In Fig.~\ref{fig:FSM_PLM} we show the excellent match between the FSM solutions and our PLM forms for the thermal gas pressures, and for the Compton-$y$ and $Y_\perp$ integrals.

{\oran{In \S~\ref{sec:PLMFSM} we describe our PLM fitting method as applied to the pressure and density curves in the analytic FSM models (and the TNG100 simulations). We derive the best-fitting PLM parameters for the FSM models, and compare the implied PLM and intrinsic FSM Compton-$y$ and $Y_\perp$ curves.}}

In \S~\ref{sec:Bregman} we review the tSZ observations presented by \citetalias{Bregman2022} with focus on their primary result, which is a radial Compton-$y$ profile (Fig.~\ref{fig:PLM_bregman}) inferred from their stack of 11 nearby ($\lesssim 10$~Mpc) star-forming galaxies. {\oran{We use an alternate method to fit PLMs to the observations.}} In \S~\ref{sec:plm-b22} we fit the observed $y$ profile {\oran{directly}} with a PLM for the mean stack virial mass of $2.75\times 10^{12}$~M$_\odot$, from the inner 29~kpc out to the virial radius of 376~kpc. Our best-fitting (MCMC) pressure slope is steep ($\apth=2.0$) and is consistent with the projected $\beta$-model fit presented by \citetalias{Bregman2022} in their analysis. They assumed isothermal conditions and found $\beta=0.6$, which to a good approximation corresponds to a PLM with  $a_n=\apth=3\beta=1.8$. However,
we argue that the gas in these halos is unlikely to be isothermal: The observed value of the Compton-$y$ parameter requires $T>1.1\times 10^6$~K. But for the gas to not be expanding outwards, one requires $T<1.1\times 10^6$~K. It seems unlikely that the typical CGM observed by \citetalias{Bregman2022} would be expanding outwards, so we conclude that it is unlikely that these CGMs are isothermal.
However, for isentropic conditions, HSE is possible for temperatures at the virial radius between $6.2\times 10^5$ and $1.1\times 10^6$~K. (In the isentropic model the gas temperature increases inward from the virial radius.) The implied hot gas fractions, $f_{\rm hCGM}$, are large, between 0.45 and 0.83. Non-thermal pressure support is also required, except at the highest boundary temperature, for which thermal pressure gradients alone are sufficient to maintain HSE. 

Next, in \S~\ref{sec:fsm-b22} we compare our three FSM model predictions 
to the \citetalias{Bregman2022} data (and to our best-fitting PLM). After scaling out the mass ratio of 2.75 between the \citetalias{Bregman2022} stack and the Milky Way based FSM models, we find that except for the innermost radii the observations are bracketed (within factors $\lesssim 2$) by the higher-pressure isothermal and lower-pressure isentropic FSM models. The isentropic models are favored in a $\chi^2$ comparison (see Fig.~\ref{fig:FSM_bregman}), as is also indicated by our PLM for which HSE is possible if the gas is isentropic. The observed $y$ parameters through the inner regions (within $\lesssim 0.1R_{\rm vir}$) are high, suggesting that the CGM thermal pressures close to the galaxies are larger than predicted by the FSM models. 

In \S~\ref{sec:tng} we turn our focus to the TNG100 simulation, and present our post-processing methodology for computing the Compton parameters. We describe our TNG100 galaxy/halo sample in \S~\ref{subsec:maps}. It consists of halos containing both star-forming (SF) and quenched galaxies from  $1.0\times 10^{11}$ to $3.0\times 10^{14}$~M$_\odot$.
We construct 2D Compton-$y$ maps, 1D radial $y$ profiles, and $Y$ integrals, given the TNG100 outputs for the CGM gas particle densities and temperatures. 

In \S~\ref{sec:fracPLM} we extract the cool ($T < 3\times 10^4$~K) versus hot ($T > 0.4T_{\rm vir}$) mass fractions, $f_{\rm cCGM}$ and $f_{\rm hCGM}$, within the simulated halos (Fig.~\ref{fig:f_cgm}). We show that the tSZ signals are dominated by the hot CGM components (Fig.~\ref{fig:Y_M_of_T}). The hot gas components are well fit with PLMs, and we compute the PLM parameters $\phi_T$, $a_{P,{\rm th}}$, and $a_n$, for each halo in our set (Figs.~\ref{fig:phi_T} and \ref{fig:TNG_alpha_beta}). {\oran{We show detailed results for two specific examples in Figs.~\ref{fig:phase} and \ref{fig:y_map}. The PLM fits may become inaccurate at small radii, $r/R_{\rm vir}\lesssim 0.1$, especially when the pressure slopes are steep with $a_{P,{\rm th}}\rightarrow 3$.}}

Our PLM fits for the TNG100 halos enable simple comparisons to the FSM model results and to the observations. For star-forming $L^*$ galaxies, with virial masses from $3\times 10^{11}$ to $3\times 10^{12}$~M$_\odot$,
we find $1.8\la \apth\la 2.3$. The values of $a_n$ are generally intermediate between the isothermal ($a_n=\apth$) and isentropic ($a_n=\frac 35\apth$) models, so that typically $a_T=\apth-a_n\simeq \frac 15 \apth\sim 0.4$. 
The value of the temperature at the virial radius for $L^*$ galaxies, is between $0.81$ and $0.96$ times $T_\vir$ (see Fig.~\ref{fig:phi_T}) consistent with our definition of the virial temperature (Eq.~[\ref{eq:Tvir}]).


In \S~\ref{sec:comparison} we compare the \citetalias{Bregman2022} to the \citetalias{Das2023} observations, with reference to our FSM, PLM and TNG100 predictions. We also discuss the tSZ data presented by \cite{Planck2013} (\citetalias{Planck2013}). The observations and theoretical predictions are presented together in various formats (Figs.~\ref{fig:TNG_full_plus_B22} through \ref{fig:as_m_vir}). We focus on observations and predictions for $y(0.2)$ (the mean Compton-$y$ at projected radius 0.2$R_{\rm vir}$) and the commonly used $Y_{\rm 500}$, the spherical pressure integral within the overdensity radius $R_{500}$. 

The \citetalias{Das2023} tSZ data are for a large set of galaxies at $z\sim 0.2$ with stellar masses from $6.3 \times 10^{10}$ to $1.6 \times10^{11}$~M$_\odot$ (or $M_{\rm vir}$ from $2.5 \times 10^{12}$ to $1.5 \times 10^{13}$~M$_\odot$), and overlap the \citetalias{Bregman2022} galaxy masses near $M_{\rm vir}=2.75\times 10^{12}$~M$_\odot$. 
At this mass scale, and within the errors, the two data sets are consistent for both $y(0.2)$ and $Y_{500}$, with nominal values of $E(z)^{-2/3}Y_{500}/M_{\rm vir,12}^{5/3}$ between $2.5\times 10^{-4}$ and $5.0\times 10^{-4}$~kpc$^2$. Remarkably, the recent \citetalias{Zhang2024} detections of CGM X-ray emissions from a large sample of $L^*$ galaxies imply $Y_{500}$ pressure integrals essentially identical to those found by the tSZ measurements, assuming isothermal conditions (see Fig.~\ref{fig:observations}).  


We also consider the higher mass \citetalias{Planck2013} tSZ measurements ($M_{\rm vir}$ from $10^{13}$ to $10^{14.5}$~M$_\odot$). Following the \cite{Hill2018} analysis of the \cite{Planck2015b} data release, including corrections for the (dominating) 2-halo terms, the observed signals can be brought into harmony with the TNG100 predictions for $Y_{500}$, and the high mass end of the \citetalias{Das2023} observations (see Fig.~\ref{fig:as_m_vir}).

The \citetalias{Bregman2022} and \citetalias{Das2023} observations for $y(0.2)$ and $Y_{500}$ are bracketed by the isothermal and isentropic \citetalias{Faerman2017} and \citetalias{Faerman2020} models that also reproduce the OVI, OVII, OVIII, emission and absorption line measurements in the CGM of $L^*$ galaxies.  However the TNG100 predictions for star-forming halos near $M_{\rm vir}\sim 3\times 10^{12}$~M$_\odot$ 
lie $\sim 0.5$~dex below the measurements. This disparity is likely due to excessive feedback mass loss from the TNG100 halos, leading to a reduction in the CGM gas fractions at this mass scale. 

For isothermal conditions, and even with large hot gas fractions, e.g.~77\% as in \citetalias{Faerman2017} or 83\% as in our isothermal PLM for the \citetalias{Bregman2022} stack, 
 the tSZ observations imply temperatures at the virial radii $\gtrsim 10^6$~K, significantly greater than the $L^*$ virial temperatures. We emphasize that there is no direct observational evidence for such high gas temperatures near the virial radii of $L^*$ galaxies. 
Non-isothermal density and pressure profiles, with $a_n<a_{P,{\rm th}}$, for which $a_T>0$ and the temperatures rise inwards, as we broadly find for TNG100, or as assumed in the \citetalias{Faerman2020} models and our isentropic \citetalias{Bregman2022} PLM, appear favored by the tSZ (and X-ray) observations. These allow lower temperatures at the outer halo edges and more readily enable gravitational binding (HSE) of the tSZ gas to the halos.  For isentropic models the hot gas CGM fractions must still be large, $\gtrsim 50\%$, not including cool and intermediate temperature gas that may contribute an additional 5-20\% to the CGM budget.

Our focus in this paper has been analysis of tSZ observations from the {\oran{hot gas}} in the CGM of $L^*$ galaxies. {\oran{Simultaneous tSZ and UV/Xray emission/absorption line observations of the CGM around individual galaxies would be very valuable, and together enable improved constraints on the gas metallicities, temperatures, and mass distributions.}} It is satisfying that three independent data sets, \citetalias{Bregman2022} and \citetalias{Das2023} for the tSZ effect, and \citetalias{Zhang2024} for the X-rays give consistent constraints on the hot gas baryon content and pressures in the CGM of such galaxies. The data are well described by our FSM models that also account for the high-ionization oxygen absorbers and emitters. Our PLM formalism provides a simple way to parameterize observations and simulation outputs, and will be useful in analyzing future tSZ observations to ever lower galaxy masses.

\begin{acknowledgements}
We thank Joel Bregman, Sanskriti Das, Drummond Fielding, Colin Hill, Ranita Jana, Fiona McCarthy, Jason Prochaska, Eliot Quataert, Kartick Sarkar, Rachel Somerville, David Spergel, Jonathan Stern, and Jessica Werk, for helpful advice and discussions. {\oran{We thank the referee for valuable comments and suggestions.}} This work was supported by the German Science Foundation via DFG/DIP grant STE/ 1869-2 GE/ 625 17-1, by the Center for Computational Astrophysics (CCA) of the Flatiron Institute, and by the Mathematical and Physical Sciences (MPS) division of the Simons Foundation, USA. The research of
CFM was supported in part by the NASA ATP grant 80NSSC20K0530 and in part by grant NSF PHY-2309135 to the Kavli Institute for Theoretical Physics (KITP). YF is supported by NASA award 19-ATP19-0023 and NSF award AST-2007012.
\end{acknowledgements}

\vspace{2cm}

\bibliographystyle{mnras}
\bibliography{references}{}

\appendix

\section{Evaluation of 
$I_{\lowercase{y},\lowercase{a}_P}(\lowercase{b})$ 
and $I_{Y,\lowercase{a}_P}(\lowercase{b})$
for special cases}
\label{sec:app_I}
    We first consider the case of small impact parameters.
    For small $b$ and $a_{P,{\rm th}}>1$, we can approximate Equation (\ref{eq:dimy}) as:
    \beq
    I_{y,a_{P,{\rm th}}}(b \rightarrow 0) = \int_0^{\infty}\frac{1}{ (1+u^2)^{a_{P,{\rm th}}/2}} \ du \ \ \ .
    \eeq
We then perform the following variable exchange:
    \beq
    u^2 = t~~\rightarrow~~dt=2udu=2\sqrt{t}du
    \eeq
to get:
    \beq
    I_{y,a_{P,{\rm th}}}(b \rightarrow 0) = \frac{1}{2} \int_0^{\infty}\frac{t^{-1/2}}{ (1+t)^{a_{P,{\rm th}}/2}} \ dt \ \ \ .
    \eeq
This expression matches the following identity for the Beta function:
    \beq
    B(a, b) = \int_0^{\infty}\frac{t^{a-1}}{ (1+t)^{a+b}} \ dt = \frac{\Gamma(a)\Gamma(b)}{\Gamma(a+b)} \ \ \ ,
\eeq
    so that for $a = 1/2$ and $b=(a_{P,{\rm th}} - 1)/2$ we obtain
    \beq
    I_{y,a_{P,{\rm th}}}(b \rightarrow 0) = \frac{1}{2}\frac{\Gamma(\frac{1}{2})\Gamma(\frac{a_{P,{\rm th}}-1}{2})}{\Gamma(\frac{a_{P,{\rm th}}}{2})} \ \ \ .
    \label{eq:iyz}
    \eeq
Note that 
$B(a,b)$ is defined only for $a,b>0$ and therefore this approximation only holds for $a_{P,{\rm th}} > 1$.
For $a_{P,{\rm th}}\rightarrow 1$, we have
\beq
I_{y,a_{P,{\rm th}}}\rightarrow \frac{1}{a_{P,{\rm th}}-1}.
\eeq

Next, for small $b$ and $a_{P,{\rm th}}<1$, equation (\ref{eq:dimy}) gives
\begin{eqnarray}
I_{y,a_{P,{\rm th}}}&\rightarrow&\int_0^{\sqrt{1/b^2-1}}u^{-a_{P,{\rm th}}}du,\\
&\rightarrow& \frac{b^{-(1-a_{P,{\rm th}})}}{1-a_{P,{\rm th}}},
\end{eqnarray}
so that 
the Compton-$y$ parameter approaches a constant in this case: 
\beq
y(b)\propto b^{1-a_{P,{\rm th}}}I_{y,a_{P,{\rm th}}}\rightarrow \frac{1}{1-a_{P,{\rm th}}}.
\eeq
    
    Analytic expressions for $I_{y,a_{P,{\rm th}}}(b)$ can be obtained for integer values of $a_{P,{\rm th}}$. For $a_{P,{\rm th}}=0$, Equation (\ref{eq:dimy}) becomes:
    \beq
    I_{y,0}(b) = \int_0^{\sqrt{1/b^2-1}}du = \sqrt{1/b^2-1}\ \ \ .
    \label{eq:iyz2}
    \eeq

For small $b$ this becomes 
    $I_{y,0}(b \rightarrow 0) = {1}/{b}$.
    For $a_{P,{\rm th}}=1$ we have
    \beq
    I_{y,1}=\ln\left(1/b+\sqrt{1/b^2-1}\right),
    \label{eq:iyo}
    \eeq
    which diverges as $b\rightarrow 0$.
    For $a_{P,{\rm th}}=2$,
    \beq
    I_{y,2}=\tan^{-1}\sqrt{1/b^2-1},
    \eeq
    which approaches $\pi/2$ as $b\rightarrow 0$.
Values of $I_{y,a_{P,{\rm th}}}(b)$ for $a_{P,{\rm th}} = 0, 1.5, 2, 2.5$ are shown in the left panel of Fig. {\ref{fig:dimensionless_y}}. 

Fewer analytic results are available for $I_{Y,a_{P,{\rm th}}}$. For $a_{P,{\rm th}}=0$, equation (\ref{eq:iyz2}) implies
\beq
I_{Y,0}=\frac{1}{b^3}\left[1-(1-b^2)^{3/2}\right].
\eeq
For $a_{P,{\rm th}}=1$ and $b\rightarrow 0$, equation (\ref{eq:iyo}) implies
\beq
I_{Y,1}\rightarrow \ln\left(\frac{2}{b}\right)+\frac 12.
\eeq
For $a_{P,{\rm th}}>1$ and $b\rightarrow 0$, equation (\ref{eq:iyz}) implies
\beq
I_{Y,a_{P,{\rm th}}}\rightarrow \frac{1}{2}\frac{\Gamma(\frac{1}{2})\Gamma(\frac{a_{P,{\rm th}}-1}{2})}{\Gamma(\frac{a_{P,{\rm th}}}{2})} b.
\eeq

\section{What if the CGM extends beyond the virial radius?}
\label{app:what}

    In our PLM, we consider the virial radius, $R_{\rm vir}$ to be the limit of the CGM, and assume that there is no contribution to the Compton-$y$ parameter from outside of it. However, the definition of $R_{\rm vir}$ can vary between works, e.g.~\citealp{Das2023}, who used an overdensity $\Delta=200$ when defining their virial masses. Even the very assumption that the CGM does not extend beyond the virial radius has been tackled. For example, \citetalias{Faerman2020} take the CGM radius to be $1.1 R_{\vir}$,  motivated by the O VI measurements from the eCGM survey \citep{Johnson15}. 

To account for the possibility of the CGM not being limited to the virial radius, we expand on our derivation in \S~\ref{sec:PLM}. Our CGM virial mass (Eq.~[\ref{eq:Mcgm}]), temperature (Eq.~[\ref{eq:Tvir}]), and hydrogen density (Eq.[\ref{eq:nHvir}]) are all defined based on their properties at the virial radius, and remain unaltered. The total CGM mass in the PLM that now extends to $R_{\rm CGM}$ becomes:
    \beq
    M_\cgm=\left(\frac{R_\cgm}{R_\vir}\right)^{3-a_n} M_\cv.
    \eeq
    
    We define our new impact parameter relative to the CGM radius instead of the virial radius: 
    \beq
    b_\cgm \equiv \frac{r_{\perp} }{R_\cgm}.
\eeq
     The Compton-$y$ parameter is generalized to:
\beq
y_\cgm(b_\cgm)=\left(\frac{R_\cgm}{R_\vir}\right)^{1-a_{P,{\rm th}}}y(b_\cgm),
\eeq
where $y(b_\cgm)$ is given by equation (\ref{eq:Compton_y}) with
$b$ replaced by $b_\cgm$. {\oran{Note that the right-hand side of this equation varies as
\beq
\left(\frac{R_\cgm}{R_\vir}\right)^{1-a_{P,{\rm th}}}\left(\frac{r_\perp}{R_\cgm}\right)^{1-a_{P,{\rm th}}}I_{y,a_{P,\rm th}}\left(\frac{r_\perp}{R_\cgm}\right),
\eeq
which increases monotonically with $R_\cgm$ since it cancels in the first two factors and $I_y(b)$ increases monotonically as $b$ decreases.}}

    The spherical ${Y}(r)$ 
    is unaffected by the change from $R_\vir$ to $R_\cgm$: 
    \beq
Y_\cgm(r)=Y(r)=Y_{\rm vir}
    \Bigl(\frac{r}{R_{\rm vir}}\Bigr)^{3-a_{P,{\rm th}}}~~~~~(r\leq R_\cgm),
    \eeq
    except that now the radius can extend to $R_\cgm\neq R_\vir$.
    For the projected ${Y}_{\perp}(r_\perp)$, we obtain:
    \beq
    Y_{\perp,\cgm}(b_\cgm)=\left(\frac{R_\cgm}{R_\vir}\right)^{3-a_{P,{\rm th}}}Y_\perp(b_\cgm),
    \eeq
    where $Y_\perp$ is given by equation (\ref{eq:Ypint}).
   For example, \citetalias{Faerman2020} adopted $R_\cgm=1.1R_\vir$ so that $b_\cgm=0.91 b$ in the above expressions.

{\oran{
\section{Comparing pressure profiles}
\label{app:p_comparison}}}

    {\oran{
    In their studies, \cite{Hill2018} and \citetalias{Das2023} assume pressure profiles that are inherently different from our simple power law model. \cite{Hill2018} and \citetalias{Das2023} both use a 
    generalized NFW profile for the pressure: }}

    {\oran{
    $$
    P(r) = P_0 (r/r_0)^{\gamma} [1+(r/r_0)^{\alpha}]^{\beta} ~~~,
    $$
    }}
    {\oran{
    with functional parameters (see Table~\ref{tab:pressure_parameters}) that were obtained from either X-ray observations \citep{Arnaud2010},  or simulations 
    \citep{Battaglia2012}.}}
    
    {\oran{For comparisons, in Fig.~\ref{fig:p_comparison} (top panel) we plot the electron pressure curves used by \cite{Hill2018} and \citetalias{Das2023} together with our best-fitting PLM ($a_{P,{\rm th}}=2.03$) for the \citetalias{Bregman2022} data (see \S~\ref{sec:plm-b22}). Each pressure curve is normalized such that it provides a best-fitting Compton $y(b)$ curve for the \citetalias{Bregman2022} data, as shown in the lower panel. We also plot the \citetalias{Bregman2022} truncated-$\beta$-model fit (Eq.~[\ref{eq:B22fit}]) to the $y$-profile. (The \citetalias{Bregman2022} fit does not have a corresponding analytic form for the pressure.)  The $\chi^2$ measures for the predicted $y$-profiles using the \citetalias{Bregman2022} $\beta$-model, \cite{Arnaud2010}, \cite{Battaglia2012} and our PLM pressures, are 9.9, 10.0, 9.6, and 8.0 respectively. The differences are small, but we conclude that our PLM provides a marginally better fit to the data.  }}

    \begin{figure}[]
        \includegraphics[width=0.45\textwidth]{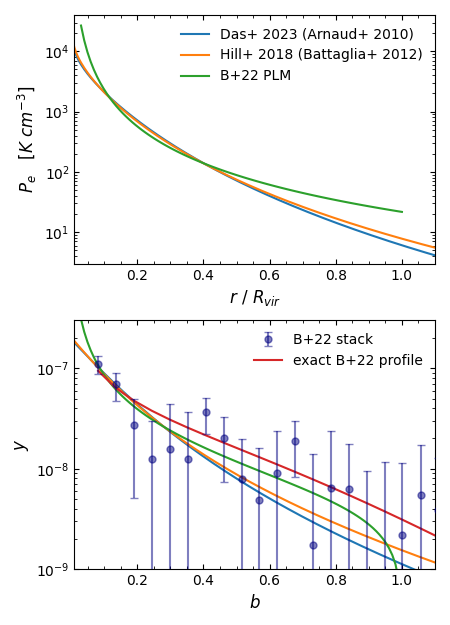}
        \caption{{\oran{Top panel: Pressure profiles used by \citetalias{Das2023} (blue), \cite{Hill2018} (orange), and our PLM with $a_{P,{\rm th}}=2.03$ (green). Each pressure curve is normalized such that it provides a best fitting Compton $y(b)$ curve for the \citetalias{Bregman2022} data. Lower panel: \citetalias{Bregman2022} Compton-$y$ data (blue points), and best-fitting $y$ profiles, as functions of impact parameter $b=r_\perp/R_{\rm vir}$, computed using the pressure curves in the upper panel. The red curve is the \citetalias{Bregman2022} truncated-$\beta$-model.}}}
        \label{fig:p_comparison}
    \end{figure}

    \input{pressure_table.tex}

\end{document}

%% file: tb_fsm.tex
\renewcommand{\arraystretch}{1.2} 

\begin{table*}
\centering
	\caption{Comparison of observables across the FSM models for $L^*$ galaxies}
	\begin{tabular}{| l || c | c | c | c | c |}
	
        \midrule
	& FSM17 & \multicolumn{2}{ c |}{FSM20} & Observations & References \\
        & Fiducial           & Fiducial             & Maximal                 &	             &            \\
        \midrule
        $M_{\rm vir}$~(\msun)  & \multicolumn{3}{ c |}{$1\times 10^{12}$} & $0.8-1.6 \times 10^{12}$ & MW: (a1) (a2) \\
        $\sigma_{\rm turb}~({\rm km~s^{-1}})$ & \multicolumn{3}{ c |}{$60$} & --- & --- \\
        \cline{2-4}
        $R_{\rm vir}$~(kpc)    & $250$ & \multicolumn{2}{ c |}{$258$} & --- & --- \\
        $R_{\rm CGM}$~(kpc)    & $250$ & \multicolumn{2}{ c |}{$283$} & --- & --- \\
        $T(R_{\rm CGM})$~(K)   & $1.14 \times 10^6$ & \multicolumn{2}{ c |}{$2.4 \times 10^5$} & --- & --- \\
        $\alpha~(\alpha_{\rm OML}$) & $1.9~(2.1)$ & \multicolumn{2}{ c |}{$2.1~(3.2)$} & --- & --- \\
        \cline{3-4}
        $Z'$ ($Z_{\odot}$)     & $0.5$ & $0.3-1.0$ & $0.15-0.5$ & --- & --- \\

	\midrule
        $P(8.5~{\rm kpc})~({\rm K~\cmv})$      & $1863$  & $1350$  & $2700$ & $500-3000$ & MW: (c) (d) (a3) \\
        $P(R_{\rm vir})~({\rm K~\cmv})$  & $116$  & $21$   & $43$       & --- & ---   \\
 

        $M_{\rm hCGM}(R_{\rm vir})$~(\msun) & $1.2 \times 10^{11}$ & $4.6 \times 10^{10}$ & $9.2 \times 10^{10}$ & $0.35-2.3 \times 10^{11}$ & MW: (a4) (l) \\


        $f_{\rm hCGM}(R_{\rm vir})$   & $0.77$ & $0.29$ & $0.59$ & --- & --- \\

        
        $DM_{\rm LMC}~({\rm \cmv~pc})$   & $17$  & $9$      & $18$  & $ 23 \pm 10$	& MW: (e) (f)  \\
        $DM_{\rm total}~({\rm \cmv~pc})$ & $\approx 25$ & $15-20$    & $30-40$ & $52-111$ & MW: (a6) (a7) (a8)   \\
	${\bar n}_{\rm H}(R_{\rm vir})~(\cmv)$ & $5.4 \times 10^{-5}$ & $1.9 \times 10^{-5}$ & $3.8 \times 10^{-5}$ & $\sim 2.5 \times 10^{-5}$ & MW: (i) \\
 
	\midrule
        $N_{\rm OVI}~(\cmc)$ & $2.5 \times 10^{14}$ & $2.9 \times 10^{14}$ & $3.4 \times 10^{14}$ & $1.9~(0.8-4.1) \times 10^{14}$ & MW: (a9) (a10) \\
	$N_{\rm OVII}~(\cmc)$ & $1.6 \times 10^{16}$ & $1.3 \times 10^{16}$ & $1.3 \times 10^{16}$ & $1.4~(1.0-2.0) \times 10^{16}$ & MW: (j) (k) (o) \\
	$N_{\rm OVIII}~(\cmc)$ & $3.8 \times 10^{15}$ & $3.1 \times 10^{15}$ & $2.7 \times 10^{15}$ & $3.6~(2.2-5.7) \times 10^{15}$ & MW: (l) (o) \\
	${N_{\rm OVII}/N_{\rm OVIII}}$      & $4.5$    &  $4.2$     & $4.9$ & 	$4.0~(2.8-5.6)$    & MW: (b) (k) (l) \\

    \midrule
    \multicolumn{6}{| l |}{(i) For FSM17 and FSM20 the assumed overdensities are $\Delta = 120$ and 110 respectively, which is why the virial radii differ}\\ 
    \multicolumn{6}{| l |}{\ \ \ \ from 267~kpc for the nominal $\Delta = 100$ we assume in this paper (see Eq.~\ref{eq:Rvir}). }\\
    \multicolumn{6}{| l |}{(ii) For FSM17, we give the mass-weighted temperature of the OVI component plus the remaining initial hot gas at $R_{\rm CGM}$ (see text).}\\ 
    \multicolumn{6}{| l |}{\ \ \ \ The pressures are as given by Eq.~(\ref{eq:warmhot}) following the formation of the OVI component. }\\
    \multicolumn{6}{| l |}{(iii)  For FSM20 the CGM extent is a free parameter, and the fiducial value is $R_{\rm CGM} = 1.1 R_{\rm vir}.$} \\
    \multicolumn{6}{| l |}{(iv) The fraction in non-thermal pressure in the FSM20 model varies with radius, and the values given here are at $R_{\rm CGM}$.} \\
    \multicolumn{6}{| l |}{$\ \ \ \ \ (\alpha-1)$ is the fraction in cosmic rays and magnetic fields, whereas $\alpha_{\rm OML}\equiv P_{\rm tot}/P_{\rm th}$ includes the turbulent pressure.} \\
    \multicolumn{6}{| l |}{(v) The metallicity profile in FSM20 is given by $Z' = Z'_0\left[ 1 + (r/r_Z)^2\right]^{-1/2}$ (see \S~2.2 in FSM20).} \\
    
    \midrule
    References:
    & \multicolumn{5}{| l |}{(a) \cite{Tumlinson11}, (b) \cite{Faerman2017} (c) \cite{Wolfire03}} \\
    & \multicolumn{5}{| l |}{(d) \cite{Dedes10} (e) \cite{Anderson10} (f) \cite{PZ19}} \\
    & \multicolumn{5}{| l |}{(g) \cite{Grcevich09} (h) \cite{Salem15} (i) \cite{Blitz00}} \\
    & \multicolumn{5}{| l |}{(j) \cite{Bregman07} (k) \cite{Fang15} (l) \cite{Gupta12} (o) \cite{Das19a}} \\
    & \multicolumn{5}{| l |}{(a1) \cite{BHG16} (a2) \cite{Posti19} (a3) \cite{Putman12}} \\
    & \multicolumn{5}{| l |}{(a4) \cite{Miller15} \cite{Platts20} (a7) \cite{Cook23} (a8) \cite{Ravi23}} \\
    & \multicolumn{5}{| l |}{(a9) \cite{Savage03} (a10) \cite{Zheng15} (a11) \cite{Johnson15}} \\

    \bottomrule
    \end{tabular}
    \vspace{0.5 cm}
  \label{tab:tb_fsm}
\end{table*}

%% file: tb_FSM_B_TNG_PLM_v2.tex
\begin{table*}[]
\centering
\caption{PLM parameters.}
\hspace*{-3cm}
\begin{tabular}{|l|c|c|c|c|c|c|c|c|}
\hline
                             & $M_{\rm vir, 12}$ & $f_{\rm hCGM}$ & $\phi_{T}$ & $a_{P,\rm th}$ & $a_n$ & $\phi_T^*$ & $\frac{y(0.2)}{M_{\rm vir, 12}} (\times 10^{-8})$ & $\frac{Y_{500}}{ M_{\rm vir, 12}^{5/3}} (\times 10^{-4}$ kpc\textsuperscript{2}) \\ \hline
FSM17                        & 1.0     & 0.77       & 3.05     & 0.78       & 0.72  & 10.62      & 2.02                     & 5.12                     \\
FSM20-maximal                      & 1.0     & 0.59       & 0.71       & 1.64       & 0.98  & 1.67       & 0.70                    & 1.98                    \\
FSM20                        & 1.0     & 0.29    &   0.71       & 1.64       & 0.98  & 0.83       & 0.35                    & 0.99                   \\
TNG100 (SF, $L^*$) & $0.3-3$  & $0.30-0.43$       & $0.81-0.96$       & $1.82-2.28$       & $1.49-1.83$  & $0.57-1.24$       & (0.35 - 0.81)                    & (1.05 - 2.73)                    \\
B+22 (isothermal)                  & 2.75     & 0.83          & 1.48          & 2.03       & 2.03     & 2.39       & 1.47                    & 4.55                    \\
B+22 (isentropic)                   & 2.75     & $0.45 - 0.83$          & $1.48-0.81$          & 2.03       & 1.22     & 2.39       & 1.47                    & 4.55                    \\
Z+24 (isothermal)                        & 1.4     & 0.51       & 2.56       & 1.26       & 1.26   & 4.54   & 1.18                      & 3.08                    \\ \hline
\end{tabular}
\tablecomments{The PLM parameters for FSM17 are for the gas distributions following the formation of the OVI component (see \S~\ref{sec:PLMFSM}). The PLM parameters for \citetalias{Bregman2022} are those for which $\eta_{\rm vir}\ge 1$ and hydrostatic equilibrium (HSE) is possible (see \S~\ref{sec:plm-b22}). }
\label{tab:PLM_parameters}
\end{table*}

%% file: tb_fsm_B22.tex
\renewcommand{\arraystretch}{1.2} 

\begin{table*}[ht]
\centering
\caption{Observed  \citetalias{Bregman2022}, \citetalias{Das2023}, \citetalias{Zhang2024}, Compton parameters, and FSM and TNG100 predictions}
\begin{tabular}{|l|c|c|c|c|c|}
\hline
              & $M_{\rm vir, 12}$& $y(0.2)~ E^{-2} / M_{\rm vir, 12}$     & $Y_{\perp, 500} ~E^{-2/3} / M_{\rm   vir, 12}^{5/3}$ & $Y_{500} ~E^{-2/3} / M_{\rm vir,   12}^{5/3}$ & $Y_{\rm vir} ~E^{-2/3} / M_{\rm vir,   12}^{5/3}$ \\
              &     &                           & $({\rm kpc^2})$                                & $({\rm kpc^2})$                        & $({\rm kpc^2})$                         \\ \hline

B+22 stack    & 2.75& $9.88 \pm 9.00 \times 10^{-9}$ & $5.99 \pm 4.33 \times 10^{-4}$                 & $4.66 \pm 3.49 \times 10^{-4}$         & $9.27 \pm 6.82 \times 10^{-4}$          \\
D+23        & 2.5         & $8.02 \pm 5.67 \times 10^{-9}$           & $ $                           & $2.70 \pm 1.49 \times10^{-4}$                  &  $ $\\
PLM Z+24    & 1.4       & $1.18 \times 10^{-8}$           & $4.85 \times 10^{-4}$                           & $2.85 \times 10^{-4}$                  & $9.53 \times 10^{-4}$                    \\ \hline     

Mean Observation & 2.22   & $9.58 \times 10^{-9}$  & $5.42 \times 10^{-4}$  & $3.77 \times 10^{-4}$  & $8.85 \times 10^{-4}$

\\ \hline

FSM17         & 1.0 & $2.00\times 10^{-8}$           & $9.54\times 10^{-4}$                           & $4.82 \times 10^{-4}$                 & $18.61\times 10^{-4}$                   \\
FSM20-maximal & 1.0 & $0.77\times 10^{-8}$           & $3.03\times 10^{-4}$                           & $2.11 \times 10^{-4}$                  & $4.68\times 10^{-4}$                    \\
FSM20         & 1.0 & $0.39\times 10^{-8}$           & $1.52\times 10^{-4}$                           & $1.06 \times 10^{-4}$                  & $2.34\times 10^{-4}$                    \\ \hline
TNG100 (SF)   & 1.0        & $7.22 \pm 2.72 \times 10^{-9}$           & $2.81 \pm 1.02 \times 10^{-4}$                           & $2.08 \pm 0.77 \times 10^{-4}$                  & $3.47 \pm 1.20 \times 10^{-4}$                    \\      
PLM TNG100 (SF) & 1.0        & $6.31 \times 10^{-9}$           & $2.60 \times 10^{-4}$                           & $2.11 \times 10^{-4}$                  & $3.82 \times 10^{-4}$                    \\     
TNG100 (SF)   & 2.75        & $5.33 \pm 2.05 \times 10^{-9}$           & $1.99 \pm 0.71 \times 10^{-4}$                           & $1.45 \pm 0.64 \times 10^{-4}$                  & $2.92 \pm 0.94 \times 10^{-4}$                    \\      
PLM TNG100 (SF) & 2.75        & $4.15 \times 10^{-9}$           & $1.68 \times 10^{-4}$                          & $1.22 \times 10^{-4}$                 & $2.69 \times 10^{-4}$                    \\      \hline
\multicolumn{6}{l}{\small * \citetalias{Bregman2022} and \citetalias{Das2023} uncertainties are derived from measurement errors, while the TNG100 uncertainties are dispersions.}

\end{tabular}
\label{tab:fsm_B22}
\end{table*}

%% file: pressure_table.tex
\renewcommand{\arraystretch}{1.2} 

\begin{table}[ht]
\centering
\caption{Parameters used for the pressure models adopted by \citetalias{Das2023} and \cite{Hill2018}.}
\begin{tabular}{|c|c|c|}
\hline
      Work    & \cite{Das2023}            & \cite{Hill2018}             \\ \hline
Reference & \cite{Arnaud2010} & \cite{Battaglia2012} \\ \hline
$r_0$     & $0.85 ~R_{500} $                & $ 0.50 ~R_{200}$                    \\
$\alpha$  & 1.05                              & 1.00                                     \\
$\beta$   & -4.93                             & -4.35                                 \\
$\gamma$  & -0.31                             & -0.30                                  \\ \hline
\end{tabular}
\label{tab:pressure_parameters}
\end{table}

%% file: text_amiel.bbl
\begin{thebibliography}{}
\makeatletter
\relax
\def\mn@urlcharsother{\let\do\@makeother \do\$\do\&\do\#\do\^\do\_\do\%\do\~}
\def\mn@doi{\begingroup\mn@urlcharsother \@ifnextchar [ {\mn@doi@}
  {\mn@doi@[]}}
\def\mn@doi@[#1]#2{\def\@tempa{#1}\ifx\@tempa\@empty \href
  {http://dx.doi.org/#2} {doi:#2}\else \href {http://dx.doi.org/#2} {#1}\fi
  \endgroup}
\def\mn@eprint#1#2{\mn@eprint@#1:#2::\@nil}
\def\mn@eprint@arXiv#1{\href {http://arxiv.org/abs/#1} {{\tt arXiv:#1}}}
\def\mn@eprint@dblp#1{\href {http://dblp.uni-trier.de/rec/bibtex/#1.xml}
  {dblp:#1}}
\def\mn@eprint@#1:#2:#3:#4\@nil{\def\@tempa {#1}\def\@tempb {#2}\def\@tempc
  {#3}\ifx \@tempc \@empty \let \@tempc \@tempb \let \@tempb \@tempa \fi \ifx
  \@tempb \@empty \def\@tempb {arXiv}\fi \@ifundefined
  {mn@eprint@\@tempb}{\@tempb:\@tempc}{\expandafter \expandafter \csname
  mn@eprint@\@tempb\endcsname \expandafter{\@tempc}}}

\bibitem[\protect\citeauthoryear{{Aiola} et~al.,}{{Aiola}
  et~al.}{2020}]{Aiola2020}
{Aiola} S.,  et~al., 2020, \mn@doi [\jcap] {10.1088/1475-7516/2020/12/047},
  \href {https://ui.adsabs.harvard.edu/abs/2020JCAP...12..047A} {2020, 047}

\bibitem[\protect\citeauthoryear{{Anderson} \& {Bregman}}{{Anderson} \&
  {Bregman}}{2010}]{Anderson10}
{Anderson} M.~E.,  {Bregman} J.~N.,  2010, \mn@doi [\apj]
  {10.1088/0004-637X/714/1/320}, \href
  {http://adsabs.harvard.edu/abs/2010ApJ...714..320A} {714, 320}

\bibitem[\protect\citeauthoryear{{Arnaud}, {Pratt}, {Piffaretti},
  {B{\"o}hringer}, {Croston}  \& {Pointecouteau}}{{Arnaud}
  et~al.}{2010}]{Arnaud2010}
{Arnaud} M.,  {Pratt} G.~W.,  {Piffaretti} R.,  {B{\"o}hringer} H.,  {Croston}
  J.~H.,   {Pointecouteau} E.,  2010, \mn@doi [\aap]
  {10.1051/0004-6361/200913416}, \href
  {https://ui.adsabs.harvard.edu/abs/2010A&A...517A..92A} {517, A92}

\bibitem[\protect\citeauthoryear{{Asplund}, {Grevesse}, {Sauval}  \&
  {Scott}}{{Asplund} et~al.}{2009}]{Asplund2009}
{Asplund} M.,  {Grevesse} N.,  {Sauval} A.~J.,   {Scott} P.,  2009, \mn@doi
  [ARA{\&}A] {10.1146/annurev.astro.46.060407.145222}, \href
  {https://ui.adsabs.harvard.edu/abs/2009ARA&A..47..481A} {47, 481}

\bibitem[\protect\citeauthoryear{{Battaglia}, {Bond}, {Pfrommer}  \&
  {Sievers}}{{Battaglia} et~al.}{2012}]{Battaglia2012}
{Battaglia} N.,  {Bond} J.~R.,  {Pfrommer} C.,   {Sievers} J.~L.,  2012,
  \mn@doi [\apj] {10.1088/0004-637X/758/2/75}, \href
  {https://ui.adsabs.harvard.edu/abs/2012ApJ...758...75B} {758, 75}

\bibitem[\protect\citeauthoryear{{Bennett} et~al.,}{{Bennett}
  et~al.}{2013}]{Bennett2013}
{Bennett} C.~L.,  et~al., 2013, \mn@doi [\apjs] {10.1088/0067-0049/208/2/20},
  \href {https://ui.adsabs.harvard.edu/abs/2013ApJS..208...20B} {208, 20}

\bibitem[\protect\citeauthoryear{{Bilicki} et~al.,}{{Bilicki}
  et~al.}{2016}]{Bilicki2016}
{Bilicki} M.,  et~al., 2016, \mn@doi [\apjs] {10.3847/0067-0049/225/1/5}, \href
  {https://ui.adsabs.harvard.edu/abs/2016ApJS..225....5B} {225, 5}

\bibitem[\protect\citeauthoryear{{Birkinshaw}}{{Birkinshaw}}{1999}]{Birkinshaw1999}
{Birkinshaw} M.,  1999, \mn@doi [\physrep] {10.1016/S0370-1573(98)00080-5},
  \href {https://ui.adsabs.harvard.edu/abs/1999PhR...310...97B} {310, 97}

\bibitem[\protect\citeauthoryear{{Bland-Hawthorn} \&
  {Gerhard}}{{Bland-Hawthorn} \& {Gerhard}}{2016}]{BHG16}
{Bland-Hawthorn} J.,  {Gerhard} O.,  2016, \mn@doi [\araa]
  {10.1146/annurev-astro-081915-023441}, \href
  {http://adsabs.harvard.edu/abs/2016ARA%26A..54..529B} {54, 529}

\bibitem[\protect\citeauthoryear{{Bleem} et~al.,}{{Bleem}
  et~al.}{2015}]{Bleem2015}
{Bleem} L.~E.,  et~al., 2015, \mn@doi [\apjs] {10.1088/0067-0049/216/2/27},
  \href {https://ui.adsabs.harvard.edu/abs/2015ApJS..216...27B} {216, 27}

\bibitem[\protect\citeauthoryear{{Blitz} \& {Robishaw}}{{Blitz} \&
  {Robishaw}}{2000}]{Blitz00}
{Blitz} L.,  {Robishaw} T.,  2000, \mn@doi [\apj] {10.1086/309457}, \href
  {https://ui.adsabs.harvard.edu/abs/2000ApJ...541..675B} {541, 675}

\bibitem[\protect\citeauthoryear{{Bregman} \& {Lloyd-Davies}}{{Bregman} \&
  {Lloyd-Davies}}{2007}]{Bregman07}
{Bregman} J.~N.,  {Lloyd-Davies} E.~J.,  2007, \mn@doi [\apj] {10.1086/521321},
  \href {https://ui.adsabs.harvard.edu/abs/2007ApJ...669..990B} {669, 990}

\bibitem[\protect\citeauthoryear{{Bregman}, {Hodges-Kluck}, {Qu}, {Pratt}, {Li}
   \& {Yun}}{{Bregman} et~al.}{2022}]{Bregman2022}
{Bregman} J.~N.,  {Hodges-Kluck} E.,  {Qu} Z.,  {Pratt} C.,  {Li} J.-T.,
  {Yun} Y.,  2022, \mn@doi [\apj] {10.3847/1538-4357/ac51de}, \href
  {https://ui.adsabs.harvard.edu/abs/2022ApJ...928...14B} {928, 14}

\bibitem[\protect\citeauthoryear{{Bryan} \& {Norman}}{{Bryan} \&
  {Norman}}{1998}]{Bryan1998}
{Bryan} G.~L.,  {Norman} M.~L.,  1998, \mn@doi [\apj] {10.1086/305262}, \href
  {https://ui.adsabs.harvard.edu/abs/1998ApJ...495...80B} {495, 80}

\bibitem[\protect\citeauthoryear{{Carlstrom}, {Holder}  \& {Reese}}{{Carlstrom}
  et~al.}{2002}]{carlstrom2002}
{Carlstrom} J.~E.,  {Holder} G.~P.,   {Reese} E.~D.,  2002, \mn@doi [\araa]
  {10.1146/annurev.astro.40.060401.093803}, \href
  {https://ui.adsabs.harvard.edu/abs/2002ARA&A..40..643C} {40, 643}

\bibitem[\protect\citeauthoryear{{Cavaliere} \& {Fusco-Femiano}}{{Cavaliere} \&
  {Fusco-Femiano}}{1976}]{Cavaliere1976}
{Cavaliere} A.,  {Fusco-Femiano} R.,  1976, \aap, \href
  {https://ui.adsabs.harvard.edu/abs/1976A&A....49..137C} {49, 137}

\bibitem[\protect\citeauthoryear{{Cook} et~al.,}{{Cook} et~al.}{2023}]{Cook23}
{Cook} A.~M.,  et~al., 2023, \mn@doi [\apj] {10.3847/1538-4357/acbbd0}, \href
  {https://ui.adsabs.harvard.edu/abs/2023ApJ...946...58C} {946, 58}

\bibitem[\protect\citeauthoryear{{Das}, {Mathur}, {Nicastro}  \&
  {Krongold}}{{Das} et~al.}{2019}]{Das19a}
{Das} S.,  {Mathur} S.,  {Nicastro} F.,   {Krongold} Y.,  2019, \mn@doi [\apjl]
  {10.3847/2041-8213/ab3b09}, \href
  {https://ui.adsabs.harvard.edu/abs/2019ApJ...882L..23D} {882, L23}

\bibitem[\protect\citeauthoryear{{Das}, {Chiang}  \& {Mathur}}{{Das}
  et~al.}{2023}]{Das2023}
{Das} S.,  {Chiang} Y.-K.,   {Mathur} S.,  2023, \mn@doi [\apj]
  {10.3847/1538-4357/acd764}, \href
  {https://ui.adsabs.harvard.edu/abs/2023ApJ...951..125D} {951, 125}

\bibitem[\protect\citeauthoryear{{Davies}, {Crain}, {McCarthy}, {Oppenheimer},
  {Schaye}, {Schaller}  \& {McAlpine}}{{Davies} et~al.}{2019a}]{Davies2019}
{Davies} J.~J.,  {Crain} R.~A.,  {McCarthy} I.~G.,  {Oppenheimer} B.~D.,
  {Schaye} J.,  {Schaller} M.,   {McAlpine} S.,  2019a, \mn@doi [\mnras]
  {10.1093/mnras/stz635}, \href
  {https://ui.adsabs.harvard.edu/abs/2019MNRAS.485.3783D} {485, 3783}

\bibitem[\protect\citeauthoryear{Davies, Crain, Oppenheimer  \& Schaye}{Davies
  et~al.}{2019b}]{Davies2020}
Davies J.~J.,  Crain R.~A.,  Oppenheimer B.~D.,   Schaye J.,  2019b, \mn@doi
  [\mnras] {10.1093/mnras/stz3201}, 491, 4462

\bibitem[\protect\citeauthoryear{{Dedes} \& {Kalberla}}{{Dedes} \&
  {Kalberla}}{2010}]{Dedes10}
{Dedes} L.,  {Kalberla} P.~W.~M.,  2010, \mn@doi [\aap]
  {10.1051/0004-6361/200912673}, \href
  {http://adsabs.harvard.edu/abs/2010A%26A...509A..60D} {509, A60}

\bibitem[\protect\citeauthoryear{{Dutton} \& {Macci{\`o}}}{{Dutton} \&
  {Macci{\`o}}}{2014}]{Dutton2014}
{Dutton} A.~A.,  {Macci{\`o}} A.~V.,  2014, \mn@doi [\mnras]
  {10.1093/mnras/stu742}, \href
  {https://ui.adsabs.harvard.edu/abs/2014MNRAS.441.3359D} {441, 3359}

\bibitem[\protect\citeauthoryear{{Faerman} \& {Werk}}{{Faerman} \&
  {Werk}}{2023}]{Faerman2023}
{Faerman} Y.,  {Werk} J.~K.,  2023, \mn@doi [\apj] {10.3847/1538-4357/acf217},
  \href {https://ui.adsabs.harvard.edu/abs/2023ApJ...956...92F} {956, 92}

\bibitem[\protect\citeauthoryear{{Faerman}, {Sternberg}  \& {McKee}}{{Faerman}
  et~al.}{2017}]{Faerman2017}
{Faerman} Y.,  {Sternberg} A.,   {McKee} C.~F.,  2017, \mn@doi [\apj]
  {10.3847/1538-4357/835/1/52}, \href
  {https://ui.adsabs.harvard.edu/abs/2017ApJ...835...52F} {835, 52}

\bibitem[\protect\citeauthoryear{{Faerman}, {Sternberg}  \& {McKee}}{{Faerman}
  et~al.}{2020}]{Faerman2020}
{Faerman} Y.,  {Sternberg} A.,   {McKee} C.~F.,  2020, \mn@doi [ApJ]
  {10.3847/1538-4357/ab7ffc}, \href
  {https://ui.adsabs.harvard.edu/abs/2020ApJ...893...82F} {893, 82}

\bibitem[\protect\citeauthoryear{{Fang}, {Buote}, {Bullock}  \& {Ma}}{{Fang}
  et~al.}{2015}]{Fang15}
{Fang} T.,  {Buote} D.,  {Bullock} J.,   {Ma} R.,  2015, \mn@doi [\apjs]
  {10.1088/0067-0049/217/2/21}, \href
  {http://adsabs.harvard.edu/abs/2015ApJS..217...21F} {217, 21}

\bibitem[\protect\citeauthoryear{{Foreman-Mackey}, {Hogg}, {Lang}  \&
  {Goodman}}{{Foreman-Mackey} et~al.}{2013}]{Foreman-Mackey2013}
{Foreman-Mackey} D.,  {Hogg} D.~W.,  {Lang} D.,   {Goodman} J.,  2013, \mn@doi
  [\pasp] {10.1086/670067}, \href
  {https://ui.adsabs.harvard.edu/abs/2013PASP..125..306F} {125, 306}

\bibitem[\protect\citeauthoryear{{Gnat}}{{Gnat}}{2017}]{Gnat2017}
{Gnat} O.,  2017, \mn@doi [\apjs] {10.3847/1538-4365/228/2/11}, \href
  {https://ui.adsabs.harvard.edu/abs/2017ApJS..228...11G} {228, 11}

\bibitem[\protect\citeauthoryear{{Grcevich} \& {Putman}}{{Grcevich} \&
  {Putman}}{2009}]{Grcevich09}
{Grcevich} J.,  {Putman} M.~E.,  2009, \mn@doi [\apj]
  {10.1088/0004-637X/696/1/385}, \href
  {https://ui.adsabs.harvard.edu/abs/2009ApJ...696..385G} {696, 385}

\bibitem[\protect\citeauthoryear{Guo et~al.,}{Guo et~al.}{2011}]{Guo2011}
Guo Q.,  et~al., 2011, \mn@doi [\mnras] {10.1111/j.1365-2966.2010.18114.x},
  413, 101

\bibitem[\protect\citeauthoryear{{Gupta}, {Mathur}, {Krongold}, {Nicastro}  \&
  {Galeazzi}}{{Gupta} et~al.}{2012}]{Gupta12}
{Gupta} A.,  {Mathur} S.,  {Krongold} Y.,  {Nicastro} F.,   {Galeazzi} M.,
  2012, \mn@doi [\apjl] {10.1088/2041-8205/756/1/L8}, \href
  {http://adsabs.harvard.edu/abs/2012ApJ...756L...8G} {756, L8}

\bibitem[\protect\citeauthoryear{{Hasselfield} et~al.,}{{Hasselfield}
  et~al.}{2013}]{Hasselfield2013}
{Hasselfield} M.,  et~al., 2013, \mn@doi [\jcap]
  {10.1088/1475-7516/2013/07/008}, \href
  {https://ui.adsabs.harvard.edu/abs/2013JCAP...07..008H} {2013, 008}

\bibitem[\protect\citeauthoryear{{Hastings}}{{Hastings}}{1970}]{Hastings1970}
{Hastings} W.~K.,  1970, \mn@doi [Biometrika] {10.1093/biomet/57.1.97}, \href
  {https://ui.adsabs.harvard.edu/abs/1970Bimka..57...97H} {57, 97}

\bibitem[\protect\citeauthoryear{{Hill}, {Baxter}, {Lidz}, {Greco}  \&
  {Jain}}{{Hill} et~al.}{2018}]{Hill2018}
{Hill} J.~C.,  {Baxter} E.~J.,  {Lidz} A.,  {Greco} J.~P.,   {Jain} B.,  2018,
  \mn@doi [\prd] {10.1103/PhysRevD.97.083501}, \href
  {https://ui.adsabs.harvard.edu/abs/2018PhRvD..97h3501H} {97, 083501}

\bibitem[\protect\citeauthoryear{{Johnson}, {Chen}  \& {Mulchaey}}{{Johnson}
  et~al.}{2015}]{Johnson15}
{Johnson} S.~D.,  {Chen} H.-W.,   {Mulchaey} J.~S.,  2015, \mn@doi [\mnras]
  {10.1093/mnras/stv553}, \href
  {http://adsabs.harvard.edu/abs/2015MNRAS.449.3263J} {449, 3263}

\bibitem[\protect\citeauthoryear{{Kaiser}}{{Kaiser}}{1986}]{Kaiser1986}
{Kaiser} N.,  1986, \mn@doi [\mnras] {10.1093/mnras/222.2.323}, \href
  {https://ui.adsabs.harvard.edu/abs/1986MNRAS.222..323K} {222, 323}

\bibitem[\protect\citeauthoryear{{Karmakar}, {Genel}  \&
  {Somerville}}{{Karmakar} et~al.}{2023}]{Karmakar2023}
{Karmakar} T.,  {Genel} S.,   {Somerville} R.~S.,  2023, \mn@doi [\mnras]
  {10.1093/mnras/stad178}, \href
  {https://ui.adsabs.harvard.edu/abs/2023MNRAS.520.1630K} {520, 1630}

\bibitem[\protect\citeauthoryear{{Khaire} \& {Srianand}}{{Khaire} \&
  {Srianand}}{2019}]{KS19}
{Khaire} V.,  {Srianand} R.,  2019, \mn@doi [\mnras] {10.1093/mnras/stz174},
  \href {https://ui.adsabs.harvard.edu/abs/2019MNRAS.484.4174K} {484, 4174}

\bibitem[\protect\citeauthoryear{{Kitayama} et~al.,}{{Kitayama}
  et~al.}{2016}]{Kitayama2016}
{Kitayama} T.,  et~al., 2016, \mn@doi [\pasj] {10.1093/pasj/psw082}, \href
  {https://ui.adsabs.harvard.edu/abs/2016PASJ...68...88K} {68, 88}

\bibitem[\protect\citeauthoryear{{Licquia} \& {Newman}}{{Licquia} \&
  {Newman}}{2016}]{Licquia2016_MW_M_star}
{Licquia} T.~C.,  {Newman} J.~A.,  2016, \mn@doi [\apj]
  {10.3847/0004-637X/831/1/71}, \href
  {https://ui.adsabs.harvard.edu/abs/2016ApJ...831...71L} {831, 71}

\bibitem[\protect\citeauthoryear{{Lim}, {Barnes}, {Vogelsberger}, {Mo},
  {Nelson}, {Pillepich}, {Dolag}  \& {Marinacci}}{{Lim} et~al.}{2021}]{Lim2021}
{Lim} S.~H.,  {Barnes} D.,  {Vogelsberger} M.,  {Mo} H.~J.,  {Nelson} D.,
  {Pillepich} A.,  {Dolag} K.,   {Marinacci} F.,  2021, \mn@doi [\mnras]
  {10.1093/mnras/stab1172}, \href
  {https://ui.adsabs.harvard.edu/abs/2021MNRAS.504.5131L} {504, 5131}

\bibitem[\protect\citeauthoryear{{Marinacci} et~al.,}{{Marinacci}
  et~al.}{2018}]{Marinacci2018}
{Marinacci} F.,  et~al., 2018, \mn@doi [MNRAS] {10.1093/mnras/sty2206}, \href
  {https://ui.adsabs.harvard.edu/abs/2018MNRAS.480.5113M} {480, 5113}

\bibitem[\protect\citeauthoryear{{Miller} \& {Bregman}}{{Miller} \&
  {Bregman}}{2015}]{Miller15}
{Miller} M.~J.,  {Bregman} J.~N.,  2015, \mn@doi [\apj]
  {10.1088/0004-637X/800/1/14}, \href
  {http://adsabs.harvard.edu/abs/2015ApJ...800...14M} {800, 14}

\bibitem[\protect\citeauthoryear{Moster, Somerville, Maulbetsch, van~den Bosch,
  Macci{\`{o}}, Naab  \& Oser}{Moster et~al.}{2010}]{Moster2010}
Moster B.~P.,  Somerville R.~S.,  Maulbetsch C.,  van~den Bosch F.~C.,
  Macci{\`{o}} A.~V.,  Naab T.,   Oser L.,  2010, \mn@doi [ApJ]
  {10.1088/0004-637x/710/2/903}, 710, 903

\bibitem[\protect\citeauthoryear{{Moster}, {Naab}  \& {White}}{{Moster}
  et~al.}{2013}]{Moster2013}
{Moster} B.~P.,  {Naab} T.,   {White} S. D.~M.,  2013, \mn@doi [\mnras]
  {10.1093/mnras/sts261}, \href
  {https://ui.adsabs.harvard.edu/abs/2013MNRAS.428.3121M} {428, 3121}

\bibitem[\protect\citeauthoryear{{Moster}, {Naab}  \& {White}}{{Moster}
  et~al.}{2018}]{Moster2018}
{Moster} B.~P.,  {Naab} T.,   {White} S. D.~M.,  2018, \mn@doi [\mnras]
  {10.1093/mnras/sty655}, \href
  {https://ui.adsabs.harvard.edu/abs/2018MNRAS.477.1822M} {477, 1822}

\bibitem[\protect\citeauthoryear{{Mroczkowski} et~al.,}{{Mroczkowski}
  et~al.}{2019}]{Mroczkowski2019}
{Mroczkowski} T.,  et~al., 2019, \mn@doi [\ssr] {10.1007/s11214-019-0581-2},
  \href {https://ui.adsabs.harvard.edu/abs/2019SSRv..215...17M} {215, 17}

\bibitem[\protect\citeauthoryear{{Nagai}, {Kravtsov}  \& {Vikhlinin}}{{Nagai}
  et~al.}{2007}]{Nagai2007}
{Nagai} D.,  {Kravtsov} A.~V.,   {Vikhlinin} A.,  2007, \mn@doi [\apj]
  {10.1086/521328}, \href
  {https://ui.adsabs.harvard.edu/abs/2007ApJ...668....1N} {668, 1}

\bibitem[\protect\citeauthoryear{{Naiman} et~al.,}{{Naiman}
  et~al.}{2018}]{Naiman2018}
{Naiman} J.~P.,  et~al., 2018, \mn@doi [\mnras] {10.1093/mnras/sty618}, \href
  {https://ui.adsabs.harvard.edu/abs/2018MNRAS.477.1206N} {477, 1206}

\bibitem[\protect\citeauthoryear{{Navarro}, {Frenk}  \& {White}}{{Navarro}
  et~al.}{1997}]{Navarro1997}
{Navarro} J.~F.,  {Frenk} C.~S.,   {White} S. D.~M.,  1997, \mn@doi [\apj]
  {10.1086/304888}, \href
  {https://ui.adsabs.harvard.edu/abs/1997ApJ...490..493N} {490, 493}

\bibitem[\protect\citeauthoryear{{Nelson} et~al.,}{{Nelson}
  et~al.}{2018}]{Nelson2018}
{Nelson} D.,  et~al., 2018, \mn@doi [MNRAS] {10.1093/mnras/stx3040}, \href
  {https://ui.adsabs.harvard.edu/abs/2018MNRAS.475..624N} {475, 624}

\bibitem[\protect\citeauthoryear{{Oppenheimer} et~al.,}{{Oppenheimer}
  et~al.}{2020}]{Oppenheimer2020}
{Oppenheimer} B.~D.,  et~al., 2020, \mn@doi [\apjl] {10.3847/2041-8213/ab846f},
  \href {https://ui.adsabs.harvard.edu/abs/2020ApJ...893L..24O} {893, L24}

\bibitem[\protect\citeauthoryear{{Osato} \& {Nagai}}{{Osato} \&
  {Nagai}}{2023}]{Osato2023}
{Osato} K.,  {Nagai} D.,  2023, \mn@doi [\mnras] {10.1093/mnras/stac3669},
  \href {https://ui.adsabs.harvard.edu/abs/2023MNRAS.519.2069O} {519, 2069}

\bibitem[\protect\citeauthoryear{{Ostriker}, {Bode}  \& {Babul}}{{Ostriker}
  et~al.}{2005}]{Ostriker2005}
{Ostriker} J.~P.,  {Bode} P.,   {Babul} A.,  2005, \mn@doi [\apj]
  {10.1086/497122}, \href
  {https://ui.adsabs.harvard.edu/abs/2005ApJ...634..964O} {634, 964}

\bibitem[\protect\citeauthoryear{{Pillepich} et~al.,}{{Pillepich}
  et~al.}{2018}]{Pillepich2018_2}
{Pillepich} A.,  et~al., 2018, \mn@doi [MNRAS] {10.1093/mnras/stx3112}, \href
  {https://ui.adsabs.harvard.edu/abs/2018MNRAS.475..648P} {475, 648}

\bibitem[\protect\citeauthoryear{{Plagge} et~al.,}{{Plagge}
  et~al.}{2010}]{Plagge2010}
{Plagge} T.,  et~al., 2010, \mn@doi [\apj] {10.1088/0004-637X/716/2/1118},
  \href {https://ui.adsabs.harvard.edu/abs/2010ApJ...716.1118P} {716, 1118}

\bibitem[\protect\citeauthoryear{{Planck Collaboration} et~al.,}{{Planck
  Collaboration} et~al.}{2013}]{Planck2013}
{Planck Collaboration} et~al., 2013, \mn@doi [\aap]
  {10.1051/0004-6361/201220941}, \href
  {https://ui.adsabs.harvard.edu/abs/2013A&A...557A..52P} {557, A52}

\bibitem[\protect\citeauthoryear{{Planck Collaboration} et~al.,}{{Planck
  Collaboration} et~al.}{2016}]{Planck2015b}
{Planck Collaboration} et~al., 2016, \mn@doi [\aap]
  {10.1051/0004-6361/201525826}, \href
  {https://ui.adsabs.harvard.edu/abs/2016A&A...594A..22P} {594, A22}

\bibitem[\protect\citeauthoryear{{Planck Collaboration} et~al.,}{{Planck
  Collaboration} et~al.}{2020a}]{Planck2020b}
{Planck Collaboration} et~al., 2020a, \mn@doi [\aap]
  {10.1051/0004-6361/201833910}, \href
  {https://ui.adsabs.harvard.edu/abs/2020A&A...641A...6P} {641, A6}

\bibitem[\protect\citeauthoryear{{Planck Collaboration} et~al.,}{{Planck
  Collaboration} et~al.}{2020b}]{Planck2020}
{Planck Collaboration} et~al., 2020b, \mn@doi [\aap]
  {10.1051/0004-6361/202038073}, \href
  {https://ui.adsabs.harvard.edu/abs/2020A&A...643A..42P} {643, A42}

\bibitem[\protect\citeauthoryear{{Platts}, {Prochaska}  \& {Law}}{{Platts}
  et~al.}{2020}]{Platts20}
{Platts} E.,  {Prochaska} J.~X.,   {Law} C.~J.,  2020, \mn@doi [\apjl]
  {10.3847/2041-8213/ab930a}, \href
  {https://ui.adsabs.harvard.edu/abs/2020ApJ...895L..49P} {895, L49}

\bibitem[\protect\citeauthoryear{{Posti} \& {Helmi}}{{Posti} \&
  {Helmi}}{2019}]{Posti19}
{Posti} L.,  {Helmi} A.,  2019, \mn@doi [\aap] {10.1051/0004-6361/201833355},
  \href {https://ui.adsabs.harvard.edu/abs/2019A&A...621A..56P} {621, A56}

\bibitem[\protect\citeauthoryear{{Prochaska} \& {Zheng}}{{Prochaska} \&
  {Zheng}}{2019}]{PZ19}
{Prochaska} J.~X.,  {Zheng} Y.,  2019, \mn@doi [\mnras] {10.1093/mnras/stz261},
  \href {https://ui.adsabs.harvard.edu/abs/2019MNRAS.485..648P} {485, 648}

\bibitem[\protect\citeauthoryear{{Prochaska} et~al.,}{{Prochaska}
  et~al.}{2017}]{Prochaska17}
{Prochaska} J.~X.,  et~al., 2017, \mn@doi [\apj] {10.3847/1538-4357/aa6007},
  \href {https://ui.adsabs.harvard.edu/abs/2017ApJ...837..169P} {837, 169}

\bibitem[\protect\citeauthoryear{{Putman}, {Peek}  \& {Joung}}{{Putman}
  et~al.}{2012}]{Putman12}
{Putman} M.~E.,  {Peek} J.~E.~G.,   {Joung} M.~R.,  2012, \mn@doi [\araa]
  {10.1146/annurev-astro-081811-125612}, \href
  {http://adsabs.harvard.edu/abs/2012ARA%26A..50..491P} {50, 491}

\bibitem[\protect\citeauthoryear{{Qu} \& {Bregman}}{{Qu} \&
  {Bregman}}{2018}]{Qu2018}
{Qu} Z.,  {Bregman} J.~N.,  2018, \mn@doi [\apj] {10.3847/1538-4357/aaafd4},
  \href {https://ui.adsabs.harvard.edu/abs/2018ApJ...856....5Q} {856, 5}

\bibitem[\protect\citeauthoryear{{Ramesh}, {Nelson}  \& {Pillepich}}{{Ramesh}
  et~al.}{2023}]{Ramesh2023}
{Ramesh} R.,  {Nelson} D.,   {Pillepich} A.,  2023, \mn@doi [\mnras]
  {10.1093/mnras/stac3524}, \href
  {https://ui.adsabs.harvard.edu/abs/2023MNRAS.518.5754R} {518, 5754}

\bibitem[\protect\citeauthoryear{{Ravi} et~al.,}{{Ravi} et~al.}{2023}]{Ravi23}
{Ravi} V.,  et~al., 2023, \mn@doi [arXiv e-prints] {10.48550/arXiv.2301.01000},
  \href {https://ui.adsabs.harvard.edu/abs/2023arXiv230101000R} {p.
  arXiv:2301.01000}

\bibitem[\protect\citeauthoryear{{Reid} \& {Spergel}}{{Reid} \&
  {Spergel}}{2006}]{Reid2006}
{Reid} B.~A.,  {Spergel} D.~N.,  2006, \mn@doi [\apj] {10.1086/507862}, \href
  {https://ui.adsabs.harvard.edu/abs/2006ApJ...651..643R} {651, 643}

\bibitem[\protect\citeauthoryear{{Salem}, {Besla}, {Bryan}, {Putman}, {van der
  Marel}  \& {Tonnesen}}{{Salem} et~al.}{2015}]{Salem15}
{Salem} M.,  {Besla} G.,  {Bryan} G.,  {Putman} M.,  {van der Marel} R.~P.,
  {Tonnesen} S.,  2015, \mn@doi [\apj] {10.1088/0004-637X/815/1/77}, \href
  {http://adsabs.harvard.edu/abs/2015ApJ...815...77S} {815, 77}

\bibitem[\protect\citeauthoryear{{Savage} et~al.,}{{Savage}
  et~al.}{2003}]{Savage03}
{Savage} B.~D.,  et~al., 2003, \mn@doi [\apjs] {10.1086/346229}, \href
  {https://ui.adsabs.harvard.edu/abs/2003ApJS..146..125S} {146, 125}

\bibitem[\protect\citeauthoryear{{Sharma}, {McCourt}, {Parrish}  \&
  {Quataert}}{{Sharma} et~al.}{2012}]{Sharma2012}
{Sharma} P.,  {McCourt} M.,  {Parrish} I.~J.,   {Quataert} E.,  2012, \mn@doi
  [\mnras] {10.1111/j.1365-2966.2012.22050.x}, \href
  {https://ui.adsabs.harvard.edu/abs/2012MNRAS.427.1219S} {427, 1219}

\bibitem[\protect\citeauthoryear{{Singh}, {Majumdar}, {Nath}  \&
  {Silk}}{{Singh} et~al.}{2018}]{Singh2018}
{Singh} P.,  {Majumdar} S.,  {Nath} B.~B.,   {Silk} J.,  2018, \mn@doi [\mnras]
  {10.1093/mnras/sty1276}, \href
  {https://ui.adsabs.harvard.edu/abs/2018MNRAS.478.2909S} {478, 2909}

\bibitem[\protect\citeauthoryear{{Singh}, {Lau}, {Faerman}, {Stern}  \&
  {Nagai}}{{Singh} et~al.}{2024}]{Singh2024}
{Singh} P.,  {Lau} E.~T.,  {Faerman} Y.,  {Stern} J.,   {Nagai} D.,  2024,
  \mn@doi [\mnras] {10.1093/mnras/stae1695}, \href
  {https://ui.adsabs.harvard.edu/abs/2024MNRAS.532.3222S} {532, 3222}

\bibitem[\protect\citeauthoryear{{Springel} et~al.,}{{Springel}
  et~al.}{2018}]{Springel2018}
{Springel} V.,  et~al., 2018, \mn@doi [MNRAS] {10.1093/mnras/stx3304}, \href
  {https://ui.adsabs.harvard.edu/abs/2018MNRAS.475..676S} {475, 676}

\bibitem[\protect\citeauthoryear{{Stern}, {Fielding}, {Faucher-Gigu{\`e}re}  \&
  {Quataert}}{{Stern} et~al.}{2019}]{Stern2019}
{Stern} J.,  {Fielding} D.,  {Faucher-Gigu{\`e}re} C.-A.,   {Quataert} E.,
  2019, \mn@doi [arXiv e-prints] {10.48550/arXiv.1909.07402}, \href
  {https://ui.adsabs.harvard.edu/abs/2019arXiv190907402S} {p. arXiv:1909.07402}

\bibitem[\protect\citeauthoryear{{Sunyaev} \& {Zeldovich}}{{Sunyaev} \&
  {Zeldovich}}{1972}]{Sunyaev1972}
{Sunyaev} R.~A.,  {Zeldovich} Y.~B.,  1972, Comments on Astrophysics and Space
  Physics, \href {https://ui.adsabs.harvard.edu/abs/1972CoASP...4..173S} {4,
  173}

\bibitem[\protect\citeauthoryear{{Tumlinson} et~al.,}{{Tumlinson}
  et~al.}{2011}]{Tumlinson11}
{Tumlinson} J.,  et~al., 2011, \mn@doi [Science] {10.1126/science.1209840},
  \href {http://adsabs.harvard.edu/abs/2011Sci...334..948T} {334, 948}

\bibitem[\protect\citeauthoryear{{Tumlinson}, {Peeples}  \& {Werk}}{{Tumlinson}
  et~al.}{2017}]{Tumlinson2017}
{Tumlinson} J.,  {Peeples} M.~S.,   {Werk} J.~K.,  2017, \mn@doi [\araa]
  {10.1146/annurev-astro-091916-055240}, \href
  {https://ui.adsabs.harvard.edu/abs/2017ARA&A..55..389T} {55, 389}

\bibitem[\protect\citeauthoryear{{Vikram}, {Lidz}  \& {Jain}}{{Vikram}
  et~al.}{2017}]{Vikram2017}
{Vikram} V.,  {Lidz} A.,   {Jain} B.,  2017, \mn@doi [\mnras]
  {10.1093/mnras/stw3311}, \href
  {https://ui.adsabs.harvard.edu/abs/2017MNRAS.467.2315V} {467, 2315}

\bibitem[\protect\citeauthoryear{{Voit}}{{Voit}}{2019}]{Voit2019}
{Voit} G.~M.,  2019, \mn@doi [\apj] {10.3847/1538-4357/ab2bfd}, \href
  {https://ui.adsabs.harvard.edu/abs/2019ApJ...880..139V} {880, 139}

\bibitem[\protect\citeauthoryear{{Weinberger} et~al.,}{{Weinberger}
  et~al.}{2017}]{Weinberger2017}
{Weinberger} R.,  et~al., 2017, \mn@doi [\mnras] {10.1093/mnras/stw2944}, \href
  {https://ui.adsabs.harvard.edu/abs/2017MNRAS.465.3291W} {465, 3291}

\bibitem[\protect\citeauthoryear{{Werk}, {Prochaska}, {Thom}, {Tumlinson},
  {Tripp}, {O'Meara}  \& {Peeples}}{{Werk} et~al.}{2013}]{Werk13}
{Werk} J.~K.,  {Prochaska} J.~X.,  {Thom} C.,  {Tumlinson} J.,  {Tripp} T.~M.,
  {O'Meara} J.~M.,   {Peeples} M.~S.,  2013, \mn@doi [\apjs]
  {10.1088/0067-0049/204/2/17}, \href
  {https://ui.adsabs.harvard.edu/abs/2013ApJS..204...17W} {204, 17}

\bibitem[\protect\citeauthoryear{{Werk} et~al.,}{{Werk} et~al.}{2014}]{Werk14}
{Werk} J.~K.,  et~al., 2014, \mn@doi [\apj] {10.1088/0004-637X/792/1/8}, \href
  {https://ui.adsabs.harvard.edu/abs/2014ApJ...792....8W} {792, 8}

\bibitem[\protect\citeauthoryear{{Wolfire}, {McKee}, {Hollenbach}  \&
  {Tielens}}{{Wolfire} et~al.}{2003}]{Wolfire03}
{Wolfire} M.~G.,  {McKee} C.~F.,  {Hollenbach} D.,   {Tielens} A.~G.~G.~M.,
  2003, \mn@doi [\apj] {10.1086/368016}, \href
  {http://adsabs.harvard.edu/abs/2003ApJ...587..278W} {587, 278}

\bibitem[\protect\citeauthoryear{{Wright}, {Somerville}, {Lagos}, {Schaller},
  {Dav{\'e}}, {Angl{\'e}s-Alc{\'a}zar}  \& {Genel}}{{Wright}
  et~al.}{2024}]{Wright2024}
{Wright} R.~J.,  {Somerville} R.~S.,  {Lagos} C. d.~P.,  {Schaller} M.,
  {Dav{\'e}} R.,  {Angl{\'e}s-Alc{\'a}zar} D.,   {Genel} S.,  2024, \mn@doi
  [arXiv e-prints] {10.48550/arXiv.2402.08408}, \href
  {https://ui.adsabs.harvard.edu/abs/2024arXiv240208408W} {p. arXiv:2402.08408}

\bibitem[\protect\citeauthoryear{{Zhang} et~al.,}{{Zhang}
  et~al.}{2024}]{Zhang2024}
{Zhang} Y.,  et~al., 2024, \mn@doi [arXiv e-prints]
  {10.48550/arXiv.2401.17308}, \href
  {https://ui.adsabs.harvard.edu/abs/2024arXiv240117308Z} {p. arXiv:2401.17308}

\bibitem[\protect\citeauthoryear{{Zheng}, {Putman}, {Peek}  \& {Joung}}{{Zheng}
  et~al.}{2015}]{Zheng15}
{Zheng} Y.,  {Putman} M.~E.,  {Peek} J.~E.~G.,   {Joung} M.~R.,  2015, \mn@doi
  [\apj] {10.1088/0004-637X/807/1/103}, \href
  {http://adsabs.harvard.edu/abs/2015ApJ...807..103Z} {807, 103}

\bibitem[\protect\citeauthoryear{{Zu} \& {Mandelbaum}}{{Zu} \&
  {Mandelbaum}}{2015}]{Zu2015}
{Zu} Y.,  {Mandelbaum} R.,  2015, \mn@doi [\mnras] {10.1093/mnras/stv2062},
  \href {https://ui.adsabs.harvard.edu/abs/2015MNRAS.454.1161Z} {454, 1161}

\bibitem[\protect\citeauthoryear{{da Silva}, {Kay}, {Liddle}  \& {Thomas}}{{da
  Silva} et~al.}{2004}]{DaSilva2004}
{da Silva} A.~C.,  {Kay} S.~T.,  {Liddle} A.~R.,   {Thomas} P.~A.,  2004,
  \mn@doi [\mnras] {10.1111/j.1365-2966.2004.07463.x}, \href
  {https://ui.adsabs.harvard.edu/abs/2004MNRAS.348.1401D} {348, 1401}

\makeatother
\end{thebibliography}
